\documentclass[12pt,preprint]{aastex}
\usepackage{amsmath,amssymb}
\usepackage{hyperref}

\usepackage{graphicx,caption,subcaption}
\usepackage{natbib}
\newcommand{\disp}{\displaystyle}
\newcommand{\mbf}[1]{\overline{\mathbf{#1}}}
\newcommand{\fpar}[2]{\frac{\partial #1}{\partial #2}}
\newcommand{\ol}[1]{\overline{#1}}
\newcommand{\ve}{\boldsymbol{\mathcal{E}}}
\newcommand{\calf}{\boldsymbol{\mathcal F}}
\newcommand{\dif}{\mathrm{d}}
\begin{document}
\title{A global galactic dynamo with a corona constrained by relative helicity}
\author{A. Prasad\altaffilmark{1, \dagger} \and A.\ Mangalam\altaffilmark{1, \ddagger}}
\altaffiltext{1}{Indian Institute of Astrophysics, Sarjapur Road, Koramangala, Bangalore, 560034, India}
\email{avijeet@iiap.res.in$^\dagger$, mangalam@iiap.res.in$^\ddagger$}

\begin{abstract}
We present a model for a global axisymmetric turbulent dynamo operating in a galaxy with a corona which treats the parameters of turbulence driven by supernovae and by magneto-rotational instability under a common formalism. The nonlinear quenching of the dynamo is alleviated by inclusion of small-scale advective and diffusive magnetic helicity fluxes, which allow the  gauge-invariant magnetic helicity to be transferred outside the disk and consequently to build up a corona during the course of dynamo action. The time-dependent dynamo equations are expressed in a separable form and solved through an eigenvector expansion constructed using the steady-state solutions of the dynamo equation. The parametric evolution of the dynamo solution allows us to estimate the final structure of the global magnetic field and the saturated value of the turbulence parameter $\alpha_m$, even before solving the dynamical equations for evolution of magnetic fields in the disk and the corona, along with $\alpha$-quenching. We then solve these equations simultaneously to study the saturation of the large-scale magnetic field, its dependence on the small-scale magnetic helicity fluxes, and the corresponding evolution of the force-free field in the corona. The quadrupolar large-scale magnetic field in the disk is found to reach equipartition strength within a timescale of 1 Gyr. The large-scale magnetic field in the corona obtained is much weaker than the field inside the disk and has only a weak impact on the dynamo operation.
\end{abstract}


\section{Introduction}
Large-scale magnetic fields with strength of the order of 1-10 $\mu$G
have been observed in disk galaxies \citep[e.g.][]{1996ARA&A..34..155B,2010ASPC..438..197F,2012SSRv..166..215B,2013pss5.book..641B,2015ApJ...799...35V}.
The origin of these fields can be explained through mean-field dynamo theory 
\citep{1988ASSL..133.....R,1996ARA&A..34..155B,2005PhR...417....1B,2008RPPh...71d6901K}.
The conservation of magnetic helicity is one of the key constraints in these models, and also leads to the suppression of the $\alpha$-effect. The operation of the mean-field dynamo automatically leads to the growth of magnetic helicity of opposite signs between
the large-scale and small-scale magnetic fields \citep{1976JFM....77..321P,1994PhRvL..72.1651G,2002PhRvL..89z5007B}.
To avoid catastrophic suppression of the dynamo action ($\alpha$-quenching), the magnetic helicity due to the small-scale magnetic field
should be removed from the system \citep{2000MNRAS.318..724B,2001PhPl....8.2407B,2000A&A...361L...5K}.
Mechanisms suggested to produce these small-scale magnetic helicity fluxes are: advection of magnetic fields by
an outflow from the disk through the galactic fountain or wind
\citep{2006A&A...448L..33S,2007MNRAS.377..874S,2014MNRAS.443.1867C},
magnetic helicity flux from anisotropy of the turbulence produced by differential rotation
\citep{2001ApJ...550..752V,2004PhRvL..93t5001S,2006ApJ...648L..71S,2007MNRAS.377..874S,2014ApJ...780..144V}, and through diffusive flux \citep{2000A&A...361L...5K,2002A&A...387..453K,2009MNRAS.398.1414B,2010AN....331..130M,2014MNRAS.443.1867C}.
The outflow of magnetic helicity from the disk through dynamo operation
leads to the formation of a corona \citep{2000MNRAS.318..724B}.
According to Taylor's hypothesis, an infinitely conducting corona
would resistively relax to force-free field configurations under the constraint of global magnetic helicity conservation \citep{1960RvMP...32..914W,1974PhRvL..33.1139T,1983PhFl...26.3540F,1984JFM...147..133B,2000JApA...21..299M}.
In this paper, we include advective and diffusive fluxes in a simple semi-analytic model of a galactic dynamo that transfers magnetic helicity outside the disk and consequently builds up a force-free corona in course of time. We first solve the time-dependent dynamo equations by expressing them as separable in variables $r$ and $z$. The radial part of the dynamo equation is solved using an eigenvector expansion constructed using the steady-state solutions of the dynamo equation. The eigenvalues of the $z$ part of the solution are obtained by solving a fourth-order algebraic equation, which primarily depends upon the turbulence parameters and the magnetic helicity fluxes. Once the dynamo solutions are written out as parametric functions of these parameters, the evolution of the mean magnetic field is computed numerically by simultaneously solving the dynamical equations for $\alpha$-quenching and the growth of large-scale coronal magnetic helicity. Since the large-scale magnetic field lines cross the boundary between the galactic disk and the corona, the magnetic helicity of the large-scale magnetic field in the disk volume is not well defined. Hence we use the concept of gauge-invariant relative helicity \citep{1983PhFl...26.3540F,1984JFM...147..133B,1985ApJS...59..433B} to estimate the large-scale magnetic helicity in the disk and the corona. Here the gauge-invariant relative helicity for the cylindrical geometry is calculated using the prescription given in \citet{2006ApJ...646.1288L,2011PhPl...18e2901L}. We then investigate the dependence of the saturated mean magnetic field strength and its geometry on the magnetic helicity fluxes within the disk and the corresponding evolution of the force-free field in the corona.

The organization of the paper is as follows. In Sections 2 and 3, we present the theoretical formulation of the nonlinear mean-field dynamo and magnetic helicity transport. The solutions for the steady-state dynamo equation are discussed in Section 4. In Section 5, we present the semi-analytic formulation of the time-dependent problem and set up the equations for the evolution of the small-scale magnetic helicity in the disk and the large-scale coronal field. The solutions of the 
time-dependent dynamo equation are presented in Section 6, where we present their parametric dependences, and discuss the strength and geometry of the saturated mean field along with its dependence on the magnetic helicity flux terms. Finally, the summary and conclusions of the paper are discussed in Section 7. In addition, the detailed equations for magnetic helicity dynamics, derivations of various equations used in the main text, and discussions on the gauge invariance of absolute magnetic helicity for cylindrical geometry and magnetic helicity balance are presented in Appendices \ref{AppA} - \ref{a:corhel}.

\section{Nonlinear mean-field dynamo and magnetic helicity dynamics}
\label{s:mfd}
The magnetic and velocity fields in the mean-field magnetohydrodynamics
\citep{1980mfmd.book.....K} can be written as the sum of their mean and fluctuating parts:
\begin{equation}
 \bf{B}=\mbf{B}+\bf{b}; \quad \bf{U}=\mbf{U}+\bf{u}
\end{equation}
 with $\mbf{u}=0$ and $\mbf{b}=0$. The overbar formally denotes ensemble averaging, but for all practical purposes it can be thought of as spatial averaging over scales greater than the turbulent scale and less
 than the scale of the system \citep{1992JFM...238..325G,2013MNRAS.430L..40G}. 
The mean magnetic field generated from small-scale turbulent motion is then described by the mean-field induction equation \citep{1978mfge.book.....M,1980mfmd.book.....K}:
\begin{equation}
 \fpar{\mbf{B}}{t}=\nabla \times\left(\mbf{U}\times \mbf{B}-\eta \mbf{J}+\ve\right),\label{dy1}
\end{equation}
where the ohmic magnetic diffusivity is given by $\eta$ and $\disp{\mbf{J}= \frac{\nabla\times\mbf{B}}{\mu_0}}$ is the current density, with $\mu_0$ being the magnetic permeability of free space (hereafter we adopt units such that $\mu_0=1$).
Also, $\ve\equiv\ol{\bf{u}\times\bf{b}}=\alpha\mbf{B}-\eta_t\mbf{J}$, is the mean turbulent emf with turbulent transport coefficients $\alpha$ and $\eta_t$. Following the closure models, such as EDQNM \citep{1976JFM....77..321P} and $\tau$-approximation
\citep{2002PhRvL..89z5007B,2003GApFD..97..249R,2005A&A...439..835B}, we represent the effect of the small-scale magnetic field on the $\alpha$-effect as $ \alpha=\alpha_k+\alpha_m$
 \citep[e.g.][]{1994PhRvL..72.1651G,2005PhR...417....1B},
where $\disp{\alpha_k=-\frac{1}{3}\overline{\tau\bf{u}\cdot\nabla\times\bf{u}}}$ represents the kinetic
$\alpha$-effect related to the mean helicity of the random flow $\overline{\bf{u}\cdot\nabla\times\bf{u}}$,
and $\disp{\alpha_m=\frac{1}{3}\rho^{-1}\overline{\tau\bf{j}\cdot\bf{b}}}$ is the magnetic contribution to the
$\alpha$-effect. The fluid density is given by $\rho$, and $\tau$ is the correlation time of the turbulent flow
$\bf{u}$. 

The magnetic helicity dynamics using the above construction can be
represented by equations for the evolution of the large-scale magnetic 
helicity $\disp{\ol{H}_d=\int_V \mbf{A}\cdot\mbf{B}~\dif V}$ and the mean small-scale magnetic helicity $\disp{h_d= \int_V \ol{\bf{a}\cdot\bf{b}} ~\dif V}$. The equations for the evolution of $\ol{H}_d$ and $h_d$ can be written as \citep[see][p. 69, also see Appendix \ref{AppA} for a derivation]{2008tdad.conf...69M}
\begin{eqnarray}
 \frac{\dif \ol{H}_d}{\dif  t}&=&2\int_V \ve\cdot\mbf{B}~ \dif V -2 \int_V \eta \mbf{J}\cdot\mbf{B}~ \dif V-
\oint_S \mathbf{F}\cdot\hat{n}~\dif S\label{H}\\
\frac{\dif h_d}{\dif t}&=&-2\int_V \ve \cdot \mbf{B}~ \dif V- 2 \int_V \eta \ol{\mathbf{j}\cdot\mathbf{b}}~ \dif V -\oint_S \mathbf{f}\cdot \hat{n}~\dif S \label{h},
\end{eqnarray}
where $\hat{n}$ represents the normal to the surface $S$ enclosing volume $V$. The surface fluxes for $\ol{H}_d$ and $h_d$ are given by ${\bf F}$ and $\bf f$ respectively, which can be written as
\begin{eqnarray}
\mathbf{F}&=&(\eta \mbf{J}-\mbf{U}\times\mbf{B}-\ve-\nabla \varphi_1)\times\mbf{A}-2\varphi_1\mbf{B}\label{F} \\  
\mathbf{f}&=&\ol{(\bf{a}\cdot\mbf{B})\bf{u}}-\ol{(\bf{a}\cdot\bf{u})\mbf{B}}-\ol{(\bf{a}\cdot\mbf{U})\bf{b}}
+\ol{(\bf{a}\cdot\bf{b})\mbf{U}}-\ol{(\bf{a}\cdot\bf{u})\bf{b}}+\ol{(\bf{a}\cdot\bf{b})\bf{u}}\label{f} \nonumber \\ 
&&+\ol{\ve\times\bf{a}}+\eta \ol{\bf{j}\times\bf{a}}-\ol{\nabla\varphi_2\times\bf{a}}-2\ol{\varphi_2\mathbf{b}}
\end{eqnarray}
 where $\varphi_1$ and $\varphi_2$ are scalar functions of space, representing the gauge freedom for the large- and small-scale magnetic vector potentials respectively. Below, we discuss some terms in Equation \eqref{f} that have been identified and found to be significant in numerical simulations and we leave the investigation of the remaining terms in Equations (\ref{F}) and (\ref{f}) for future studies aided by numerical simulations. The relative contribution from each term in Equation \eqref{f} to the small-scale magnetic helicity transport equation has been explored recently through numerical simulations 
\citep{2014PhRvL.112l5003E,2014ApJ...780..144V}. \citet{2014ApJ...780..144V} found that the advective flux, $\ol{(\bf{a}\cdot\bf{b})\mbf{U}}$, is the most dominant term in Equation \eqref{f}, contributing about 80\% of the helicity flux. The next most dominant term in their analysis was $\ol{(\bf{a}\cdot\mbf{B})\bf{u}}$, which is part of the Vishniac--Cho flux \citep{2001ApJ...550..752V}, arising from the anisotropy of the turbulence. Apart from this, a term relating to a Fickian diffusion, $\sim\kappa\nabla\alpha_m$ \citep{2002A&A...387..453K,2009MNRAS.398.1414B}, has been argued to exist on physical and phenomenological grounds. It has been found in direct numerical simulations that $\kappa\approx0.3 \eta_t$ \citep{2010AN....331..130M,2011PhPl...18a2903C,2011A&A...535A..48H}. In this paper, we consider only the advective and diffusive flux terms. The effect of inclusion of the other flux terms from Equation \eqref{f} will be taken up in later studies. 

Usually, $\alpha_m$ is amplified in the dynamo action with a sign opposite to $\alpha_k$, which 
balances the kinetic $\alpha$-effect leading to saturation of the mean magnetic field.
To constrain $\alpha_m$, we write the
transport equation for small-scale magnetic helicity density $\chi$ using the magnetic helicity conservation equation given by
\citep{2006A&A...448L..33S,2006ApJ...648L..71S,2007MNRAS.377..874S,2014MNRAS.443.1867C}
\begin{equation}
 \fpar{\chi}{t}=-2 \ve\cdot \mbf{B}-2\eta\ol{\bf{j}\cdot\bf{b}}-\nabla\cdot\bf{f}, \label{chieq}
\end{equation}
where $\chi$ is approximately equal to $\ol{\bf{a}\cdot{b}}$, and $\bf{a}$ is the vector potential 
for $\bf{b}$ in the Coulomb gauge. The small-scale magnetic flux density is given by $\bf{f}$ (Equation \ref{f}) and $\bf{j}=\nabla\times\bf{b}$.
We can relate $\chi$ to $\alpha_m$ by arguing that $\alpha_m$ is mainly 
contributed by the integral scale of turbulence, $\disp{l_0=\frac{2\pi}{k_0}}$ 
\citep{2006A&A...448L..33S,2007MNRAS.377..874S}, which gives
$ \ol{\bf{j}\cdot{b}}\simeq l_0^{-2} \ol{\bf{a}\cdot{b}}$ and 
$\disp{ \alpha_m\simeq \frac{1}{3}\tau \frac{\chi}{\rho l_0^2}}$.
Introducing a reference (equipartition) magnetic field $B_{eq}^2\equiv\rho\mbf{u}^2$ and the magnetic Reynolds number
as $\disp{R_m=\frac{\eta_t}{\eta}}$, gives 
$\disp{\alpha_m\simeq\frac{\eta_t}{l_0^2 B_{eq}^2}\chi}$,
where $\disp{\eta_t\simeq \frac{1}{3}\ol{\tau \bf{u}^2}}$. 
We can rewrite Equation (\ref{chieq}) in terms of $\alpha_m$ \citep{2007MNRAS.377..874S} as
\begin{equation}
 \fpar{\alpha_m}{t}=-\frac{2\eta_t}{l_0^2}\left(\frac{\ve\cdot\mbf{B}}{B_{eq}^2}+\frac{\alpha_m}{R_m}\right)-\nabla\cdot\calf.\label{quench}
\end{equation}
Here $\disp{\boldsymbol{\mathcal{F}}=\frac{\eta_t}{l_0^2 B_{eq}^2}\mathbf{f}}$ is flux density of $\alpha_m$ taken as \citep{2014MNRAS.443.1867C}:
\begin{equation}
 \calf=\calf_a+\calf_d, \label{calf}
\end{equation}
where $\calf_a$ is the advective flux density given by \citep{2006A&A...448L..33S,2007MNRAS.377..874S,2012ApJ...754L..35H}
\begin{equation}
 \calf_a=\mbf{U}\alpha_m, \label{ru}
\end{equation}
and $\calf_d$ is the diffusive flux density given by \citep{2002A&A...387..453K,2009MNRAS.398.1414B}
\begin{equation}
 \calf_d=-\kappa\nabla\alpha_m, \quad \kappa\approx 0.3 \eta_t. \label{rk}
\end{equation}

As the dynamo operates within the disk, we allow for the large-scale magnetic helicity flux to be redistributed by advection in the disk but not escape (see Section \ref{s:alp} for details). The small-scale magnetic helicity flux on the other hand escapes through the vertical efflux and diffusion. As the adjustment timescale in the corona is small due to high conductivity, the corona is expected to be in a relaxed force-free state according to Taylor's hypothesis \citep{1974PhRvL..33.1139T,1994ApJ...434..509M}. This is also motivated by the corona of the Sun, where the magnetic field structure is dominated by nonlinear force-free fields \citep{2014ApJ...786...81P}. We use the term `corona' instead of the more commonly used term `halo'  to emphasize that we are geometrically dividing the region into parts where the dynamo does and does not operate. In our formulation, we consider an extended disk dynamo with a corona where the large-scale magnetic field is built entirely through reconnection of the small-scale magnetic field fluxes emerging from the galactic disk. 

There have been previous attempts, in which the galactic disk is considered to be embedded in a spherical halo (of radius $\sim$ 15 kpc), where the dynamo operation takes place in both the disk and the halo \citep{1992A&A...259..453B, 1993A&A...271...36B,2008A&A...487..197M,2010A&A...512A..61M}. \citet{1992A&A...259..453B} find that a turbulent dynamo can generate a magnetic field on the scale of the halo, but these fields generally have a dominant toroidal field and do not attain a steady state during the Hubble time (due to the large turbulent diffusivity considered for the halo). In order to obtain a global dominance of the poloidal field above the galactic disk, \citet{1993A&A...271...36B} include turbulent diamagnetism, anisotropy of the $\alpha-$effect and galactic winds in their model and obtain fields that are compatible with observations. In more recent simulations, \citet{2008A&A...487..197M} study the coexistence of odd and even parities in the magnetic fields of the disk--halo system. They find that, in cases where the dynamo action in the disk is dominant, the magnetic fields are symmetric in the disk as well as the halo, whereas in cases where the halo is more active, both the disk and the halo favor antisymmetric fields. However, by including a galactic wind, \citet{2010A&A...512A..61M} obtain an approximate even-parity magnetic field in the disk and odd-parity magnetic field in the halo. We plan to consider the more complete halo models in the future. But this would entail further detailed treatment of turbulence in the halo, which is beyond the scope of this paper.

In this work, however, we  adopt an \textit{ansatz}, in which the coronal magnetic field can be described by a linear force-free field with a dynamic force-free parameter $\mu(t)$. The strength of field in the corona is much smaller than in the disk and we find that the chosen prescription of the coronal field does not affect the overall results within the disk. Since the small-scale magnetic helicity in the corona grows with the advective flux of the magnetic helicity generated within the disk, it has the same sign as that of the small-scale magnetic helicity in the disk $h_d$, but opposite to that of the mean-field helicity in the disk, $\ol{H}_d$. Due to magnetic reconnection events occurring in the corona, a fraction, $R_c$ of this small-scale magnetic helicity gets converted to the large-scale magnetic helicity of the corona given by $\ol{H}_c$ (see Section \ref{s:alp} for details). Thus the total magnetic helicity of the corona is given by $\ol{H}_c/R_c$. The conservation of total magnetic helicity for disk and corona combined together can be written as
\begin{equation}
H_0=\ol{H}_d+h_d+\frac{\ol{H}_c}{ R_c}\label{e:htot1}
\end{equation}
were $H_0$ is the initial magnetic helicity of the system contributed entirely by the mean field in the disk.
Equation \eqref{e:htot1} can be differentiated with respect to time to obtain an equation for the rate of change of large-scale magnetic helicity in the corona. Using Equations (\ref{H}) and (\ref{h}), in the absence of large-scale magnetic helicity fluxes, this gives
\begin{equation}
 \frac{\dif \ol{H}_c}{\dif t}=-R_c\left(\frac{\dif \ol{H}_d}{\dif t}+\frac{\dif h_d}{\dif t}\right)= R_c\int_V \left(\nabla\cdot \mathbf{f}\right)\dif V= R_c \int_V \left(\frac{l_0^2 B_{eq}^2}{\eta_t}\nabla\cdot\calf\right)\dif V \label{hceq}
\end{equation}
where $V$ represents the volume of the corona.

\section{The dynamo equations}
We represent the axisymmetric mean magnetic field $\mbf{B}$, in terms of its poloidal $\mbf{B}_P$ and toroidal $\mbf{B}_T$ components, using the scalar stream functions $\ol{\psi}$ and $\ol{T}$ in cylindrical
coordinates as 
\begin{equation}
 \mbf{B}_P=\ol{B}_r\hat{r}+\ol{B}_z\hat{z}=\frac{1}{r}\nabla \ol{\psi}\times\hat{\phi}= \left(\frac{-1}{r}\fpar{}{z} \hat{r}+\frac{1}{r}\fpar{}{r}\hat{z}\right)\ol{\psi}\equiv \mathbf{\hat{P}}\ol{\psi} \label{poleq}
\end{equation}
and 
\begin{equation}
\mbf{B}_\phi=\frac{\ol{T}}{r}\hat{\phi}. \label{toreq}
\end{equation}
Upon substituting Equations (\ref{poleq}) and (\ref{toreq}) in Equation (\ref{dy1}), we get \citep{1994ApJ...434..509M}
\begin{eqnarray}
 \left(\fpar{}{t}+\mbf{U}_P\cdot \nabla-\eta_t \Lambda\right)\ol{\psi}&=&\alpha \ol{T}\label{poleq1}\\
 \left(\fpar{}{t}+\mbf{U}_P\cdot \nabla-\eta_t \Lambda\right)\ol{T}&=&-\alpha \Lambda \ol{\psi}-
 \nabla\alpha\cdot\nabla\ol{\psi}+r\nabla\left(\frac{1}{r}\ol{U}_\phi\right)\times\nabla\ol{\psi}\nonumber\\
 &-&r^2\ol{T}\nabla\cdot\left(\frac{\ol{U}_P}{r^2}\right)+\nabla\eta_t\cdot\nabla\ol{T}\label{toreq1}
\end{eqnarray}
where the operator $\Lambda$ is defined as
\begin{equation}
\Lambda\equiv r^2\nabla\cdot\left(\frac{\nabla}{r^2}\right)=r\fpar{}{r}
\left(\frac{1}{r}\fpar{}{r}\right)+\frac{\partial^2}{\partial z^2} \label{e:lam}
\end{equation}
and $\ol{U}_P$, $\ol{U}_\phi$ are the poloidal and toroidal components of velocity. The right-hand side (rhs) of Equation (\ref{poleq1}) represents the generation of poloidal fields from toroidal fields and the rhs of Equation (\ref{toreq1}) contains terms representing the generation of toroidal fields from poloidal fields through the $\alpha$-effect, shear, compression, transport, and advection of $\ol{T}$ due to varying $\eta_t$. The term representing field transport is on the left-hand side (lhs) of both the equations. We consider a mean flow consisting of differential rotation and vertical advection given as
$ \mbf{U}=(0,U_\phi,U_z)$,
where 
\begin{equation}
\ol{U}_\phi=r\Omega(r); \quad \Omega(r)=\frac{r_0 \Omega_0}{r}. \label{e:uphi}
\end{equation}
For the inputs $r_0=4$ kpc and $\Omega_0$=62.5 km s$^{-1}$ kpc$^{-1}$, this gives $\ol{U}_\phi=250$ km s$^{-1}$ = constant. Since there is no radial component of velocity, the fourth term on the rhs of Equation \eqref{toreq1} becomes $\disp{r^2\ol{T}\nabla\cdot\left(\frac{\ol{U}_P}{r^2}\right)}$ $\disp{=\ol{T} \fpar{\ol{U}_z}{z}}$. We neglect the first and second terms on the rhs of Equation \eqref{toreq1} as they are much smaller than the shear term, i.e., we take the dynamo to be of the $\alpha-\omega$ type. For mathematical simplification, we also neglect the last term on the rhs of Equation \eqref{toreq1} as it is of the order $(z/r)^2$ times smaller than the $z$ diffusion terms. Thus keeping only the dominant terms in the rhs of Equations (\ref{poleq1}) and (\ref{toreq1}), we get a simplified set of equations as
\begin{eqnarray}
 \left(\fpar{}{t}+\ol{U}_z\fpar{}{z}-\eta_t \Lambda\right)\ol{\psi}&=&\alpha \ol{T}\label{poleq2}\\
 \left(\fpar{}{t}+\ol{U}_z\fpar{}{z}-\eta_t \Lambda\right)\ol{T}
 &=&-r\frac{\rm{d}\Omega}{\rm{d}r} \fpar{\ol{\psi}}{z}-\ol{T} \fpar{\ol{U}_z}{z}\label{toreq2}.
\end{eqnarray}
In order to estimate the turbulence parameters $\alpha$ and $\eta_t$, we investigate two possible scenarios for turbulence in the disk: magneto-rotational instability (MRI)-driven turbulence and supernovae (SNe)-driven turbulence.
The details for these cases are given below.
\begin{enumerate}
 \item \textit{MRI-driven turbulence:} weak magnetic fields can generate turbulence in a differentially rotating 
 disk \citep{velikhove.p.1959,1960PNAS...46..253C,1991ApJ...376..214B}. Such MRI-driven turbulence can be responsible for the amplification of magnetic field in the outer parts of the galaxy \citep{1999ApJ...511..660S}.
 The turbulence parameters in this case can then be defined as \citep[see][]{1981MNRAS.195..897P,1994ApJ...434..509M,1999A&A...349..334A}
\begin{equation}
\eta_t=\frac{\mathcal{M}^2 h^2}{\tau_{MRI}},\quad \alpha_0=\frac{\mathcal{M}^2 h}{\tau_{MRI}}, \label{eamri}
\end{equation}
where $\disp{\tau_{MRI}= 2\pi/\Omega(r)=\frac{2\pi r}{r_0 \Omega_0}}$ is the rotational time period at radius $r$. Here, the Mach number, $\mathcal{M}$ is calculated as $\disp{\mathcal{M}=\frac{u}{c_s}\sim\frac{u}{h \Omega(r=r_0)}}$ \citep{1981MNRAS.195..897P,1994ApJ...434..509M},
with $u$ and $c_s$ being the velocities of turbulence and sound respectively, and $h$ being the half-width of the galactic disk. From Equation \eqref{eamri}, we note that both the turbulence parameters $\eta_t$ and $\alpha_0$ vary as $1/r$ over the disk.

\item \textit{SNe-driven turbulence:} the turbulence parameters $\eta_t$ and $\alpha$ are defined in this case as \citep[see][]{1988ASSL..133.....R,2004astro.ph.11739S}
\begin{equation}
 \eta_t=\frac{\mathcal{M}^2 h^2}{\tau_{SN}},\quad \alpha_0=\frac{l^2 \Omega_0}{h}, \label{easn}
\end{equation}
where the correlation time $\tau_{SN}$ is taken as the time interval between supernova shocks
(\citealp{1977ApJ...218..148M}; \citealp[][pp. 181-200]{1990ASSL..161..181C}; \citealp{2004astro.ph.11739S}). The expression for $\alpha_0$ given in Equation \eqref{easn} assumes that Rossby's number $\disp{R_o\equiv\frac{u}{l\Omega}}$ exceeds unity (which is satisfied for $r>$ 2.5 kpc in our case). If $R_o<1$, then the expression for $\alpha_0$ is scaled by a factor of $R_o^{1/2}$ \citep{1988ASSL..133.....R}. However, for mathematical simplification, we use Equation \eqref{easn} for the entire disk. In order to estimate the spatial dependence of $\tau_{SN}$, we proceed as follows.
The locations of the SN stars tend to cluster in regions of intense star formation (known as OB associations). The occurrence of SNe is thus related to the star formation rate (SFR) and $\tau_{SN}\propto \mathrm{SFR}$, \citep{2004astro.ph.11739S,2015MNRAS.450.3472R}.
The SFR depends on the density and the dynamics of the interstellar gas, and is represented by a Schmidt power-law relation $\mathrm{SFR}\propto \Sigma_g^p$ with the index $p=1.3\pm0.3$ \citep{1959ApJ...129..243S,1989ApJ...344..685K}. For mathematical simplification, we take $p\sim1$, which is true for most of the galaxies that fall in the $1\sigma$ range of this distribution.
The mean gas surface density, $\Sigma_g$, is related to the threshold surface density for gravitational stability, $\Sigma_c$, as $\Sigma_g\sim0.7~\Sigma_c$ \citep{1989ApJ...344..685K}. For a flat rotation curve, the stability condition gives $\disp{\Sigma_c\propto \frac{r_0 \Omega_0}{r}}$ \citep{1964ApJ...139.1217T,1981ApJ...245...66C}.
This implies that $\Sigma_g$ and hence the SFR can be expected to vary as $1/r$ over the galactic disk.
Thus, we write the expression for SNe frequency as
\begin{equation}
1/\tau_{SN}=\nu_{SN}(r)=\frac{r_0 \nu_{SN0}}{r} \label{tsn}
\end{equation}
where $\nu_{SN0}=2.5 \Omega_0$ \citep[see][]{2004astro.ph.11739S} is the corresponding frequency at $r_0$=4 kpc. Substituting
Equation \eqref{tsn} into Equation \eqref{easn}, we again find (similar to the case for MRI-driven turbulence) 
that both the turbulence parameters $\eta_t$ and $\alpha_0$ vary as $1/r$ across the disk.
\end{enumerate}
In order to estimate the vertical advection, we note that the energy input from the SNe produces a hot super-bubble that can break away from the galactic disk \citep{1988ARA&A..26..145T}.
This gives rise to a vertical outflow of gas, known as the galactic fountain \citep{1976ApJ...205..762S,2004astro.ph.11739S}.
The radial variation of this advective flow is dependent on the SNe distribution and thus $\propto$ SFR \citep{2015MNRAS.450.3472R}.
The vertical advection, having the same radial dependence, can then be written as
\begin{equation}
\ol{U}_z=\frac{U_0 r_0}{r}, \label{e:uz}
\end{equation}
where $U_0$ can vary between 0 and 2 km s$^{-1}$ \citep[see][]{2006A&A...448L..33S,2015MNRAS.450.3472R}. Note that the last term on the rhs of Equation \eqref{toreq2} goes to zero for this choice of $U_z$. We now write $\alpha(r,z,t)$ as
\begin{equation}
\alpha(r,z,t)= [\alpha_k(r,t)+\alpha_m(r,t)]\Theta(z),
\end{equation}
 where $\Theta(z)=[\theta(z)+\theta(-z)]$ and $\theta(z)$ is the step function. The terms $\alpha_k(r,t)$ and $\alpha_m(r,t)$ can be further split into $r$ and $t$ dependent parts as $\alpha_k(r,t)=\alpha_0(r)~\tilde{\alpha_k}(t)$ and $\alpha_m(r,t)=\alpha_0(r)~\tilde{\alpha_m}(t)$. Following \citet{2007MNRAS.377..874S}, we assume that $\alpha_k$ is only modestly affected by the magnetic field and take $\tilde{\alpha_k}(t)=1$. Thus the time dependence of $\alpha$ is completely ascribed to $\tilde{\alpha_m}(t)$. Thus we can write
\begin{equation}
 \alpha(r,z,t)=\alpha_0(r)[1+\tilde{\alpha_m}(t)]\Theta(z).
\end{equation}
 In the steady state, the time-dependent part is a constant and can be written generally as $\tilde{\alpha_m}=\alpha_m^s$. It is then convenient to define the following dimensionless parameters:
\begin{equation}
 R_\alpha=\frac{\alpha_0 h}{\eta_t},\quad R_\omega=\frac{h^2\Omega}{\eta_t},\quad R_U=\frac{\ol{U}_zh}{\eta_t}. \label{dim}
\end{equation}
Since the quantities $\alpha_0$, $\eta_t$, $\Omega$, and $U_z$ have similar $1/r$ radial dependence, \textit{all the parameters defined in Equation \eqref{dim} are nearly independent of $r$}. This greatly 
simplifies our formulation. We now rewrite Equations (\ref{poleq2}) and (\ref{toreq2}) in dimensionless form through the following substitutions:
\begin{eqnarray}
 \tilde{r}&=&\frac{r}{h},\quad \tilde{z}= \frac{z}{h}, \quad \tau= \frac{t}{t_d}, \quad \tilde{\alpha}=\frac{\alpha}{\alpha_0}\nonumber\\
 \tilde{T}&=&\frac{h \ol{T}}{\psi_0},\quad \tilde{\Lambda} = h^2\Lambda,\quad \tilde{\psi}=\frac{\ol{\psi}}{\psi_0}, \label{scale}
\end{eqnarray}
where $\psi_0=h^2~B_{eq}$ and $\disp{t_d=\frac{h^2}{\eta_t(r=h)}}$ is the diffusion timescale. Here $h=$ 400 pc is the half-width of the disk and the radius of the galactic disk is taken as $r_d=$ 16 kpc. The equipartition field strength is taken as $B_{eq}=$ 5 $\mu$G and the amplitude of $\alpha-$effect is set by $\alpha_0$ given in Table \ref{t:comp}. Dropping the tilde for the sake of clarity, we get the dynamo equations in dimensionless form as (see Appendix \ref{a:dim} for a detailed derivation)
\begin{subequations}
\begin{eqnarray}
 \left(r\fpar{}{\tau}+R_U\fpar{}{z}-\Lambda\right)\psi&=&R_\alpha \alpha(z,t)T\label{poleq3}\\
 \left(r\fpar{}{\tau}+R_U\fpar{}{z}-\Lambda\right)T &=&R_\omega \fpar{\psi}{z}\label{toreq3}.
\end{eqnarray}
\label{e:dim} 
\end{subequations}

A comparison of timescales of operation and the dynamo parameters for both MRI- and SNe-driven turbulence scenarios is presented in Table \ref{t:comp}. As the turbulent parameters in both the cases have similar radial dependence, the two processes can be contribute toward the dynamo operation simultaneously. The combined treatment of both the scenarios, however, is beyond the scope of this paper. We plan to address this in future studies as it likely that a joint operation will make the amplification more effective. However, we note that the MRI-driven dynamo operates at a much slower rate than the SNe-driven dynamo and has a lower dynamo number (see Table \ref{t:comp}). Thus, the SNe-driven dynamo is likely to be the dominant source of magnetic field generation within the galactic disk and hence we present the subsequent calculations only for the case of SNe-driven turbulence.
\begin{table}[h]
\centering
\resizebox{\textwidth}{!}{
\begin{tabular}{|c|c|c|c|c|c|c|c|c|c|c|c|c|}
\hline
Source of  & $\tau_{MRI}$ or $\tau_{SN}$ & $u$    & $c_s$     & $\mathcal{M}$      &$l_0$& $\eta_t$ & $\alpha_0$        & $R_\alpha $ & $R_\omega$ & $R_U$ & $t_d$ \\ 
turbulence & (Myr)                     & (km/s) &  (km/s) &                      & (pc)       &(10$^{26}$cm$^2$/s) & (km/s)      &    &
           &       & (Myr)  \\ \hline
MRI        & 98   &10  & 25     & 0.4    &100       & 0.786                 & 0.64     & 1           & 6.25         & 0-3.14     & 61  \\ \hline
SN         & 6.25  &10  & 80     & 0.125  &100       & 1.2                & 1.56     & 1.6        & 64         & 0-2   & 40  \\ \hline
\end{tabular}
}
\caption[A comparison of parameters for MRI and SNe driven turbulence.]{
A comparison of parameters for MRI and SNe driven turbulence. The 
characteristic time scale for MRI and SNe are given by $\tau_{MRI}$ and $\tau_{SN}$ respectively. The turbulent velocity and sound speed are denoted by $u$ and $c_s$ respectively, while $\mathcal{M}$ gives the Mach number. The length scale of turbulence and turbulent diffusivity are given by $l_0$ and $\eta_t$ respectively, while the strength of $\alpha$ effect is set by $\alpha_0$. The dimensionless dynamo parameters $R_\alpha$, $R_\omega$ and $R_U$ are defined in Equation \eqref{dim}. The range in values of $R_U$ is shown for $U_0 = 0-2$ km s$^{-1}$. The diffusion time scale is given by $t_d$.
}
\label{t:comp}
\end{table}

\section{Solutions to the steady-state dynamo equation}
\label{s:steady}
In this section, we first solve the global dynamo equations for the steady state. The full time-dependent solutions are presented in the next subsection. The steady-state solutions are written assuming a separable form such that
\begin{equation}
 \psi^s(r,z)=Q^s(r) a^s(z), \quad T^s(r,z)=Q^s(r) b^s(z) \label{e:st0}
\end{equation}
where the superscript $s$ denotes steady-state solutions.
Substituting Equation \eqref{e:st0} into Equations (\ref{poleq3}) and (\ref{toreq3}) with the time derivative term dropped, we get the following equations upon simplification for the upper half of the galactic disk:
\begin{subequations}
\begin{align}
 \left[ r\frac{\dif}{\dif r}\left(\frac{1}{r}\frac{\dif Q^s(r)}{\dif r}\right)\right] a^s(z)+\left[\frac{\dif^2 a^s(z)}{\dif z^2}-R_U \frac{\dif a^s(z)}{\dif z}+ R_\alpha \left(1+\alpha_m^s\right)b^s(z)\right] Q^s(r)  &= 0 \label{e:st1a}\\
 \left[ r\frac{\dif}{\dif r}\left(\frac{1}{r}\frac{\dif Q^s(r)}{\dif r}\right)\right] b^s(z)+\left[\frac{\dif^2 b^s(z)}{\dif z^2}-R_U \frac{\dif b^s(z)}{\dif z}+R_\omega \frac{\dif a^s(z)}{\dif z}\right] Q^s(r)&= 0 \label{e:st1b}
\end{align}
\end{subequations}
where $\alpha_m^s$ represents the steady-state value of $\alpha_m$. Dividing Equation \eqref{e:st1a} by $Q^s(r)a^s(z)$ and Equation \eqref{e:st1b} by $Q^s(r) b^s(z)$, and combining the resulting equations, we obtain
\begin{eqnarray}
 \frac{r}{Q^s(r)} \frac{\dif}{\dif r}\left(\frac{1}{r}\frac{\dif Q^s(r)}{\dif r}\right)&=&-\frac{1}{a^s(z)}\left[\frac{\dif^2 a^s(z)}{\dif z^2}- R_U \frac{\dif a^s(z)}{\dif z}+R_\alpha \left(1+\alpha_m^s\right)b^s(z)\right]\nonumber\\
 &=&-\frac{1}{b^s(z)}\left[\frac{\dif^2 b^s(z)}{\dif z^2}- R_U \frac{\dif b^s(z)}{\dif z}+R_\omega \frac{\dif a^s(z)}{\dif z}\right]=-\gamma^s.  \label{e:st2}
\end{eqnarray}
Since the lhs of Equation \eqref{e:st2} is the function of only variable $r$ while its rhs is function of only the variable $z$, the equality can hold only when both sides are actually equal to a constant (taken to be $- \gamma^s$). Rearranging the terms in Equation \eqref{e:st2}, we obtain the following set of Equations \citep[see][]{1994ApJ...434..509M}:
\begin{eqnarray}
 \frac{\dif ^2 Q_n^s(r) }{\dif  r^2}-\frac{1}{r}\frac{\dif Q_n^s(r) }{\dif  r}&=&-\gamma_n^s Q_n^s(r) \label{qeq}\\
 \frac{\dif ^2 a_n^s(z)}{\dif  z^2}-R_U\frac{\dif  a_n^s(z)}{\dif  z}+R_\alpha(1+\alpha_m^s)b_n^s(z)&=& \gamma_n^s a_n^s(z)\label{aneq}\\
 \frac{\dif ^2 b_n^s(z)}{\dif  z^2}-R_U\frac{\dif  b_n^s(z)}{\dif  z}+R_\omega \frac{\dif a_n^s(z)}{\dif z} &=& \gamma_n^s b_n^s(z) \label{bneq}
\end{eqnarray}
where we have introduced the subscript $n$ to represent a set of solutions \{$Q_n^s(r) $, $a_n^s(z)$, $b_n^s(z)$\} for a given value of local growth rate $\gamma_n^s$ that satisfies the radial and vertical boundary conditions.
Upon substituting $Q_n^s(r) =r f_n(r)$, Equation (\ref{qeq}) becomes
\begin{equation}
 \frac{\dif ^2 f_n}{\dif  r^2}+\frac{1}{r}\frac{\dif  f_n}{\dif r}+\left(\gamma_n^s-\frac{1}{r^2}\right)f_n=0,
\end{equation}
which is the well known Bessel differential equation, and the general solution for $\gamma_n^s>0$ is given by
\begin{equation}
 Q_n^s(r) =r J_1\left(\sqrt{\gamma_n^s} r\right)\equiv r \mathcal{J}_n^s(r). \label{bes}
\end{equation}
From Equation \eqref{bneq}, we obtain
\begin{equation}
b^s_n(z)=-\left(\frac{\dif^2}{\dif z^2}-R_U \frac{\dif}{\dif z}-\gamma_n^s\right)^{-1} R_\omega \frac{\dif a_n^s}{\dif z}, \label{e:bsn}
\end{equation}
 which is substituted into Equation \eqref{aneq} to obtain the following differential equation for $a_n^s(z)$ \citep[see][]{1994ApJ...434..509M}:
\begin{equation}
 \frac{\dif ^4 a_n^s}{\dif  z^4}-2 R_U \frac{\dif ^3 a_n^s}{\dif  z^3}+(R_U^2-2\gamma_n^s)\frac{\dif ^2 a_n^s}{\dif  z^2}
 +\left[2R_U\gamma_n^s-(1+\alpha_m^s)R_\alpha R_\omega\right]\frac{\dif  a_n^s}{\dif  z}+(\gamma_n^s)^2 a_n^s=0. \label{aneq2}
\end{equation}
The above fourth-order differential equation can be solved by expanding $a_n^s$ in terms of its four eigenfunctions with eigenvalues $\lambda_{nj}$, written as
\begin{equation}
a_n^s(z)=\sum_{j=1}^4 c_{nj}\exp({\lambda_{nj} z}). \label{e:ans1}
\end{equation}
Substituting Equation \eqref{e:ans1} into Equation (\ref{aneq2}), we obtain a fourth-order equation for $\lambda_{nj}$ given by
\begin{equation}
 \lambda_{nj}^4-2R_U \lambda_{nj}^3+(R_U^2-2\gamma_n^s)\lambda_{nj}^2 +\left[2R_U\gamma_n^s-(1+\alpha_m^s)R_\alpha R_\omega\right]\lambda_{nj}+(\gamma_n^s)^2=0. \label{lameq}
\end{equation}

The dynamo solutions within the galactic disk depend critically on the boundary conditions and the eigenfunctions present in the corona. Here, we consider a scenario in which a corona forms continuously around the galactic disk during the course of dynamo action, due to the contributions from the small-scale magnetic helicity fluxes as given in Equations (\ref{ru}) and (\ref{rk}). We assume that the magnetic field topology in the infinitely conducting corona quickly relaxes into a force-free field, which minimizes the energy while conserving the global magnetic helicity \citep{1960RvMP...32..914W,1974PhRvL..33.1139T,1983PhFl...26.3540F,1984JFM...147..133B,2000JApA...21..299M}. Following the treatment in \citet{1994ApJ...434..509M}, we consider that the coronal magnetic field follows the linear force-free field configuration with a parameter $\mu(t)$ (which has no spatial dependence). Thus, we write the following equations for the coronal magnetic field:
\begin{equation}
 \nabla\times\mathbf{B}=\mu \mathbf{B},\quad \nabla\cdot\mathbf{B}=0. \label{e:cor1}
\end{equation}
Here $\mu=0$ corresponds to a vacuum field outside the disk, which is a likely initial condition. In the course of dynamo action, as the corona builds up, we expect $|\mu(t)|$ to take higher non-zero values. Taking the curl of Equation \eqref{e:cor1}, we obtain 
\begin{equation}
 \nabla^2 \mathbf{B}=-\mu^2 \mathbf{B}.\label{e:nab1}
\end{equation}
Splitting Equation \eqref{e:nab1} into poloidal, $\psi_c$, and toroidal, $T_c$, components using the definitions given in Equations (\ref{poleq}) and (\ref{toreq}), we can write \citep{1994ApJ...434..509M}
\begin{eqnarray}
 \Lambda \psi_c &=& -\mu^2 \psi_c \label{e:psic}\\
 \Lambda T_c &=& -\mu^2 T_c \label{e:tc},
\end{eqnarray}
where $\Lambda$ is defined in Equation \eqref{e:lam}. Here the subscript $c$ denotes coronal fields. The general solution to these equations is given by \citep{1994ApJ...434..509M}
\begin{eqnarray}
 \psi_c (r,z)&=&\int a (p) \exp (-\sqrt{p^2-\mu^2}|z|)r J_1(p r)\dif p, \label{psicor1}\\
 T_c(r,z)&=& \int b(q) \exp (-\sqrt{q^2-\mu^2}|z|)r J_1(q r)\dif q, \label{tceq1}
\end{eqnarray}
where the amplitudes are related by $b(k)=\mu a (k)$, which follows from the force-free condition given in Equation \eqref{e:cor1}. For a galactic disk of radius $r_d$, under the condition that the solution goes to zero at $r=r_d$, the functions $\psi_c$ and $T_c$ can be written as \citep{1994ApJ...434..509M}
\begin{equation}
 \psi_c(r,z)=\sum_{n=1}^N e_n r J_1(k_n r)\exp\left(-\sqrt{k_n^2-\mu^2 }|z|\right),\quad T_c(r,z)=\mu\psi_c(r,z), \label{coreq}
\end{equation}
where $e_n$ are the coefficients to be evaluated from the boundary conditions and $k_n r_d$ are the zeros of Bessel function $J_1$.
Due to the symmetry of the solutions about the mid-plane of the disk, we solve the equation only for the upper half of the disk and use the symmetry to generate the solution for the lower half. A description of the boundary conditions written for the top surface of the disk and the mid-plane is given as follows. The poloidal flux $\psi$ and the radial component of the magnetic field, $\disp{B_r=-\frac{1}{r}\fpar{\psi}{z}}$, are continuous at the top boundary ($z=1$), which means that
\begin{equation}
 [\psi]_{z=1}=0; \quad \textrm{and} ~\left[\fpar{\psi}{z}\right]_{z=1}=0, \label{ebc0}
\end{equation}
where the square bracket represents the continuity of the field. We have investigated the resulting solutions numerically and found that Equation \eqref{ebc0} can be approximated by
\begin{equation}
  \fpar{\psi}{z}(1)=0,  \label{e:bc12}
 \end{equation}
 which also means $B_r =0$ at $z=1$. This is true because $\disp{\fpar{\psi_c}{z}=-\sqrt{k_n^2-\mu^2 }\psi_c}\approx\frac{\psi_c}{r_d}\approx 0$ for the current choice of parameters. Since the magnetic field generated in the galactic disk matches with the linear force-free field of the corona at the top surface, the amplitudes of $a_n$ and $b_n$ (at the top surface) satisfy the same conditions as given in Equations (\ref{psicor1}) and (\ref{tceq1}). Thus
 \begin{equation}
 b_n(1)=\mu^s a_n(1),\label{e:bc3}
 \end{equation}
 where $\mu^s$ denotes the steady-state value of $\mu$. The equatorial boundary conditions specify the symmetry of the solution. For the quadrupolar mode, we write
\begin{subequations}
\begin{eqnarray}
\psi (0)&=&0\label{e:bc4}\\
\fpar{T}{z}(0)&=&0\label{e:bc5}. 
\end{eqnarray}
\label{e:bc45}
\end{subequations}
and for the dipolar mode we write
\begin{subequations}
\begin{eqnarray}
 \fpar{\psi}{z}(0)&=&0\\
 T(0)&=&0
 \end{eqnarray}
\end{subequations}
From Equations (\ref{bes}) and (\ref{coreq}), we find that the radial part of $\psi$ has the same functional form for both the disk and the corona. The requirement of continuity of $\psi$ at the boundary, which is valid even as $r\rightarrow0$, implies that $k_n=\sqrt{\gamma_n^s}$. 
Thus, Equation \eqref{coreq} can now take the form 
\begin{equation}
 \psi_c(r,z)= \sum_{n=1}^N e_n r J_1\left(\sqrt{\gamma_n^s} r\right)\exp\left(-\sqrt{\gamma_n^s-\mu^2 }|z|\right),\quad T_c(r,z)=\mu^s \psi_c(r,z). \label{coreq2}
\end{equation}
Upon substituting the expressions for $\psi$ and $T$ for the disk (Equations \eqref{e:st0}, \eqref{bes}, and \eqref{e:ans1}) and the corona (Equation \eqref{coreq2}) in the boundary conditions for quadrupolar symmetry (Equations \eqref{e:bc12}-\eqref{e:bc45}), we find that the radial part of the solution cancels out and the following equations are obtained for the four eigenfunctions of $a_n^s$ (see Appendix \ref{a:bound} for details):
\begin{subequations}
 \begin{eqnarray}
 \sum_{j=1}^4 \lambda_{nj} c_{nj}\exp({\lambda_{nj}})&=&0\\
 \sum_{j=1}^4 \mu^s\left[ R_\alpha (1+\alpha_m^s)+\lambda_{nj}^2-R_U \lambda_{nj} -\gamma_n^s\right]c_{nj} \exp({\lambda_{nj}})&=&0\\
 \sum_{j=1}^4 c_{nj}&=&0\\
  \sum_{j=1}^4 \left(\gamma_n^s \lambda_{nj} +R_U \lambda_{nj}^2- \lambda_{nj}^3\right)c_{nj}&=&0.
    \end{eqnarray}
\label{quadbc}
\end{subequations}
The above set of equations can be written in a compact form as $\tilde{O}\tilde{c}=0$, where $\tilde{O}$ is a
$4\times4$ matrix comprising the coefficients $c_{nj}$ in Equation (\ref{quadbc}) and $\tilde{c}$ is a $4\times1$
column vector comprising of $c_{nj}$. The condition for non-trivial solutions demands that the determinant of $\tilde{O}$
vanishes, Det $\tilde{O}=0$ \citep[see Appendix B in][]{1994ApJ...434..509M}. This condition is used to evaluate $\gamma_n^s$ as a function of inputs $\alpha_m^s$ and $\mu^s$. Since the scale of coefficients is arbitrary, we can set $c_{n4}=1$, without loss of generality and solve for the other coefficients using the first three Equations in (\ref{quadbc}).

\begin{figure}[hp]
  \centering
  \begin{subfigure}[]{0.45\textwidth}
    \centering
    \includegraphics[width=1\linewidth]{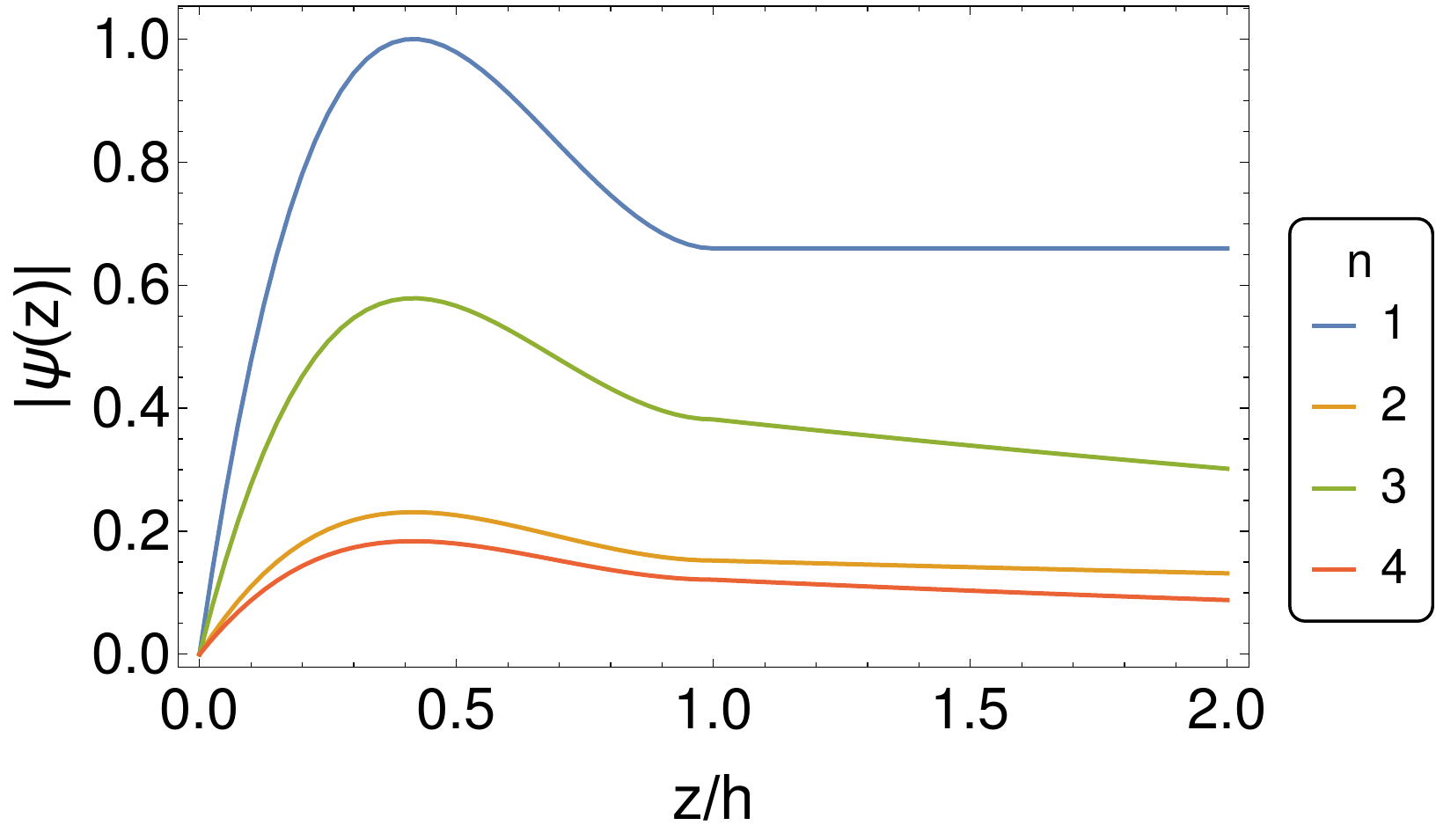}
    \caption{}
    \label{pzplt}
  \end{subfigure}
\quad
  \begin{subfigure}[]{0.45\textwidth}
    \centering
    \includegraphics[width=1\linewidth]{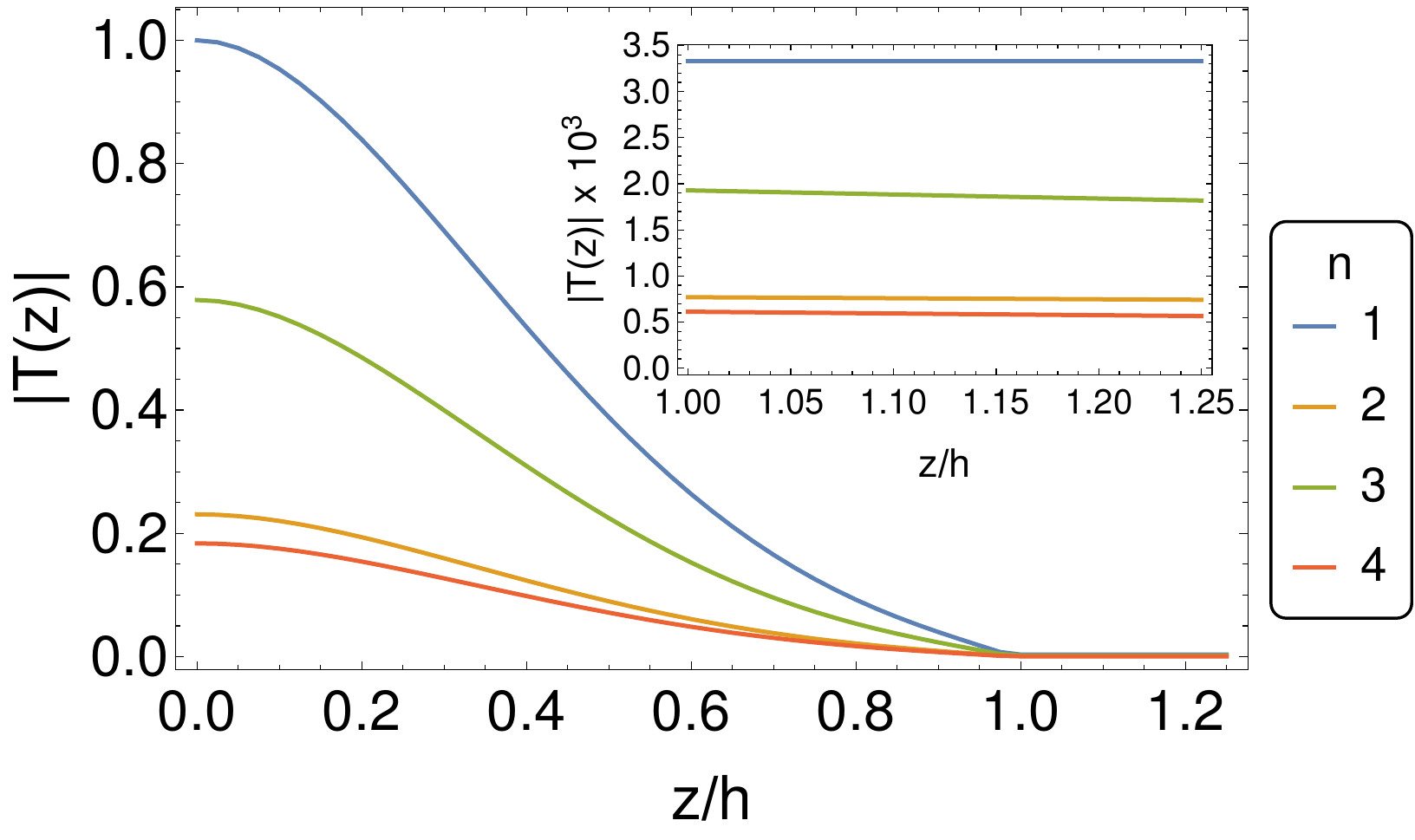}
    \caption{}
    \label{tzplt}
  \end{subfigure}
\quad
  \begin{subfigure}[]{0.4\textwidth}
    \centering
    \includegraphics[width=1\linewidth]{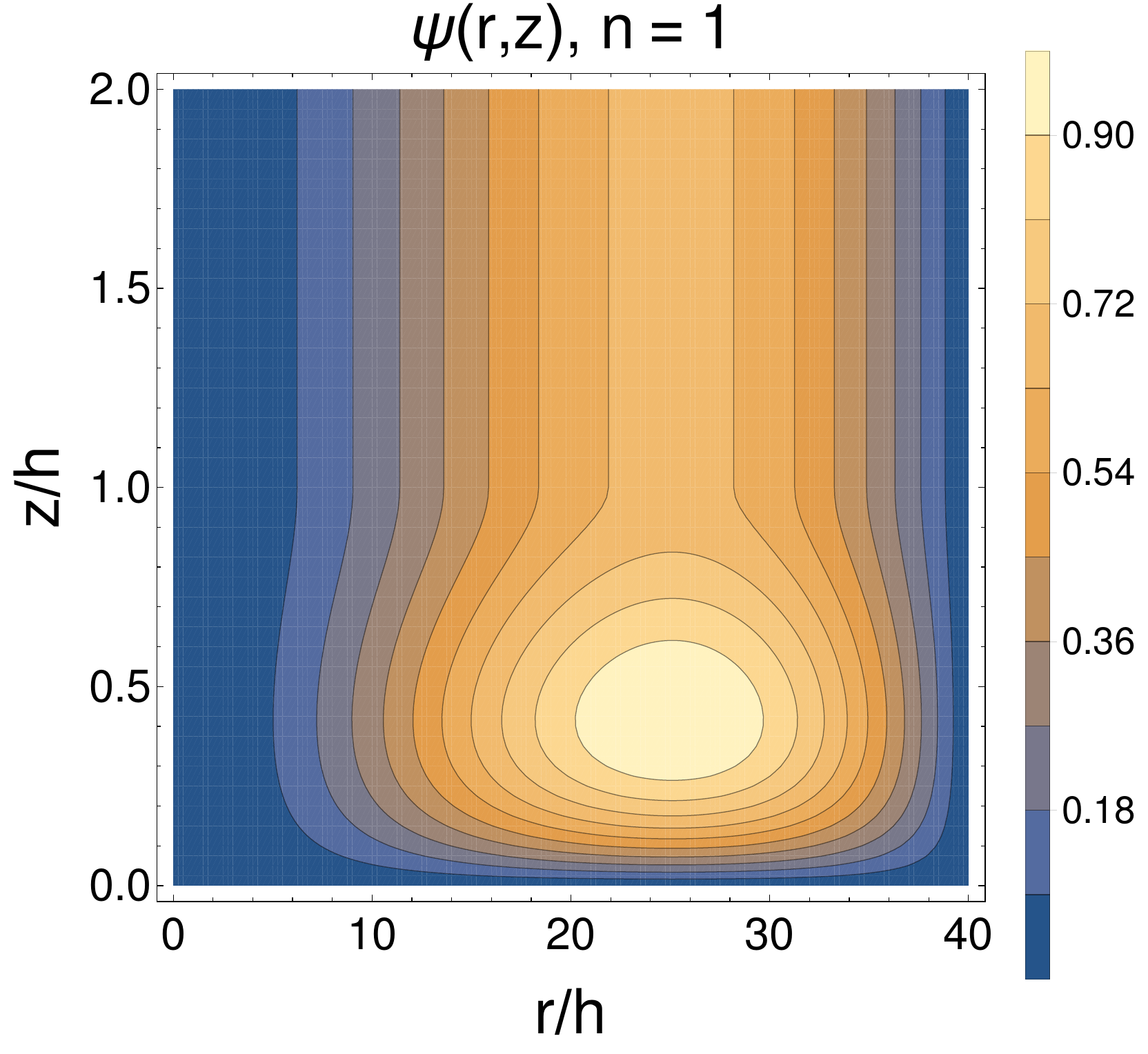}
    \caption{}
    \label{pcp1}
  \end{subfigure}
\quad
  \begin{subfigure}[]{0.4\textwidth}
    \centering
    \includegraphics[width=1\linewidth]{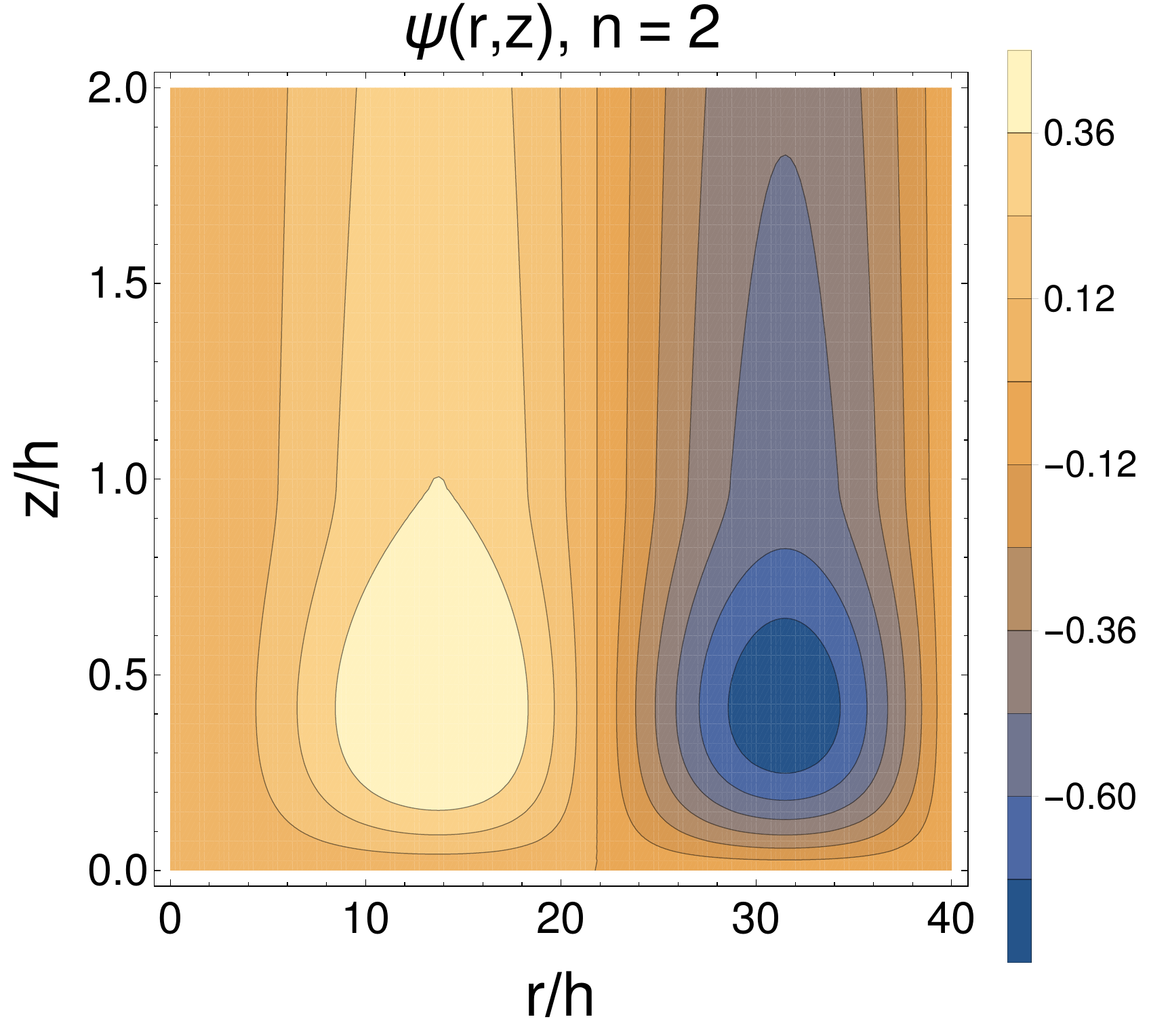}
    \caption{}
    \label{pcp2}
  \end{subfigure}
\quad
\begin{subfigure}[]{0.4\textwidth}
    \centering
    \includegraphics[width=1\linewidth]{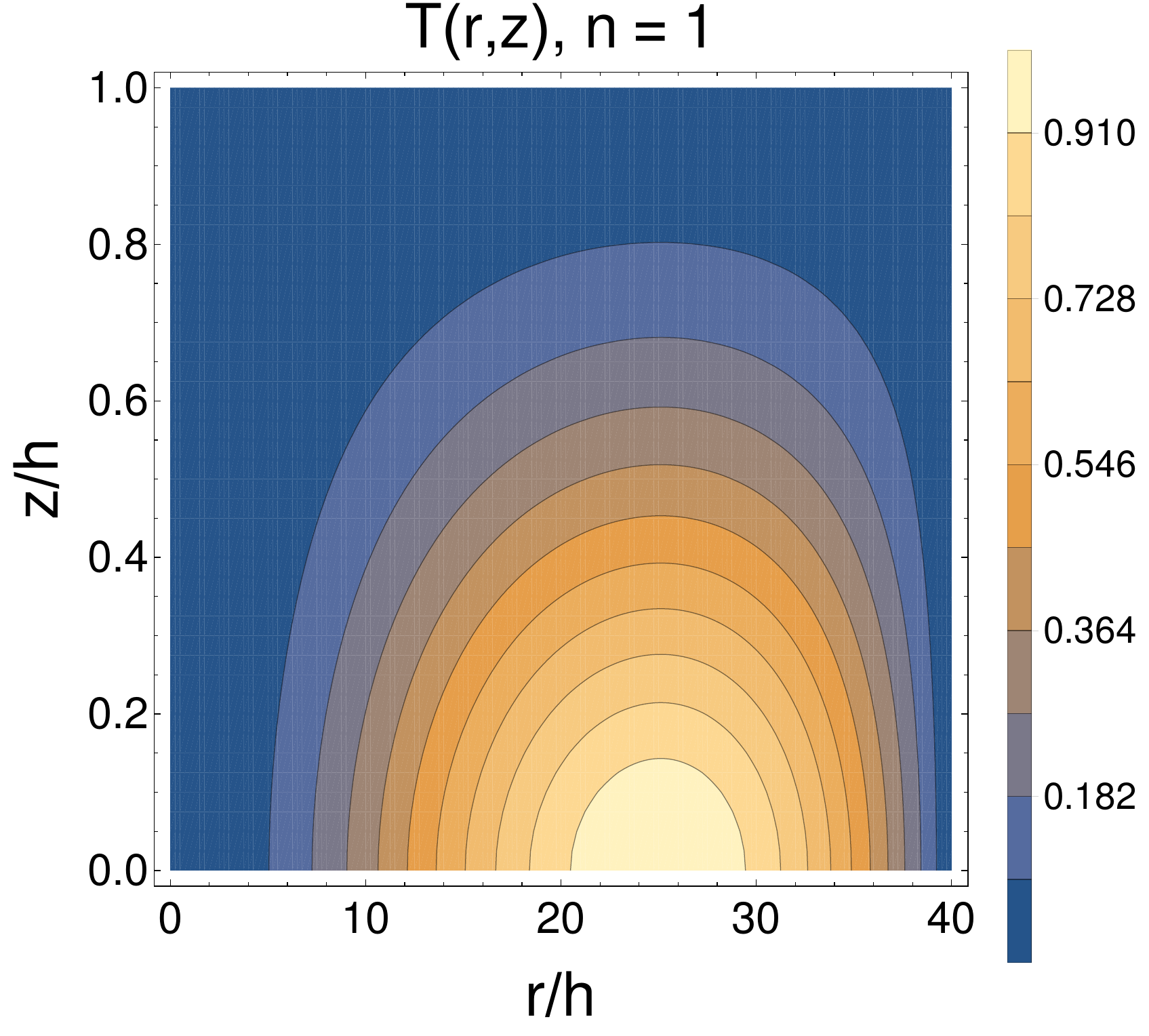}
    \caption{}
    \label{tcp1}
  \end{subfigure}
\quad
  \begin{subfigure}[]{0.4\textwidth}
    \centering
    \includegraphics[width=1\linewidth]{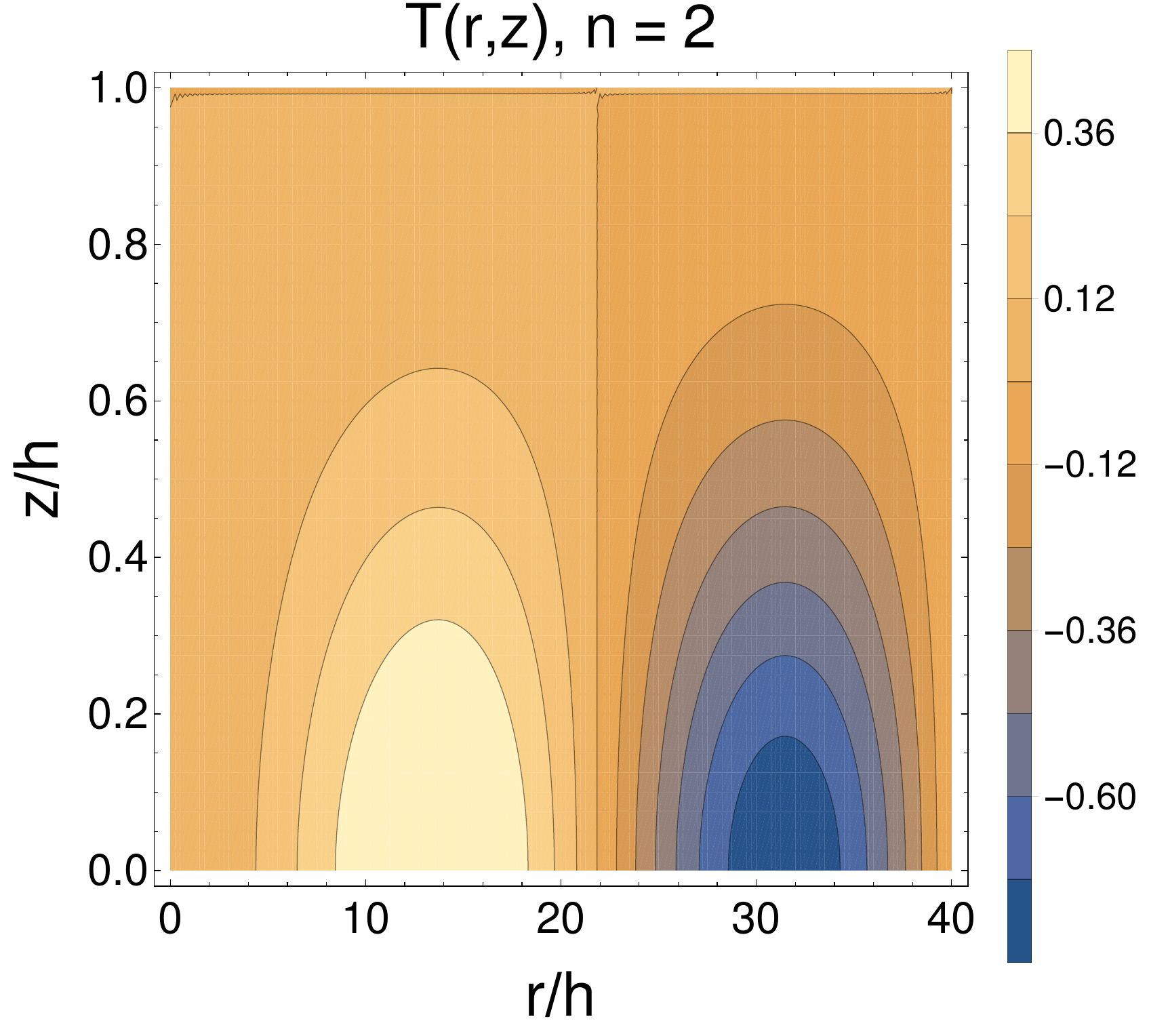}
    \caption{}
    \label{tcp2}
  \end{subfigure}
  \caption{Normalized vertical cross sections of the stream functions $\psi$ and $T$ for different values of 
  $n$ and advective flux $R_U=2$ at a radius of 8 kpc ($r/h =20$) are shown in panels (a) and (b).
  Panels (c) and (d) represent the meridional contour plots of $\psi$ for $n$= 1 and 2 respectively, and
  panels (e) and (f) represent the same for $T$. The contour plots are normalized with respect to the 
  corresponding value of $n$=1. Here $h=$ 400 pc is the half-width of the disk.}
  \label{f:steady}
\end{figure}

The steady-state solutions for advective flux $R_U=2$ and $\mu^s \approx\sqrt{\gamma_1^s}=-0.0958$
 (which is close to the final value of $\mu$, as presented later in Table \ref{t:sat}) are shown in Figure \ref{f:steady} for illustration. A formulation for time-dependent dynamo solutions constructed using these
steady-state solutions is presented in the next section. In Figures \ref{pzplt} and \ref{tzplt}, we show the vertical distribution of $\psi$ and $T$ for different values of parameter $n$ at a radius $r=8$ kpc. The plots are scaled with respect to the maximum values of $\psi$ and $T$, so as to compare the relative strengths of the different modes. The field lines are evidently continuous across the vertical boundary ($z=1$).
For all the cases in Figure \ref{pzplt}, the stream function $\psi$ peaks around $z=0.4~h$ and then falls off with increasing height. The strengths of the different radial modes are comparable, with $n=1$ being the most dominant mode in this case. The poloidal current $T$, shown in Figure \ref{tzplt} starts with its maximum strength at the mid-plane, and then falls off sharply with increasing height. The value of $T$ is negligible outside the disk as the force-free parameter $\mu^s$ is very small in the corona (i.e. the force-free fields in the corona are very close to potential fields). The variations of $\psi$ and $T$ with both $r$ and $z$
for $n$ =1 and 2 are shown in Figures \ref{pcp1} - \ref{tcp2}. The contour plots have been normalized with respect to their corresponding maximum value for $n=1$ in order to compare the strength of the
two modes. As is clearly seen in the plots, the mode $n$ corresponds to the number of oscillations in the radial direction. In both cases the strength of the $n=1$ mode is higher than that of the $n=2$ mode. The quadrupolar nature of the fields are also quite evident from these contour plots.

\section{Time-dependent formulation}
\label{s:time}
In order to set up the time-dependent formulation of the dynamo equation, we use the closure principle offered by the Sturm--Liouville theory and the completeness of the Bessel functions to express the radial part of the time-dependent solution as a linear combination of the various radial modes obtained in the steady-state case. We express time-dependent $\psi$ and $T$ with an implicit dependence on $\alpha$ and $\mu$  as
\begin{subequations}
\begin{eqnarray}
 \psi(r,z,\tau;\alpha,\mu)&=& w(\tau;\alpha,\mu)Q(r;\alpha,\mu)a(z;\alpha,\mu)\label{e:psit1}\\
 T(r,z,\tau;\alpha,\mu)&=& w(\tau;\alpha,\mu)Q(r;\alpha,\mu)b(z;\alpha,\mu),
\end{eqnarray}
 \label{tds} 
\end{subequations}
where 
\begin{equation}
 Q(r;\alpha,\mu)=\sum_{m=1}^\infty X_m(\alpha,\mu) Q_m^s(r)=\sum_{m=1}^\infty X_m(\alpha,\mu) r \mathcal{J}_m^s(r) \label{qsum}
\end{equation}
is a linear combination of the steady-state radial functions $Q_m^s(r)$ given in Equation \eqref{bes}. For our calculations, we have truncated the summation in Equation \eqref{qsum} to $N=6$. Substituting Equation \eqref{tds} in Equation (\ref{poleq3}), we obtain
\begin{equation}
 r \dot{w}Q a +R_U w Q a'-w (\Lambda_r Q) a - w Q (\Lambda_z a)-R_\alpha(1+\alpha_m)wQ b =0. \label{e:tim1}
\end{equation}
where $\disp{\dot{w}=\frac{\dif w}{\dif \tau}}$,
$\disp{a'(z)=\frac{\dif a}{\dif z}}$; $\Lambda_r$ and $\Lambda_z$ are the $r-$ and $z-$dependent parts of operator $\Lambda$ defined in Equation \eqref{e:lam}.
 We have neglected the terms containing partial derivatives with respect to $\alpha$ and $\mu$ in Equation \eqref{e:tim1}, as they are small compared to the derivatives with respect to $z$ (we have checked this by evaluating these coefficients numerically from the steady solutions and also a posteriori from the time-dependent solutions). Dividing Equation \eqref{e:tim1} throughout by $w Q a$, we obtain upon rearranging
\begin{align}
  \frac{r \dot{w}(\tau)}{w(\tau)}-\frac{[\Lambda_r Q(r)]}{Q(r)} &=\frac{[\Lambda_z a(z,\tau)]}{a(z,\tau)}-\frac{R_U a'(z,\tau)}{a(z,\tau)}+R_\alpha[1+\alpha_m(z,\tau)]\frac{b(z,\tau)}{a(z,\tau)} &= \gamma(\tau). \label{e:an1}
\end{align}
Since the lhs of the above equation is a function of $(r,\tau)$ and the rhs is a function of $(z,\tau)$, the equality is satisfied only if both are equal to $\gamma(\tau)$, which depends only on $\tau(\alpha,\mu)$.
Following similar steps with Equation (\ref{toreq3}) for the solutions $(w,Q,b)$, we obtain
\begin{equation}
  \frac{r \dot{w}(\tau)}{w(\tau)}-\frac{[\Lambda_r Q(r)]}{Q(r)} =\frac{[\Lambda_z b(z,\tau)]}{b(z,\tau)}-\frac{R_U b'(z,\tau)}{b(z,\tau)}+R_\omega\frac{a'(z,\tau)}{b(z,\tau)} = \gamma(\tau). \label{e:bn1}
\end{equation}
Combining Equations (\ref{e:an1}) and (\ref{e:bn1}), we obtain the following set of equations:
\begin{subequations}
\begin{eqnarray}
  \frac{r \dot{w}(\tau)}{w(\tau)}- \frac{[\Lambda_r Q(r)]}{Q(r)}&=&\gamma(\tau) \label{e:wn}\\
 \frac{[\Lambda_z a(z,\tau)]}{a(z,\tau)}-R_U\frac{a'(z,\tau)}{a(z,\tau)}+R_\alpha [1+\alpha_m(z,\tau)]\frac{b(z,\tau)}{a(z,\tau)}&=&\gamma(\tau)\label{e:an11}\\
  \frac{[\Lambda_z b(z,\tau)]}{b(z,\tau)}-R_U\frac{b'(z,\tau)}{b(z,\tau)}+R_\omega\frac{a'(z,\tau)}{b(z,\tau)}&=&\gamma(\tau)\label{e:bn11}
 \end{eqnarray} 
 \label{tdeq}
\end{subequations}
The functional form of Equations (\ref{e:an11}) and (\ref{e:bn11}) is same as that of Equations (\ref{aneq}) and (\ref{bneq}), except that the functions $\alpha_m$ and $\gamma$ now vary with time, and can be solved in the same manner as done for the steady-state case as discussed in Section \ref{s:steady}. 
In order to solve the radial equation, we assume 
\begin{equation}
\disp{\frac{\dot{w}(\tau)}{w(\tau)}=\Gamma(\tau)}\quad\Rightarrow w(\tau)=\exp\left[\int_0^{\tau}\Gamma (\tau')\dif \tau'\right],
\label{wneq}
\end{equation}
so that $w(0)=1$. Multiplying equation (\ref{e:wn}) by $Q$, we obtain
\begin{equation}
r \Gamma(\tau) Q -\Lambda_r Q - \gamma(\tau) Q=0.
 \label{e:rw}
\end{equation}
Substituting $\disp{Q =\sum_{m=1}^N X_m Q_m^s}$ from Equation \eqref{qsum} in Equation \eqref{e:rw}, we find
\begin{equation}
 \sum_{m=1}^N \Bigl(r \Gamma X_m Q_m^s- X_m\left(\Lambda_r  Q_m^s - \gamma Q_m^s\right) \Bigr)=0.\label{e:wn1}
\end{equation}
Substituting $\Lambda_r Q_m^s =-\gamma_m^s Q_m^s=-\gamma_m^s r\mathcal{J}_m^s$ using Equations (\ref{qeq}) and (\ref{bes}) in Equation \eqref{e:wn1}, we obtain
\begin{equation}
 \sum_{m=1}^N X_m \Bigl(r^2 \mathcal{J}_m^s \Gamma+r \mathcal{J}_m^s (\gamma_m^s-\gamma)\Bigr)=0.\label{e:wn2}
\end{equation}
Multiplying both sides by $\mathcal{J}_l^s$ for a given index $l$ and integrating over $r$ (represented by angular brackets), we obtain
\begin{equation}
 \sum_{m=1}^N X_m\Bigl(\langle \mathcal{J}_l^s|r^2 \mathcal{J}_m^s\rangle \Gamma+ \langle \mathcal{J}_l^s|r \mathcal{J}_m^s\rangle(\gamma_m^s-\gamma)\Bigr)=0. \label{jeq}
\end{equation}
The orthogonality property of Bessel functions gives
\begin{align}
 \langle \mathcal{J}_l^s|r^2 \mathcal{J}_m^s\rangle&=\int_0^{r_d}r^2 \mathcal{J}_l^s \mathcal{J}_m^s \dif r\equiv G_{lm}\nonumber\\
\langle \mathcal{J}_l^s|r\mathcal{J}_m^s\rangle&=\int_0^{r_d}r\mathcal{J}_l^s \mathcal{J}_m^s \dif r=\frac{r_d^2}{2}J_2^2(\sqrt{\gamma_l^s}r_d)\equiv \delta_{lm} K_m. \label{besf}
\end{align}
For different values of $l$, we obtain a set of $N$ equations similar to Equation \eqref{jeq}, which can be compactly written as
\begin{equation}
 \sum_{m=1}^N X_m\Bigl(\Gamma G_{lm}+ \delta_{lm} K_m  (\gamma_m^s-\gamma)\Bigr)=0 \label{req1}
\end{equation}
where $G_{lm}$ and $K_m$ are defined in Equation \eqref{besf}. Equation \eqref{req1} can be written as an eigenvalue problem by reframing it in terms of matrices in the following manner:
\begin{align}
  \sum_{m=1}^N G_{lm} \Gamma X_m  &= \sum_{m=1}^N \delta_{lm} K_m (\gamma-\gamma_m^s)X_m\nonumber\\
\Rightarrow \tilde{G}~\Gamma \tilde{X}&=\tilde{K}\tilde{X} \label{mat1}
\end{align}
where $\tilde{G}$ and $\tilde{K}$ are $N\times N$ square matrices defined as $(\tilde{G})_{lm}= G_{lm}$ and $(\tilde{K})_{lm}= \delta_{lm} K_m (\gamma-\gamma_m^s)$ respectively, and $\tilde{X}$ is a column vector of length $N$ defined as $(\tilde{X})_m = X_m$ . Multiplying Equation \eqref{mat1} by $\tilde{G}^{-1}$, we obtain
\begin{equation}
 (\tilde{G}^{-1} \tilde{K}) \tilde{X}=\Gamma \tilde{X}. \label{mat2}
\end{equation}
Solving Equation \eqref{mat2} gives us $N$ real eigenvalues for $\Gamma$. We now denote the $n$th eigenvalue of $\Gamma$ as $\Gamma_n$ and the corresponding components of the eigenvectors as $C_{nm}$, which are defined to be orthonormal with the sign determined by $C_{n1}$. Since $\tilde{G}$ and $\tilde{K}$ are known, $C_{nm}$ are uniquely determined. We then define an $N\times N$ square matrix $\tilde{C}$ such as $(\tilde{C})_{nm}=C_{nm}$. We can write the general solution as the linear combination of all these modes (since $w(0)=1$), i.e.,
\begin{subequations}
\begin{eqnarray}
 \psi(r,z,\tau)&=& \sum_{n=1}^N q_n Q_n(r)a(z)w_n =\sum_{n,m=1}^N q_n C_{nm}Q_m^s(r)a(z)w_n(\tau)\label{e:psit1a}\\
 T(r,z,\tau)&=& \sum_{n=1}^N q_n Q_n(r)b(z)w_n =\sum_{n,m=1}^N q_n C_{nm}Q_m^s(r)b(z)w_n(\tau),
\end{eqnarray}
 \label{tdsb} 
\end{subequations}
where $q_n$ are constants that set the initial ratios for various modes $Q_n$ at $\tau=0$. We estimate the coefficients $q_n$ by writing the initial condition for the radial part of the solution as
\begin{equation}
 \sum_{n,m=1}^N q_n C_{nm} (\tau=0) ~Q_m^s = \sum_{m=1}^N s_m Q_m^s \label{qn1}
\end{equation}
where the coefficients $s_m$ are the seeds that set the relative strength of each steady-state mode $Q_m^s$. Comparing the coefficient of $Q_m^s$ from both sides in Equation \eqref{qn1}, we find
\begin{equation}
\sum_{n=1}^N q_n C_{nm} (\tau=0)=s_m \quad \Rightarrow \quad  \tilde {C_0}^T\tilde{q} =\tilde{s} \label{qn2}
\end{equation}
where $\tilde{q}$ and $\tilde{s}$ are column vectors of length $N$ defined as $(\tilde{q})_n= q_n$,  $(\tilde{s})_m=s_m$ respectively, and $\tilde {C_0^T}$ is an $N\times N$ square matrix defined as $(\tilde {C_0}^T)_{nm}=C_{mn} (\tau=0)$. Equation \eqref{qn2} can then be rewritten as $\disp{\tilde{q}= \left(\tilde {C_0^T}\right)^{-1} \tilde{s}}$, which can be solved to obtain the values of constants $q_n$. Henceforth, for clarity, we use $\mathcal{C}_{nm}=q_l\delta_{ln}C_{nm}=q_n C_{nm}$ (where no summation is implied on $n$); $\mathcal{\tilde{C}}=\tilde{C^T}\tilde{q}$.

As discussed in Section \ref{s:mfd}, during the course of dynamo operation, a corona builds up around the disk due to the release of small-scale magnetic helicity fluxes across the boundary.
The boundary conditions given in Equation \eqref{quadbc} are still valid with $\alpha_m^s$, $\mu^s$, and $\gamma_n^s$ now replaced by their time-dependent counterparts. 
 Using Equations (\ref{bes}), (\ref{coreq2}) and (\ref{e:psit1a}), the continuity of $\psi$ at the top surface can be written as
\begin{equation}
 \sum_{n,m=1}^N w_n(\tau) \mathcal{C}_{nm} r J_1\left(\sqrt{\gamma_m^s} r\right) a(1)= \sum_{m=1}^N e_m r J_1\left(\sqrt{\gamma_m^s} r\right)\exp \left(-\sqrt{\gamma_m^s-\mu^2}\right). \label{ent1}
\end{equation}
From Equation \eqref{ent1}, we can write
\begin{equation}
 e_m= \sum_{n=1}^N w_n(\tau)\mathcal{C}_{nm} a(1) \exp\left(-\sqrt{\gamma_m^s-\mu^2}\right).
\end{equation}
The expressions for $\psi_c$ and $T_c$ are now given as
\begin{equation}
 \psi_c=\sum_{n,m=1}^N w_n \mathcal{C}_{nm} a(1)r J_1\left(\sqrt{\gamma_m^s} r\right)\exp\left(\sqrt{\gamma_m^s-\mu^2}(1-z)\right), 
\quad T_c=\mu \psi_c. \label{e:psicor}
\end{equation}

\subsection{Time dependence of $\alpha_m$ and coronal helicity}
\label{s:alp}
The equation for the evolution of $\alpha_m$ is derived in Appendix \ref{a:alp} and can be written as
\begin{equation}
 r \frac{\dif  \alpha_m}{\dif  \tau}= -C\left[(1+\alpha_m)\ol{B^2}-R_\alpha^{-1}\ol{\bf{J}\cdot\bf{B}}\right]-(R_U+R_\kappa)\alpha_m, \label{alp2}
\end{equation}
where 
\begin{equation}
C=2\left(\frac{h}{l_0}\right)^2, \quad R_\kappa=\frac{\kappa}{\eta_t}. \label{cru}
\end{equation}
For $R_m=10^5$ \citep{2006A&A...448L..33S}, the ratio $\disp{\frac{\alpha_m}{R_m}}$ is very small compared to other terms in Equation \eqref{alp2} and is hence neglected. In order to obtain an equation for dynamical evolution of $\alpha_m$, we take a spatial average
of Equation (\ref{alp2}) over the entire volume of the disk, and the resulting equation can now be written as 
\begin{equation}
\langle r \rangle \frac{\dif  \alpha_m}{\dif \tau}= -C\left[(1+\alpha_m)\langle\ol{B^2}\rangle-R_\alpha^{-1}
\langle\ol{\bf{J}\cdot\bf{B}}\rangle\right]- (R_U+R_\kappa)\alpha_m, \label{alp3}
\end{equation}
where the angular brackets in the above equation represent volume averaging.

 Now $\disp{\langle r \rangle =\frac{2}{r_d^2}\int_0^{r_d} r^2 \dif r =\frac{2r_d}{3 }}$, and the expressions for $\langle\ol{B^2}\rangle$ and $\langle\ol{\bf{J}\cdot\bf{B}}\rangle$ are given by (see Appendix \ref{a:avg} for details)
\begin{align}
\langle \ol{B^2}\rangle =\sum_{n,m,l=1}^N & \mathcal{C}_{nl} \mathcal{C}_{ml} w_n w_m \Bigl(J_2^2(\sqrt{\gamma_l^s} r_d) \langle {a'}^2 + b^2 \rangle+ \gamma_l^s J_0^2 (\sqrt{\gamma_l^s} r_d) \langle a^2 \rangle \Bigr)\label{bbeq}\\
\langle\ol{\mathbf {J}\cdot \mathbf{B}}\rangle= \sum_{n,m,l=1}^N & \mathcal{C}_{nl} \mathcal{C}_{ml} w_n w_m\Bigl(J_2^2(\sqrt{\gamma_l^s} r_d)\langle a' b' + a'' b -\gamma_l^s a b\rangle 
+\gamma_l^s J_0^2 (\sqrt{\gamma_l^s} r_d) \langle a b\rangle \Bigr)\label{jbeq}.
\end{align}
Equation \eqref{alp2} can now be written compactly as
\begin{equation}
 \frac{\dif  \alpha_m}{\dif \tau}= -\frac{3 C}{2 r_d}\left[(1+\alpha_m)\langle\ol{B^2}\rangle-R_\alpha^{-1}
\langle\ol{\bf{J}\cdot\bf{B}}\rangle\right]-\frac{3}{2 r_d} (R_U+R_\kappa)\alpha_m.\label{alp4}
\end{equation}

As mentioned in Section \ref{s:mfd}, during the dynamo operation inside the disk, we allow for the large-scale magnetic helicity flux to be redistributed by advection in the disk but not escape. The justification for this can be given as follows. In disks of spiral galaxies, the hot gas produced by SNe can rise to large scale heights above the disk surface. Upon  cooling and due to thermal instabilities, these gases form  discrete dense clouds and fall back to the disk. This is known as the galactic fountain \citep{1976ApJ...205..762S,1995MNRAS.276..651B}. The magnetic fields carried away by these hot gases are typically of scales smaller than the size of the hot cavities (0.1-1 kpc). Hence the fields carried outside the disk mostly represent the small-scale turbulent magnetic fields \citep{2006A&A...448L..33S}. The scale of the hot cavities is greater than that of the turbulent magnetic field but smaller than the scale of the mean magnetic field. So, the Lorentz force resists the advection of the mean field at the disk surface more efficiently than that of the small-scale turbulent magnetic field \citep{2006A&A...448L..33S}. Also reconnection can remove the loops in the large-scale magnetic field arising due to the fountain flow from their parent magnetic field lines. Thus the galactic fountain flow is more likely to carry only the small-scale magnetic field \citep{2006A&A...448L..33S}. For large values of the turbulent magnetic Reynolds number $R_U~ ( >20)$, \citet{1995MNRAS.276..651B} argue that the large-scale magnetic field can be transported from the disk into the halo by topological pumping. However, in our case $R_U \leq 2$ (see Table \ref{t:comp}), and thus the small-scale magnetic fields are expected to be removed from the disc more efficiently than the large-scale magnetic fields.

In order to calculate the mean magnetic helicity of the coronal field, we use the prescription given in \citet{2006ApJ...646.1288L,2011PhPl...18e2901L},
which gives the measure of mean magnetic helicity as (see Appendix \ref{AppB} for details)
\begin{equation}
 \ol{H}_c=\int_V \frac{2 \psi_c T_c}{r^2}~ \dif V= \int_V \frac{2 \mu \psi_c^2}{r^2}~ \dif V. \label{corhel}
\end{equation}
The mean magnetic helicity in the corona is given by Equation \eqref{e:hc2} as
\begin{equation}
 \ol{H}_c=\sum_{n,m,l=1}^N \pi \mu r_d^2 J_2^2(\sqrt{\gamma_l^s}r_d)w_n w_m \mathcal{C}_{nl}\mathcal{C}_{ml}\frac{a^2(1)}{\sqrt{\gamma_l^s-\mu^2}}.\label{a:Hcf}
\end{equation}
The equation for the rate of change of large-scale magnetic helicity in the corona is given by Equation \eqref{hceq}:
\begin{equation}
 \frac{\dif \ol{H}_c}{\dif t}= R_c \int_V \left(\frac{l_0^2 B_{eq}^2}{\eta_t}\nabla\cdot\calf\right)\dif V.\label{e:hce1}
\end{equation}
Combining Equations (\ref{calf}), (\ref{e:ruf}) and (\ref{e:rkf}) with the above equation, and reducing the above equation in dimensionless form using the transformations given in Equation \eqref{scale}, we obtain the final equation as (see Appendix \ref{a:corhel} for details)
\begin{equation}
\frac{\dif \ol{H}_c}{\dif \tau}=\frac{4\pi R_c r_d}{C} R_\alpha (R_U + R_\kappa) \alpha_m. \label{hc7}
\end{equation}
 The above equation gives the dynamical evolution of the large-scale coronal helicity $H_c$. The fraction of small-scale magnetic helicity flux getting converted into the large-scale helicity in the corona, $R_c$, can be estimated as follows. The small-scale magnetic field escaping into the corona, $\mathbf{b}$, gets converted into the large-scale magnetic field, $\mathbf{B}_c$ through random reconnection events. Thus  $B_c=b/\sqrt{N_c}$, where $N_c=k_f/k_m$; $k_m=\mu$ and $k_f=1/l_0$ are the wavenumbers for the mean and turbulent fields in the corona respectively. We now estimate fraction $R_c$ through
\begin{equation}
\disp{R_c \equiv \frac{\dif \ol{H}_c}{\dif \tau}\Big/ \left[ \frac{\dif h_d}{\dif \tau} \right ]_{flux} \approx \Bigl( \frac{B_c^2}{k_m\tau_R}R_V\Bigr)\Big/ \Bigl(\frac{b^2 }{k_f\tau_a}\Bigr)}
\end{equation}
where $\disp{\tau_R=\frac{R_x(S)}{ k_f {\cal M}_A c_s }}$ is the reconnection timescale and the factor $R_x(S)$ is a theory-dependent function of the Lundquist number $S= (R_m l_0 c_s {\cal M}_A)/\eta_t$. The alfv\'{e}nic Mach number is given by $\cal M_A$, the advection timescale by $\tau_a= h/U_z$, and $R_V \approx 1/( \sqrt{\gamma_n^s-\mu^2} h) \simeq 1/(\mu h)$ is the ratio of the two effective volumes. Upon 
simplification, we obtain the fraction as
\begin{equation}
R_c = \frac{k_f{\cal M}_A c_s}{k_m R_x(S) U_z}=\frac{k_f{\cal M}_A}{k_m R_x(S){\cal M}_z} \label{Rc}
\end{equation}
where ${\cal M}_z=U_z/c_s$ is assumed to be of the order of unity. We then obtain $\disp{R_c \approx \frac{k_f{\cal M}_A }{k_m R_x(S)}}$. Now $S = 1.6 \times 10^6 {\cal M}_A$, and $R_x(S) = \sqrt{S}$ for Sweet--Parker reconnection or $ R_x(S)=  8 \ln{S}/\pi$ for the Petschek process \citep{2005ppfa.book.....K}. For ${\cal M}_A\approx 1$, which may be reasonable to assume, given that corona is being filled by plasma containing small-scale helicity flux, that is near equipartition. This leads to a range for $R_c$ of $10^{-3}-10^{-1}$. We investigate this entire range of $R_c$, given the uncertainties in this parameter, but find that the final solution (discussed in Section \ref{s:sol}) is insensitive to $R_c$.

Below, we summarize the important equations from this section for quick reference and discuss the solutions in the next section. To obtain the radial solutions $Q(r)$, we solve Equation (\ref{e:rw}) given by
\begin{equation}
r \Gamma(\tau) Q -\Lambda_r Q - \gamma(\tau) Q=0,
\end{equation}
and the coefficients $w(\tau)$ are obtained from Equation (\ref{wneq}) given by
\begin{equation}
\disp{\frac{\dot{w}(\tau)}{w(\tau)}=\Gamma(\tau)}\quad\Rightarrow w(\tau)=\exp\left[\int_0^{\tau}\Gamma (\tau')\dif \tau'\right],
\end{equation}
where $w(0)=1$. The solutions for $Q(r)$ and the global growth rate $\Gamma$ are obtained from the eigenvectors and eigenvalues obtained from Equation \eqref{mat2} given by
 \begin{equation}
 (\tilde{G}^{-1} \tilde{K}) \tilde{X}=\Gamma \tilde{X}.
\end{equation}
with the initial conditions set by Equation \eqref{qn2}. The $z$ part of the solution is obtained from solving the equations (\ref{e:an11}) and \ref{e:bn11}) given by
\begin{align} 
 \frac{[\Lambda_z a(z,\tau)]}{a(z,\tau)}-R_U\frac{a'(z,\tau)}{a(z,\tau)}+R_\alpha [1+\alpha_m(z,\tau)]\frac{b(z,\tau)}{a(z,\tau)}&=\gamma(\tau)\\
  \frac{[\Lambda_z b(z,\tau)]}{b(z,\tau)}-R_U\frac{b'(z,\tau)}{b(z,\tau)}+R_\omega\frac{a'(z,\tau)}{b(z,\tau)}&=\gamma(\tau)
\end{align}
with the boundary conditions given in Equation \eqref{quadbc} for time-dependent $\alpha_m$ and $\mu$. The dynamical equations for $\alpha_m$ and $\mu$ are solved from Equations (\ref{alp4}) and (\ref{hc7}) given by
\begin{align}
\frac{\dif  \alpha_m}{\dif \tau}&= -\frac{3 C}{2 r_d}\left[(1+\alpha_m)\langle\ol{B^2}\rangle-R_\alpha^{-1}
\langle\ol{\bf{J}\cdot\bf{B}}\rangle\right]-\frac{3}{2 r_d} (R_U+R_\kappa)\alpha_m\\
\frac{\dif \ol{H}_c}{\dif \tau}&=\frac{4\pi R_c r_d}{C} R_\alpha (R_U + R_\kappa) \alpha_m,
\end{align}
with $\langle\ol{B^2}\rangle$, $\langle\ol{\bf{J}\cdot\bf{B}}\rangle$ and $\ol{H}_c$ defined by Equations (\ref{bbeq}), (\ref{jbeq}), and (\ref{e:hc2}) respectively, for the initial conditions $\alpha_m=10^{-3}$ and $\mu=0$ at $\tau=0$.

 \section{Solutions of time-dependent dynamo equations}
\label{s:sol}
In this section, we present a summary of our simulations for the global nonlinear dynamo using the turbulence parameters for the SNe-driven scenario presented in Table \ref{t:comp}. The time-dependent dynamo equation given by Equations (\ref{tdeq}) and (\ref{wneq}) can be solved parametrically as a function of local and global growth rates, $\gamma$ and $\Gamma$ respectively. These parameters in turn depend upon the values of input $\alpha_m$ and $\mu$. In Section \ref{s:para} , we discuss the dependence of the dynamo solutions on parameters $\alpha$ and $\mu$, for fixed values of $R_\alpha$, $R_\omega$, and $R_c$. The dynamical equations for $\alpha_m$ and $\mu$ given by Equations (\ref{alp4}) and (\ref{hc7}) respectively, which specify the evolution of the amplitude and structure of the large-scale magnetic field with time, are discussed in Section \ref{s:evol}-\ref{s:dist}.
\subsection{Parametric study of time-dependent dynamo solutions}
\label{s:para}
\begin{figure}[h]
  \centering
  \begin{subfigure}[]{0.45\textwidth}
    \centering
    \includegraphics[width=1\linewidth]{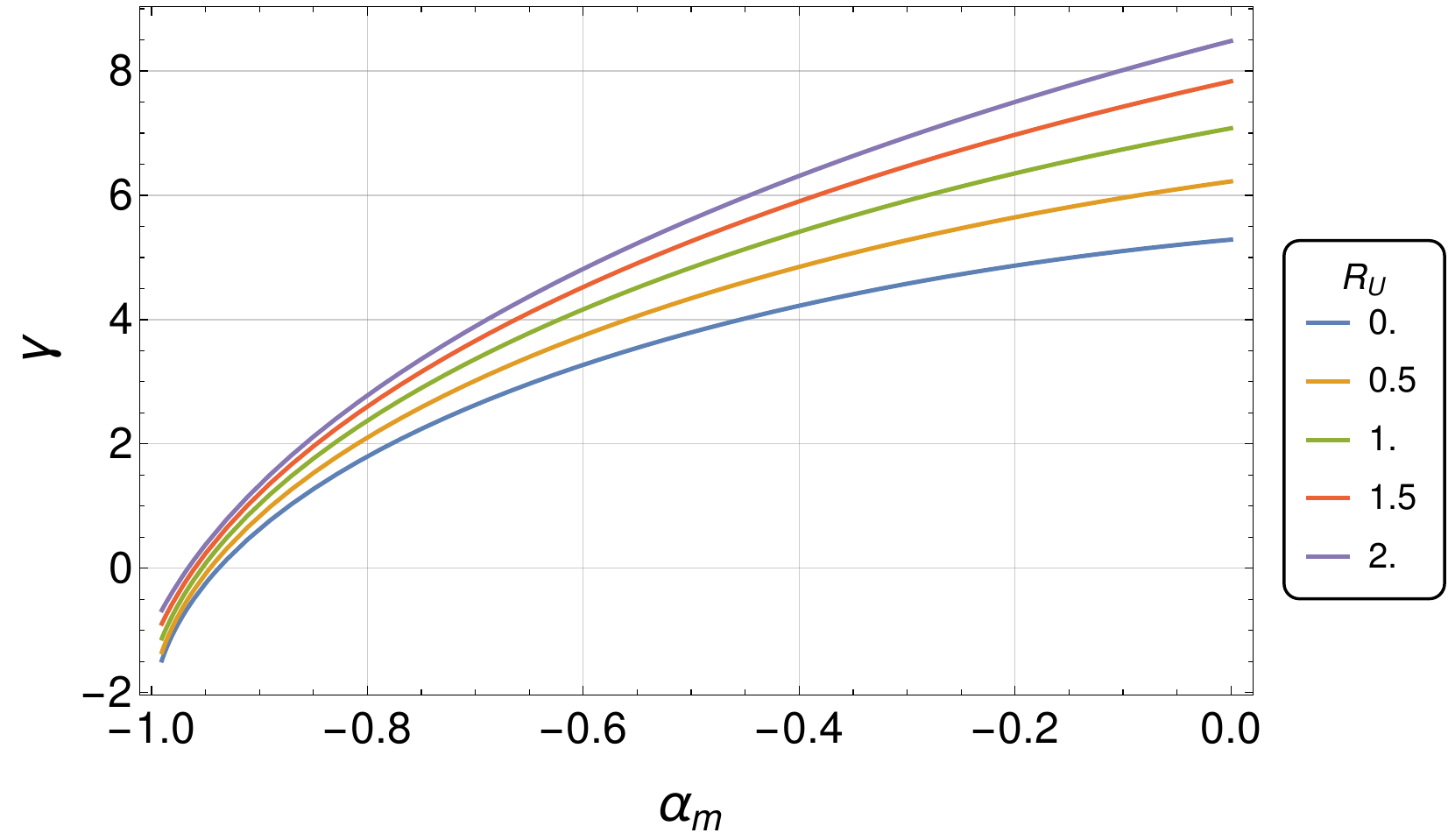}
    \caption{}
    \label{gamuplt}
  \end{subfigure}
\quad
  \begin{subfigure}[]{0.45\textwidth}
    \centering
    \includegraphics[width=1\linewidth]{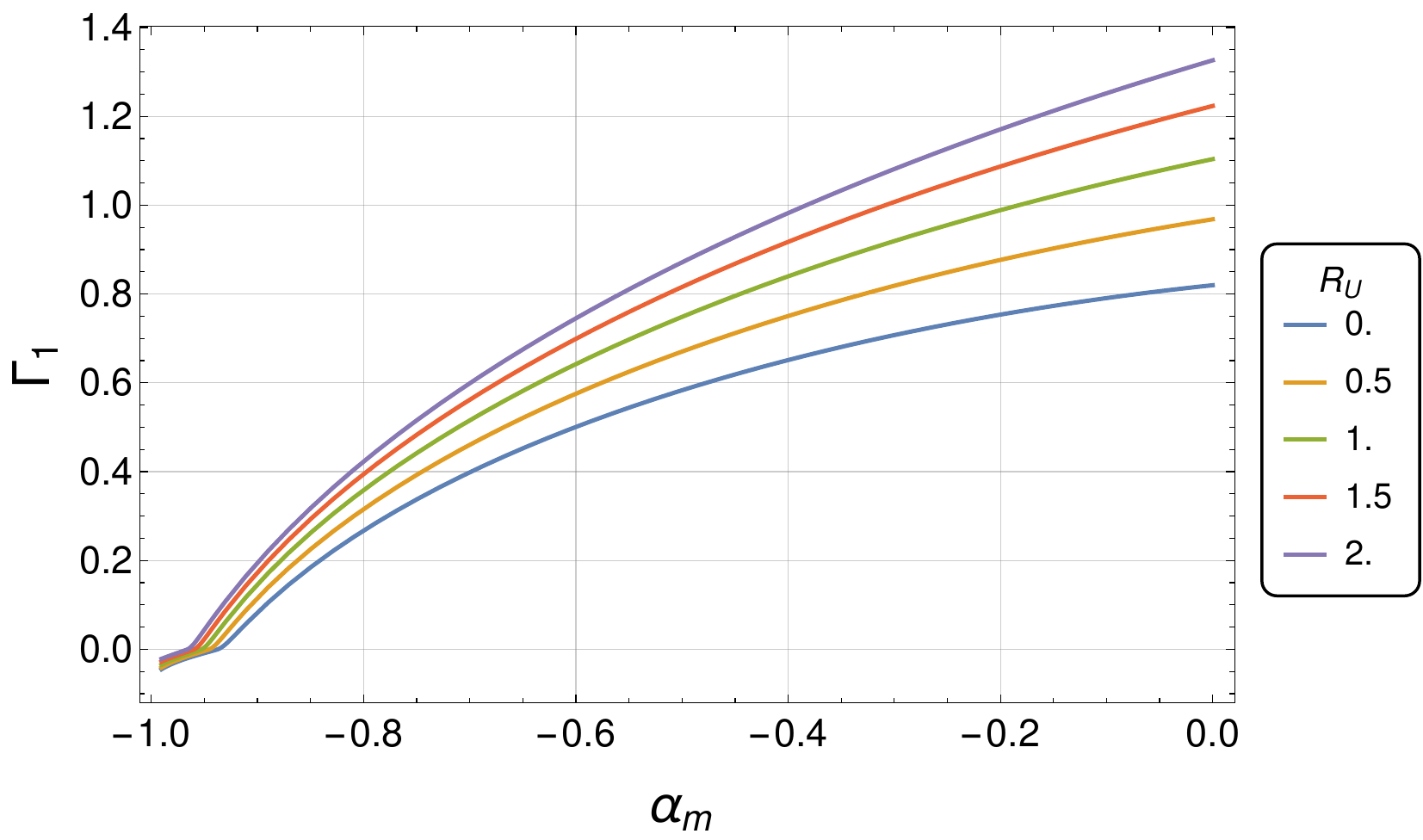}
    \caption{}
    \label{G1amuplt}
  \end{subfigure}
  \caption{Variations of (a) $\gamma$ and (b) $\Gamma_1$ with respect to $\alpha_m$ for different values of $R_U$ at $\mu =-0.09579$.}
  \label{f:para1}
\end{figure}

\begin{figure}[h]
  \centering
  \includegraphics[width=0.9\linewidth]{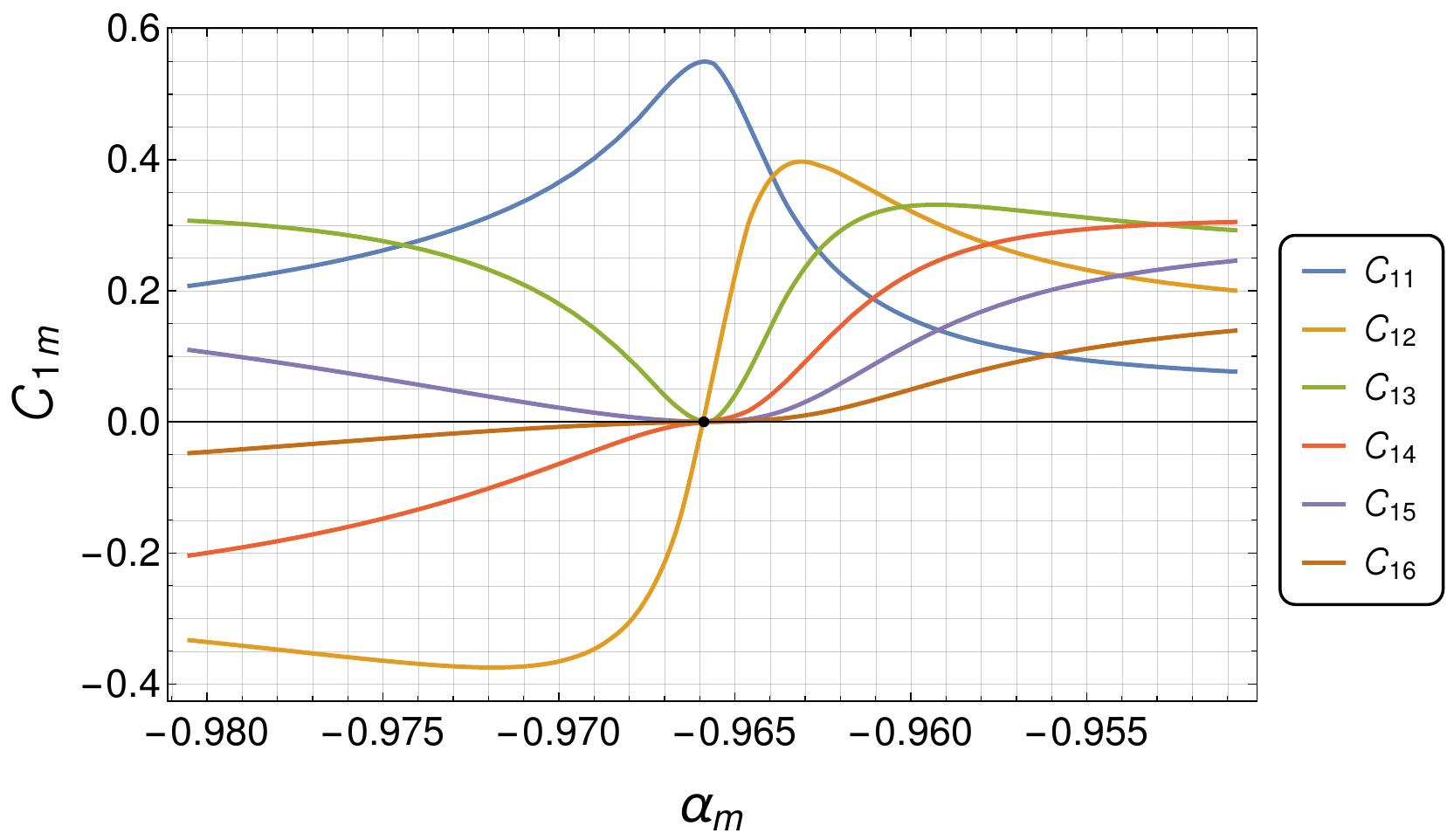}
  \captionof{figure}{Variation of different elements of the eigenvector $\mathcal{C}_{1m}$, corresponding to the largest eigenvalue $\Gamma_1$ with $\alpha_m$. The black dot corresponds to the value of $\alpha_m$ where $\Gamma_1=0$ and $\gamma=\gamma_1^s$.}
  \label{c1mplt}
\end{figure}

The dependence of $\gamma$ on $\alpha_m$ and $\mu$ can be obtained by demanding non-trivial solutions for $a$ in Equation \eqref{tdeq}; see the discussion following Equation \eqref{quadbc} in Section \ref{s:steady}. The resulting solution of $\gamma$ as a function of $\alpha_m$  is shown in Figure \ref{gamuplt} ($\gamma$ was found to be nearly independent of variation in $\mu$). We find that $\gamma$ decreases monotonically with decreasing $\alpha_m$ and increases with increasing $R_U$.
 For given values of $\gamma(\mu,~\alpha_m)$, we can solve Equation \eqref{mat2} to obtain the various eigenvalues $\Gamma_n$ and their corresponding eigenvectors $\mathcal{C}_{nm}$ as functions of $\alpha_m$ and $\mu$. 
\begin{figure}[hp]
  \centering
\begin{subfigure}[]{0.65\textwidth}
    \centering
    \includegraphics[width=1\linewidth]{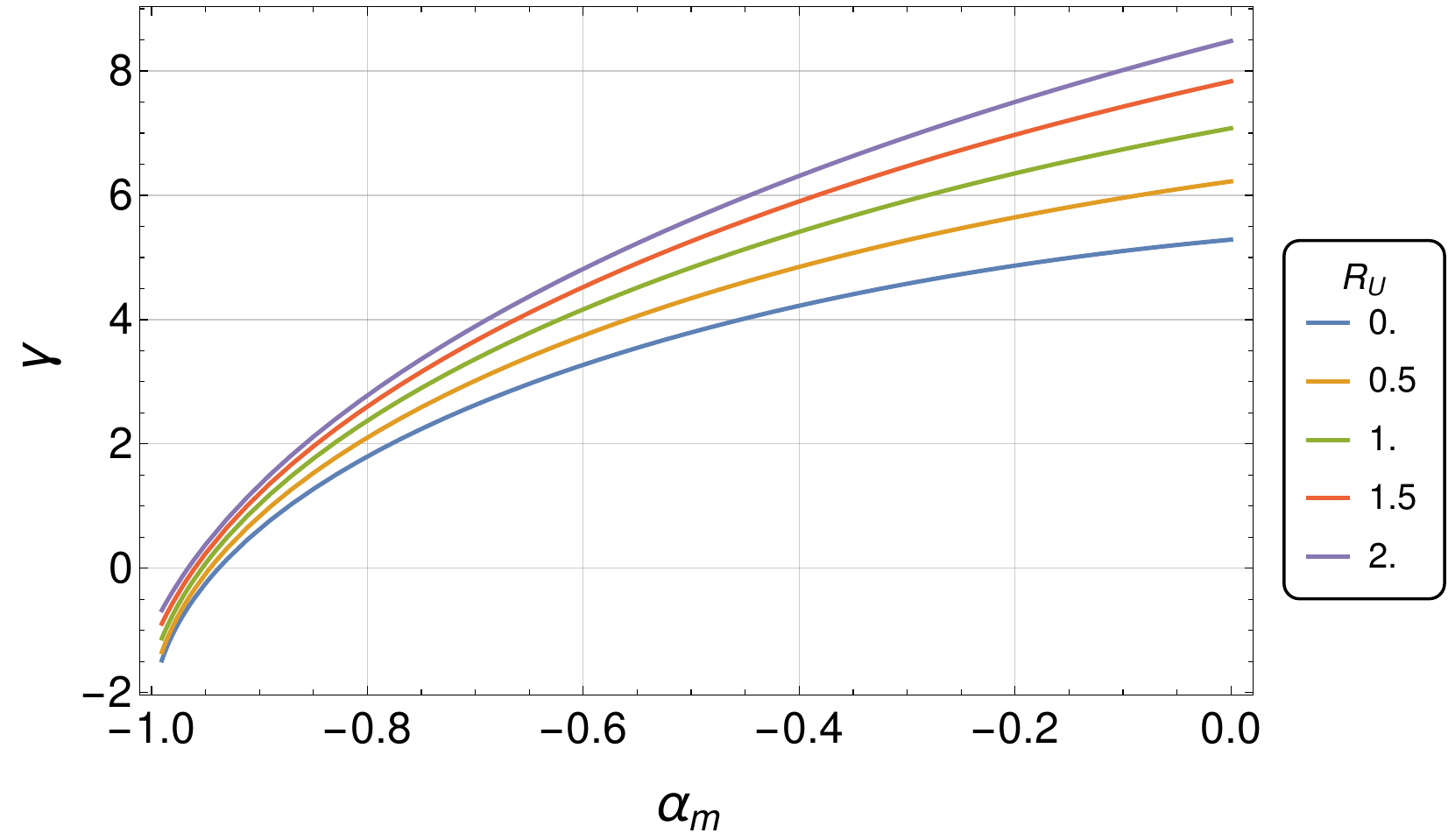}
    \caption{}
    \label{gplt}
  \end{subfigure}
\quad
  \begin{subfigure}[]{0.65\textwidth}
    \centering
    \includegraphics[width=1\linewidth]{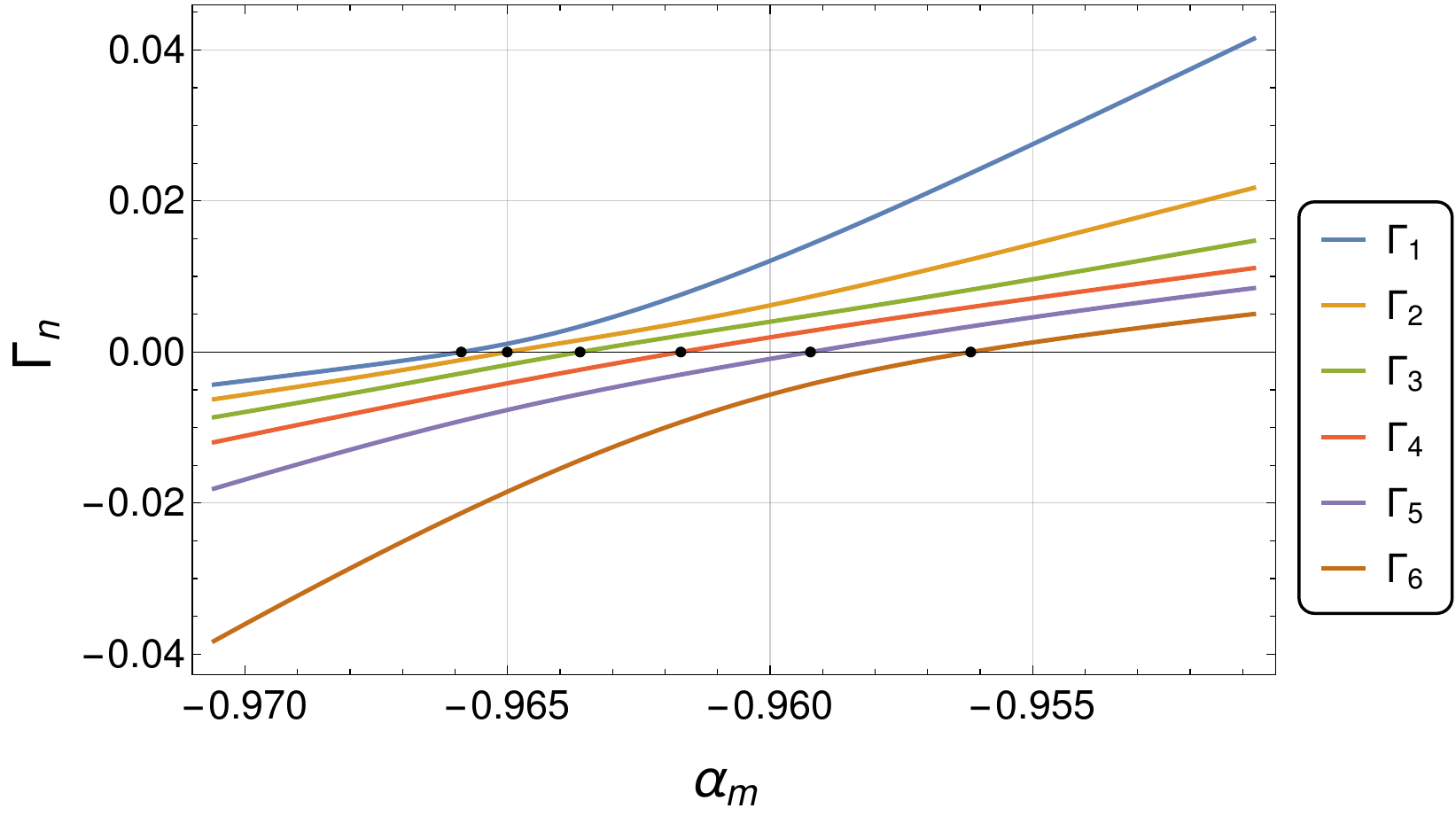}
    \caption{}
    \label{Gaplt}
  \end{subfigure}
\quad
  \begin{subfigure}[]{0.65\textwidth}
    \centering
    \includegraphics[width=1\linewidth]{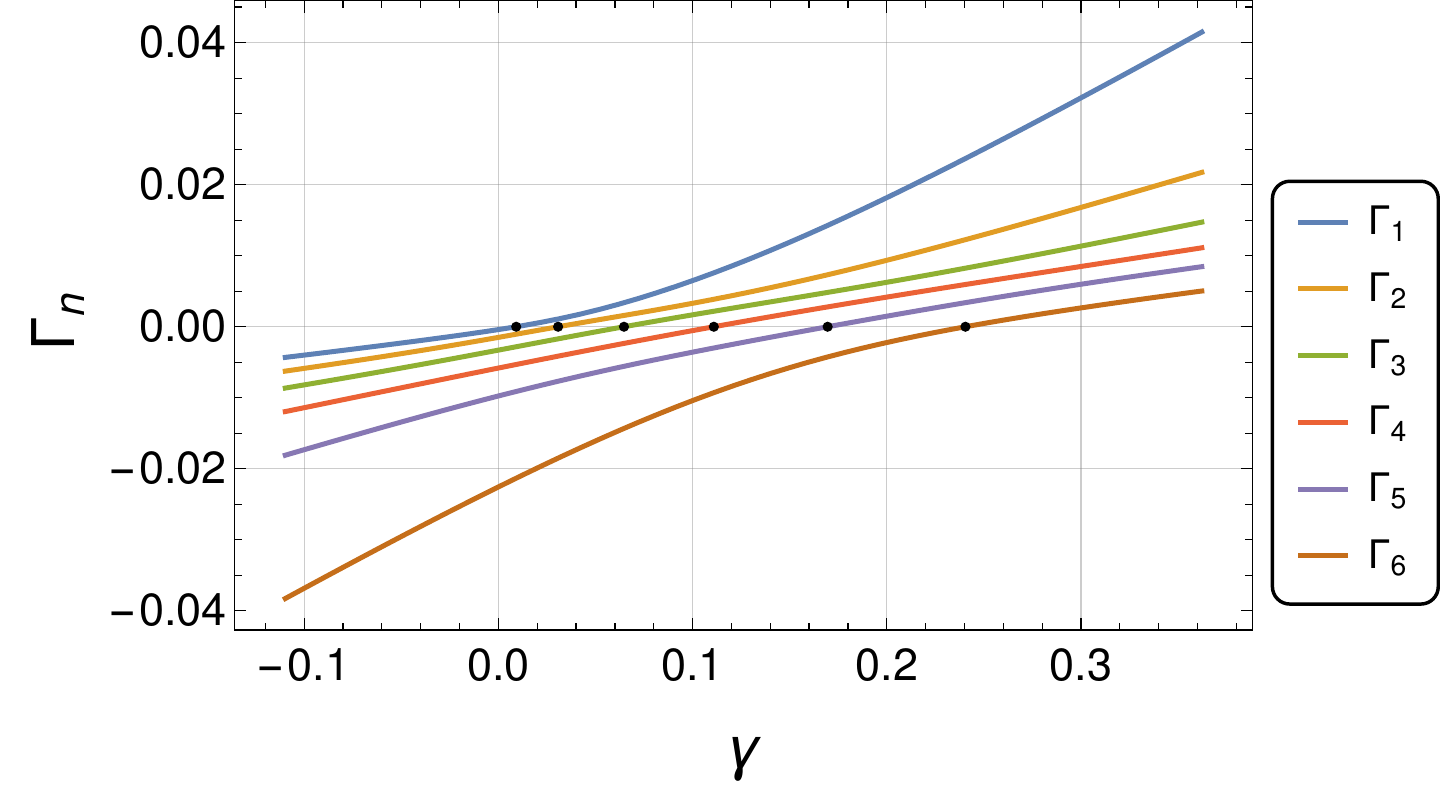}
    \caption{}
    \label{Ggpplt}
  \end{subfigure}
  \caption{(a) The variation of $\gamma$ with $\alpha_m$ for different values of $R_U$. (b) The dependence of different values of $\Gamma_n$ on $\alpha_m$. The parametric plot between $\Gamma_n$ and $\gamma$. The black dots in the figure represent the positions where $\Gamma_n=0$ for different values of $n$. The plots in panels (b) and (c) correspond to the case of $R_U=2$.}
  \label{f:para2}
\end{figure}
The variation of the highest eigenvalue $\Gamma_1$ as a function of $\alpha_m$ is shown in Figure \ref{G1amuplt}. Similar to the case of $\gamma$ as shown in Figure \ref{gamuplt}, we find that $\Gamma_1$ is nearly independent of variation in $\mu$, decreases monotonically with decreasing $\alpha_m$ and increases with increasing $R_U$. Since we find that both $\gamma$ and $\Gamma_n$ (in general) show very weak dependence on $\mu$, so for rest of the analysis in this section we focus only on the dependence of $\gamma$ and $\Gamma_n$ on $\alpha_m$. For a fixed value of $\disp{\mu ~(=-\sqrt{\gamma_1^s}=-0.09579)}$, the variation of $\gamma$ with $\alpha_m$ for different values of $R_U$ is shown in Figure \ref{gplt}. The variation of the different eigenvalues $\Gamma_n$ as a function of $\alpha_m$ is shown in Figure \ref{Gaplt}, and the parametric dependence of $\Gamma_n$ with $\gamma$ is shown in Figure \ref{Ggpplt}. We find that the roots of $\Gamma_n$ (marked by black dots in Figure \ref{f:para2}) occur at values of $\alpha_m$ at which $\gamma=\gamma_m^s$, where $\sqrt{\gamma_m^s}r_d$ is the zero of the Bessel function $\mathcal{J}_m^s$ defined in Equation \eqref{bes}. The functional dependence of various elements of the eigenvector $\mathcal{C}_{1m}$ (corresponding to the eigenvalue $\Gamma_1$) on $\alpha_m$ is shown in Figure \ref{c1mplt}. In this case ($R_U=2$), we find that $\alpha_m^s=-0.966$ (marked by the black dot in Figure \ref{c1mplt}), where $\gamma=\gamma_1^s$ and $\Gamma_1=0$; the coefficient corresponding to $\mathcal{J}_1^s$, $\mathcal{C}_{11}$, attains its maximum value, whereas all other coefficients of $\mathcal{C}_{1m}$ go to zero. This is because, when $\gamma=\gamma_m^s$, from Equation \eqref{mat1}, $\mathrm{Det}~ \tilde{K}=0$ and $\Gamma~ \mathrm{Det}~\tilde{G}=0$. Since $\mathrm{Det}~\tilde{G}\neq0$, this implies $\Gamma =0$ and $\mathcal{C}_{mm}\propto \delta_{mm}$, with $\mathcal{C}_{mm}$ reaching its peak value. At this value of $\alpha_m$, the radial mode of the time-dependent solution is a pure Bessel mode ($\mathcal{J}_1^s$), which also satisfies the steady-state dynamo equation. Hence $\alpha_m^s=-0.966$ represents the value of $\alpha_m$ for which the dynamo solution reaches its steady-state values for $R_U=2$. During the evolution of $\alpha_m$, from its initial value $\alpha_0$ to final value $\alpha_m^s$, $\sqrt{\gamma}$ goes through all Bessel roots $\sqrt{\gamma_m^s}$ (see Figure \ref{f:para2}), which forces $\Gamma_n$ to zero and then to negative values. Subsequently the expression $\disp{w_n=\exp\int_0^\tau \Gamma_n(\tau')\dif \tau'}$ decays for all values $n>1$. Only $w_1$ survives and is amplified before $\Gamma_1$ goes through zero near $\alpha_m=\alpha_m^s$, where $w_1$ saturates to a constant value (see Section \ref{s:evol} for details). This, however, does not give us an estimate of the time required for the solution to reach the steady-state condition and the final magnetic field strength. To find this, we study the dynamical solutions of $\alpha_m$ and $\mu$ in the next subsection.
\subsection{Evolution and saturation of the dynamo with time}
\label{s:evol}
To study the effect of advective and diffusive fluxes on the saturation of the volume-averaged large-scale magnetic field strength $\langle B\rangle$, we vary the vertical advective flux, $R_U$, in the range $0-2$ corresponding to $U_0$ of $0-2$ kms$^{-1}$ and the diffusive flux $R_\kappa$ as 0 or 0.3. The saturated volume-averaged values of $B$, $\alpha_m$ and $\mu$ are denoted as $\langle B_{sat} \rangle$, $\alpha_m^s$ and $\mu^s$ respectively. The results are shown for a choice of $N=6$ in Equation \eqref{qsum}, with the initial condition $q_3=1$, $q_1=q_2=q_4=q_5=q_6=0$ in  Equation \eqref{qn2}, $\alpha_m=10^{-3}$, $\mu=0$ and $R_c=0.01$ at $\tau=0$, corresponding to a seed field of 1 nG. The following are the key results: 
\begin{enumerate}

\begin{figure}[hp]
  \centering
  \begin{subfigure}[b]{0.55\textwidth}
    \centering
    \includegraphics[width=1\linewidth]{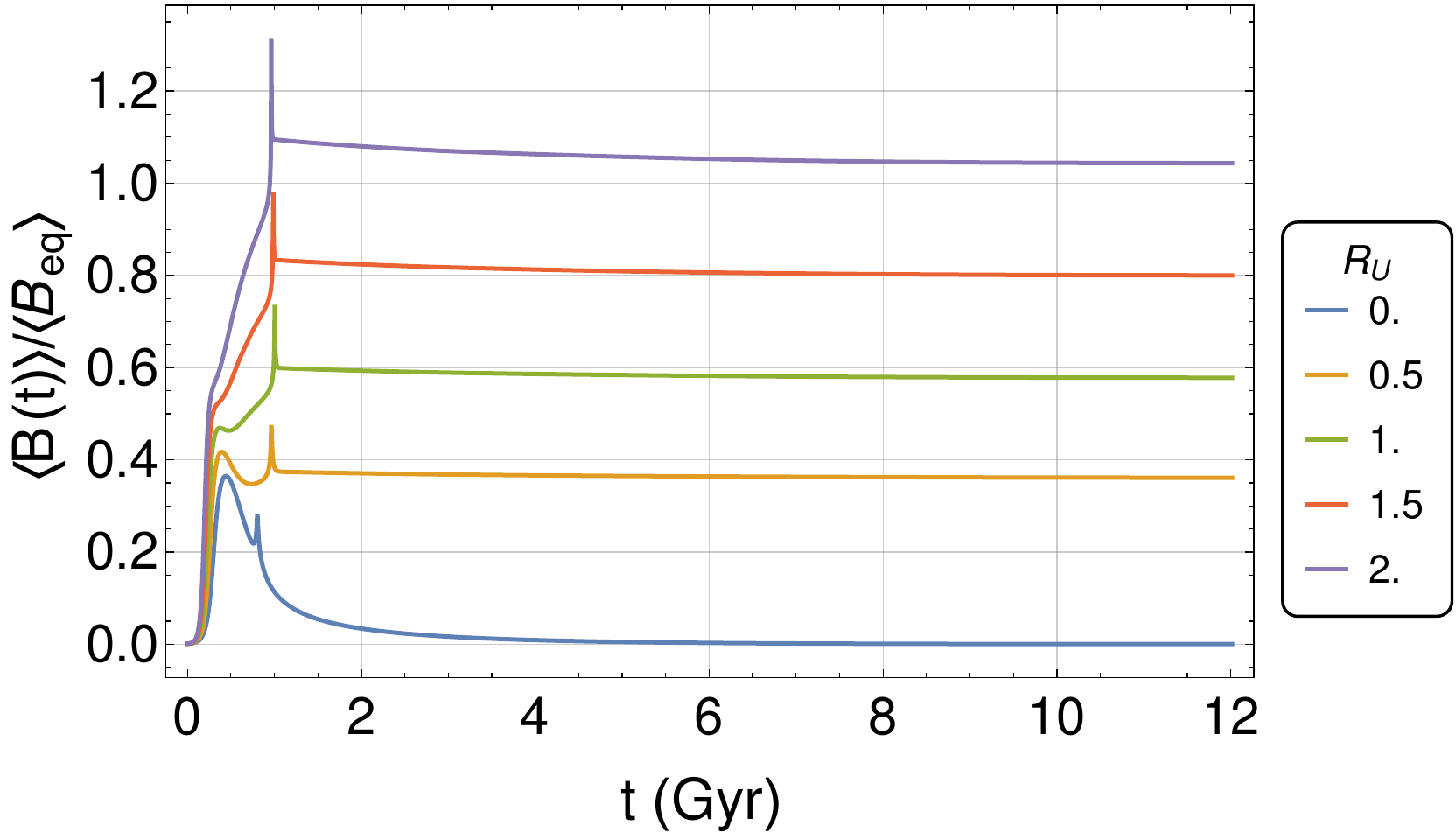}
    \caption{}
    \label{bbfiga}
  \end{subfigure}
\quad
 \begin{subfigure}[b]{0.55\textwidth}
    \centering
    \includegraphics[width=1\linewidth]{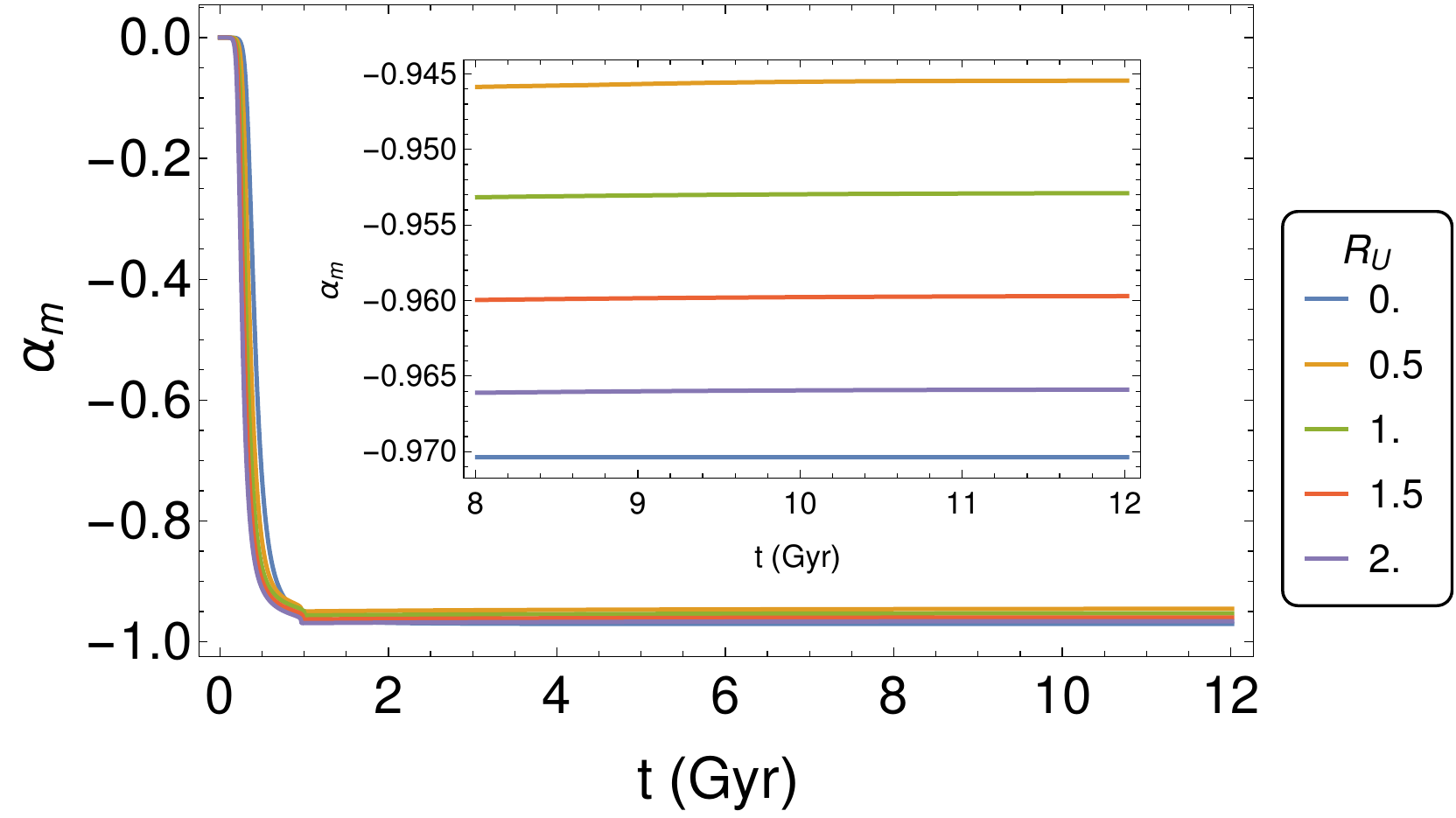}
    \caption{}
    \label{alpfiga}
  \end{subfigure}
\quad
\begin{subfigure}[b]{0.55\textwidth}
    \centering
    \includegraphics[width=1\linewidth]{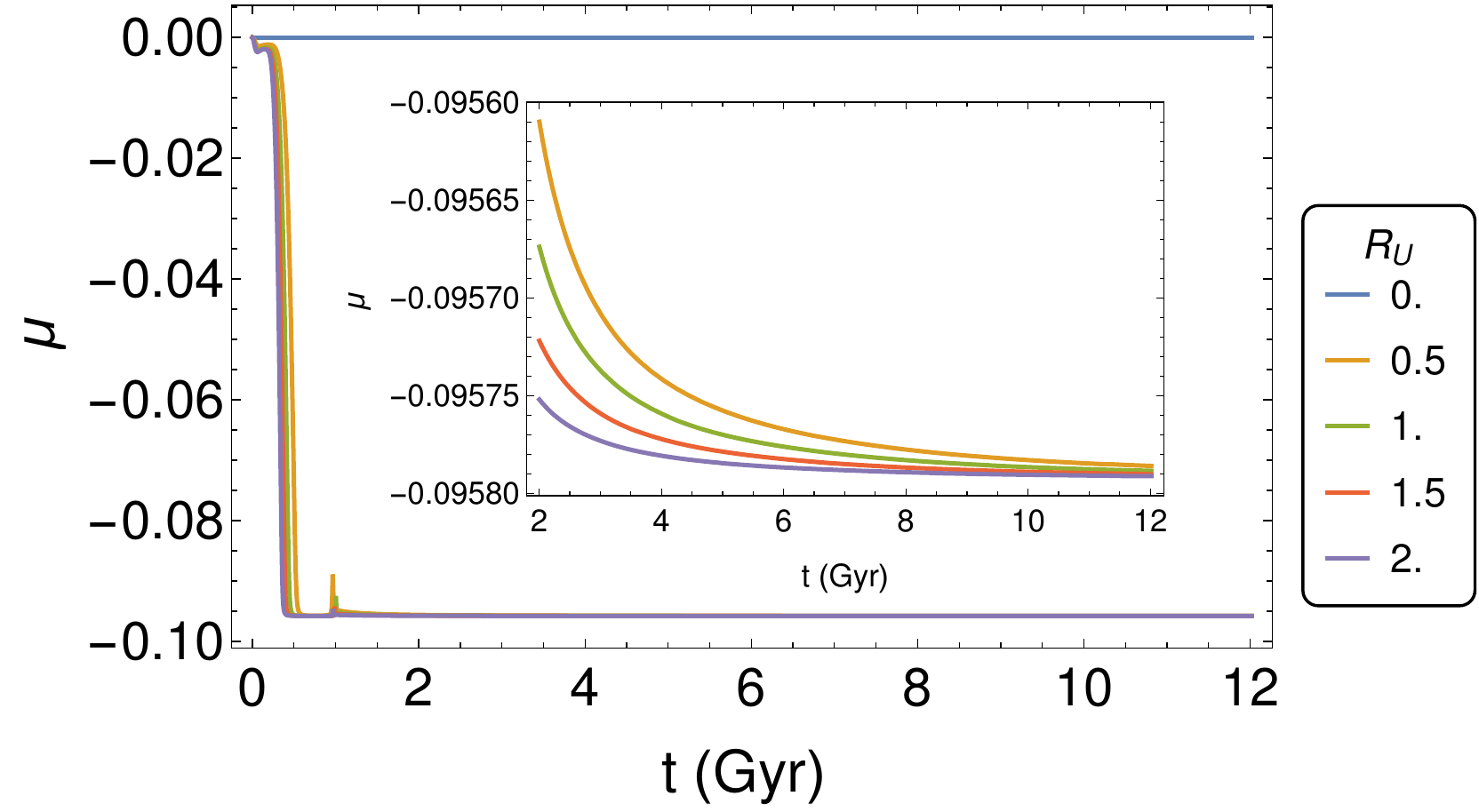}
    \caption{}
    \label{mufiga}
  \end{subfigure}
\quad
  \caption{The evolution of (a) the magnetic field (normalized with respect to the equipartition field strength), (b)
$\alpha_m$, and (c) the force-free parameter $\mu$ with time for different values of $R_U$ and $R_\kappa$ = 0.}
  \label{balphcfig}
\end{figure}

\item We find that in all cases there is initially a brief phase of rapid growth ($t =0-0.5$ Gyr), when the magnetic field grows exponentially with time (see Figure \ref{bbfiga}). This represents the kinematic regime of dynamo operation where $\alpha_m\approx0$ (see Figure \ref{alpfiga}). After that $\alpha_m$ decreases rapidly during the period $t=0.5-1$ Gyr and reaches close to its saturation strength. Since $\alpha_k$ = 1 is assumed to be constant throughout, the total $\alpha-$effect also decreases during this phase. The dynamo starts getting quenched and the mean magnetic field reaches saturation around $9-10$ Gyr. The timescale for saturation, $t_{sat}$, is defined as the time when $\alpha_m$ reaches $99.99\%$ of $\alpha_m^s$ (obtained from the eigen mode analysis discussed in Section \ref{s:para}, also see Table \ref{t:sat}). The saturated mean magnetic field strength is in equipartition with the turbulent kinetic energy ($\sim \langle B_{eq} \rangle$) and reaches $99\%$ of its final value within a period of 1 Gyr. These values (see Table \ref{t:sat}) are in agreement with the results from numerical simulations previously presented in \citet{2013MNRAS.429..967G} and \citet{2014MNRAS.443.1867C}. The final saturation value of $\alpha_m$ is between $-0.93$ to $-0.97$, depending upon the value of $R_U$, which corresponds to a net $\alpha\sim0.03-0.07~(3\%-7\%$ of the initial value).

\item Due to the transport of the small-scale magnetic helicity fluxes from the disk to the corona, the large-scale magnetic helicity of the corona grows with time, carrying the same sign as that of the small-scale fields. This is reflected in the evolution of the force-free parameter $\mu$ in the corona, as shown in Figure \ref{mufiga}. In the kinematic phase, when $\alpha_m=0$, there is no magnetic helicity flux and hence the magnetic helicity of the coronal field is zero. The parameter $\mu$ then grows rapidly for non-zero $\alpha_m$ between $t$ = 0.5 and 1 Gyr. When $\alpha_m$ saturates (around $t\geq 1$ Gyr), $\mu$ also saturates at a value that is very close to $-\sqrt{\gamma_1^s}$. The final saturation values for $\langle B \rangle$, $\alpha_m$, and $\mu$ corresponding to the different values of $R_U$ and $R_\kappa$ are presented in Table \ref{t:sat}.

\begin{figure}[hp]
 \centering
\begin{subfigure}[]{0.55\textwidth}
    \centering
    \includegraphics[width=1\linewidth]{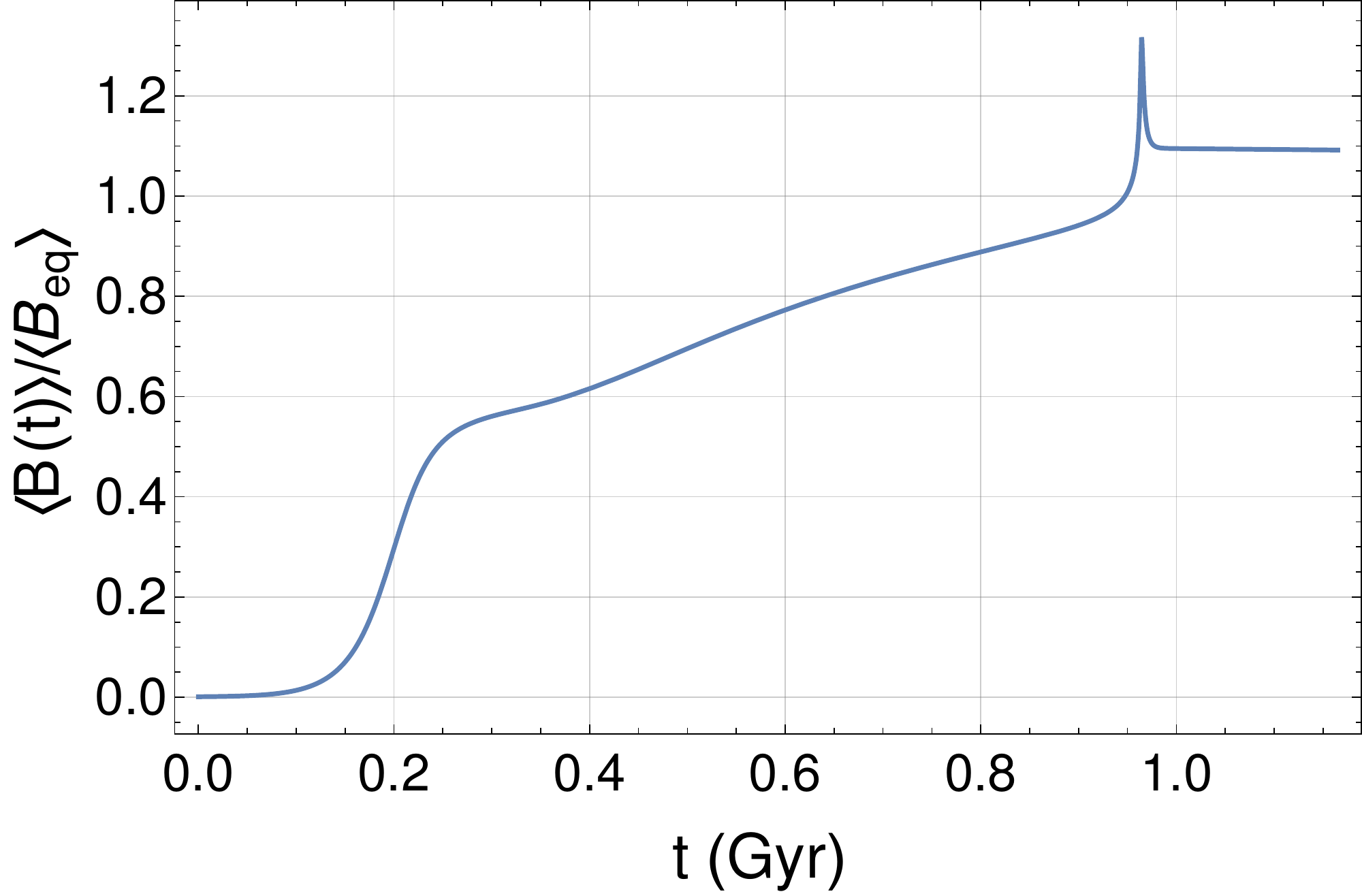}
    \caption{}
    \label{BBplt0}
  \end{subfigure}
\quad
\begin{subfigure}[]{0.55\textwidth}
    \centering
    \includegraphics[width=1\linewidth]{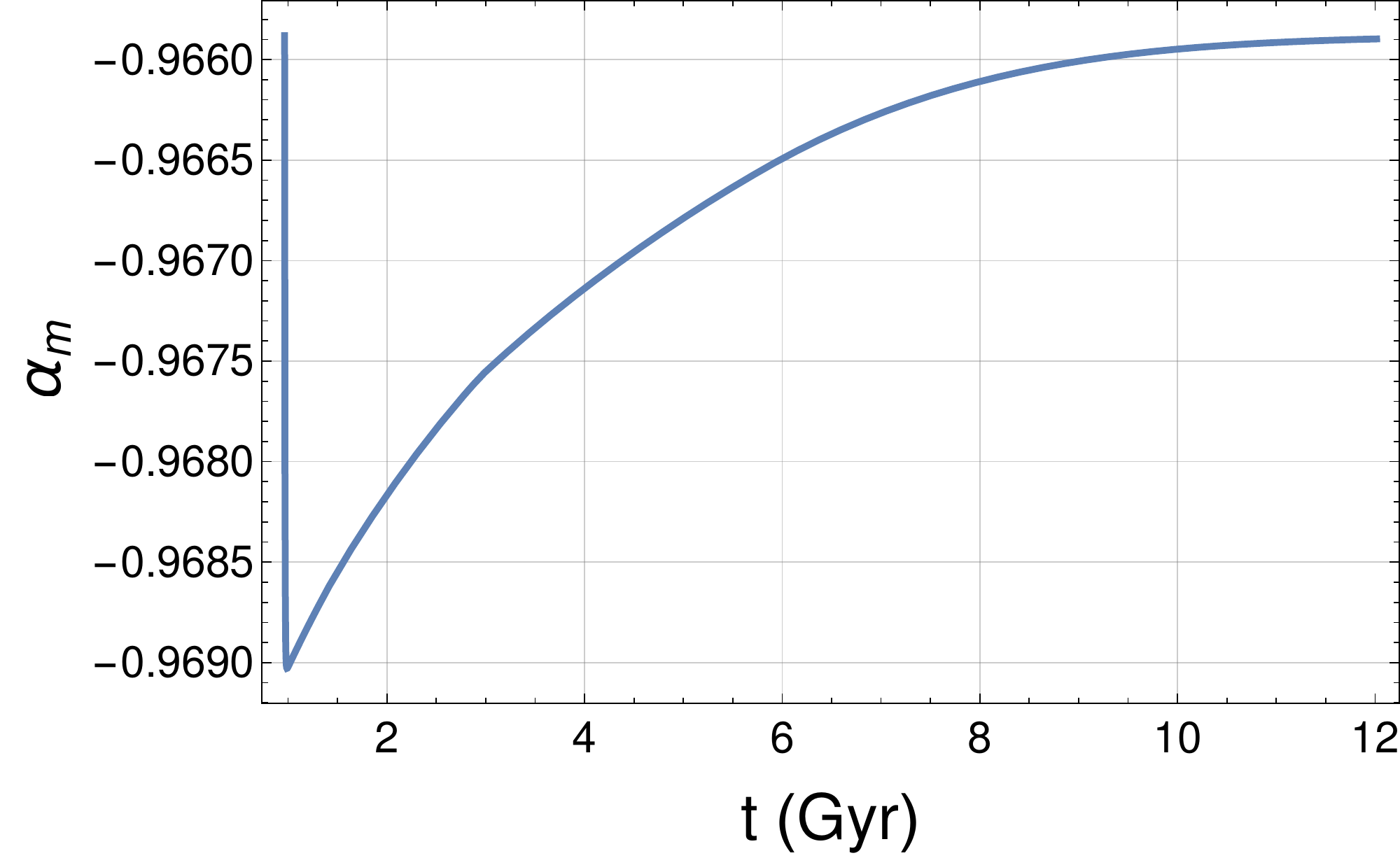}
    \caption{}
    \label{aplt}
  \end{subfigure}
\quad
  \begin{subfigure}[]{0.55\textwidth}
    \centering
    \includegraphics[width=1\linewidth]{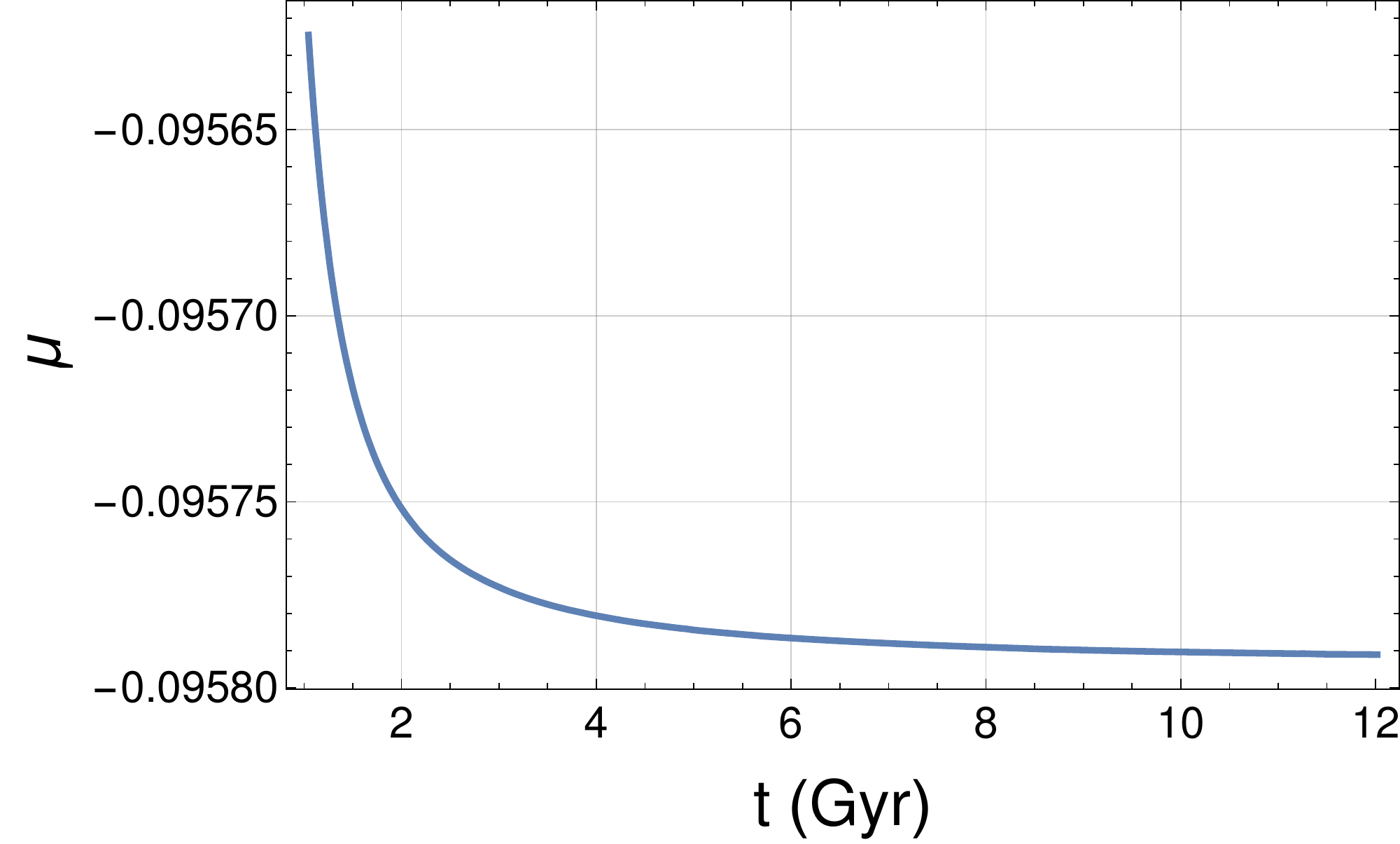}
    \caption{}
    \label{muplt}
  \end{subfigure}
  \caption{(a) The evolution of the volume-averaged magnetic field strength for the case of $R_U=2$ and $R_\kappa=0$ during the period $t=0-1$ Gyr. (b), (c) The variation of $\alpha_m$ and $\mu$ with time respectively for the case of $R_U=2$ and $R_\kappa=0$. For (b) and (c), we have plotted the functions beginning at $t=0.96$ Gyr, when $\Gamma_1=0$, in order to highlight the small variation. }
  \label{f:amc}
\end{figure}

\begin{table}[h]
\centering
\resizebox{\textwidth}{!}{
\begin{tabular}{|c|c|c|c|c|c|c|c|c|c|}
\hline
$R_U$ & $R_\kappa$ & $t_{sat}$ (Gyr) & $\langle B_{sat} \rangle$/$\langle B_{eq} \rangle$ & $\alpha_m^s$ & $\mu^s$ & $t_{99}$ (Gyr) & $\langle B_{99}\rangle$ & $\alpha_{99}$ & $\mu_{99}$ \\ \hline
0 & 0 & 4.35288 & 0.00709 & -0.97026 & 0 & 0.12661 & 0.00702 & -0.00008 & 0 \\ 
0 & 0.3 & 10.73855 & 0.24704 & -0.93758 & -0.09578 & 0.32034 & 0.24457 & -0.11043 & -0.00157 \\ \hline
0.5 & 0 & 9.76078 & 0.3614 & -0.94555 & -0.09578 & 0.87414 & 0.35779 & -0.93575 & -0.09572 \\ 
0.5 & 0.3 & 9.50699 & 0.45716 & -0.94555 & -0.09578 & 0.99903 & 0.45259 & -0.93834 & -0.09569 \\ \hline
1 & 0 & 9.59684 & 0.57857 & -0.95299 & -0.09579 & 0.97572 & 0.57279 & -0.94709 & -0.09571 \\ 
1 & 0.3 & 9.53675 & 0.65967 & -0.95299 & -0.09579 & 1.00783 & 0.65308 & -0.94743 & -0.09571 \\ \hline
1.5 & 0 & 9.51527 & 0.80103 & -0.95981 & -0.09579 & 0.97181 & 0.79301 & -0.95485 & -0.09572 \\ 
1.5 & 0.3 & 9.48252 & 0.87748 & -0.95981 & -0.09579 & 0.98711 & 0.8687 & -0.95498 & -0.09572 \\ \hline
2 & 0 & 9.14301 & 1.04499 & -0.966 & -0.09579 & 0.95521 & 1.03454 & -0.96156 & -0.09574 \\ 
2 & 0.3 & 9.12089 & 1.12063 & -0.966 & -0.09579 & 0.96441 & 1.10942 & -0.96163 & -0.09574 \\ \hline
\end{tabular}
}
\caption{The volume-average steady-state values of $\langle B \rangle$, $\alpha_m$ and $\mu$ along with the corresponding timescale $t_{sat}$ are listed for different input values of $R_U$ and $R_\kappa$ are listed in the first six columns. The last four columns represent the time at which $\langle B\rangle$ reached 99\% of its saturated value, given by $t_{99}$, the corresponding magnetic field strength given by $\langle B_{99} \rangle$ and the corresponding values of $\alpha_m$ and $\mu$ denoted by $\alpha_{99}$ and $\mu_{99}$.}
\label{t:sat}
\end{table}

\item In the absence of the magnetic helicity fluxes ($R_U$ = $R_\kappa$ = 0), we find that the large-scale magnetic field initially grows in the kinematic regime to a strength of $\sim 0.4~ \langle B_{eq}\rangle$ \citep{2007MNRAS.377..874S}, but once the $\alpha$-quenching becomes operative, it decays catastrophically to nearly zero field strength (see Figure \ref{bbfiga}). As expected, the force-free parameter in the corona, $\mu$, remains zero throughout for this case. In all other cases, the saturated value of the field proportional to the net flux, i.e. the saturated mean-field strength, is higher for higher values of $R_U$ and $R_\kappa$.

\item For the case of $U_0=2$ kms$^{-1}$ ($R_U=2$), we obtain a saturated field strength of $\sim$ 1.05-1.12 $\langle B_{eq}\rangle$ depending upon the value of $R_\kappa$ used (see Table \ref{t:sat} and Figure \ref{BBplt0}). This is in good agreement with the field strengths reported in observations \citep{2012SSRv..166..215B,2015ApJ...799...35V}.
The timescale needed for reaching saturation in this case is about 9 Gyr. 

\item At the point when $\Gamma_1=0$, where $\gamma=\gamma_1^s$, the radial function becomes a pure $r \mathcal{J}_1^s$ mode, causing a ``resonance'', and there is sudden transfer of energy from all the other modes to the fundamental mode $\mathcal{J}_1^s$ (seen as a peak in Figure \ref{BBplt0} at $t=0.96$ Gyr). From Equation \eqref{bbeq}, we find that at this point, $B^2\propto \mathcal{C}_{11}^2 w_1^2$. This resonance condition manifests as a sharp jump in the magnetic field strength and sharp decrease in the values of $\alpha_m$ and $\mu$ (as seen in Figure \ref{balphcfig}). For $\sqrt{\gamma}\geq\sqrt{\gamma_1^s}$, $\Gamma_1<0$ (see Figure \ref{Ggpplt}) and $w_1$ slowly saturates (see Figure \ref{wnplt}), and the saturation of $B^2$ follows. Both $\alpha_m$ and $\mu$ change rapidly before this period and very gradually afterwards. In Figures \ref{aplt} and \ref{muplt}, we show the variation of $\alpha_m$ and $\mu$ with time, beginning from $t=0.96$ Gyr (when $\Gamma_1=0$) to $t=12$ Gyr, for the case of $R_U=2$ and $R_\kappa=0$. It can be seen from Figure \ref{aplt} that $\alpha_m$ first overshoots its saturation value near $\alpha_m=-0.9660$ and then asymptotically tends toward this value over the period of the next 10 Gyr. This variation, though small, is still significant, because from Figure \ref{c1mplt} we find that the relative ratio of various radial Bessel modes $\mathcal{C}_{1m}$ is very sensitive to the value of $\alpha_m$ in this region. So, a small difference in the value of $\alpha_m$ can result in a vastly different radial profile for the magnetic field (as shown later in Figures \ref{cnmplt} and \ref{jplt}). This also tells us that even though the magnetic field initially begins with some random combination of Bessel modes, the dynamo equation drives it asymptotically toward a single Bessel mode that is a solution of the steady-state dynamo equation. From Figure \ref{muplt}, we find that $\mu$ asymptotically tends toward the value $-\sqrt{\gamma_1^s}=-0.09579$, which is its maximum permissible value (see Equation \ref{coreq2}).

\begin{figure}[p]
  \centering
\begin{subfigure}[]{0.45\textwidth}
    \centering
    \includegraphics[width=1\linewidth]{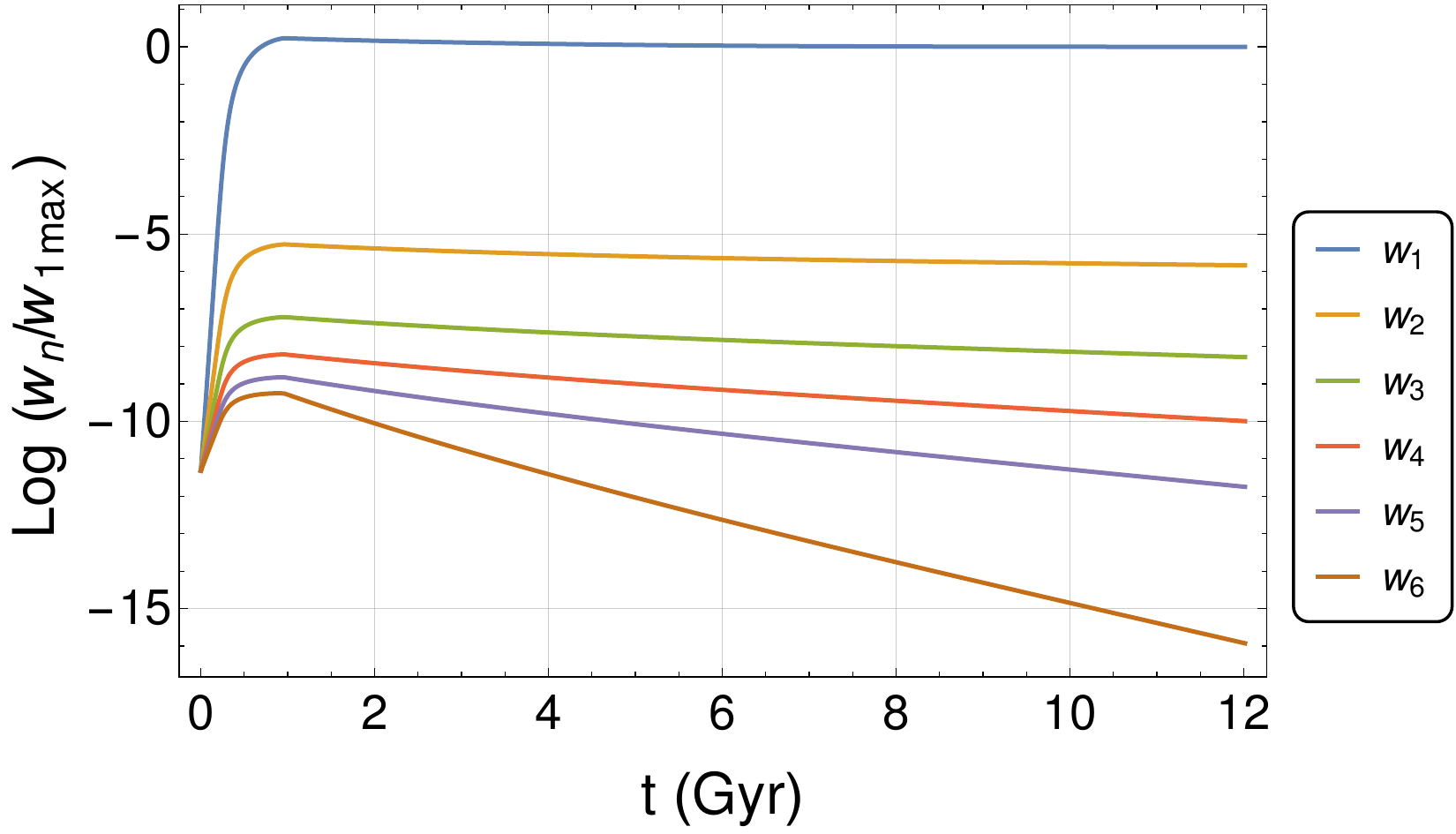}
    \caption{}
    \label{wnplt}
  \end{subfigure}
\quad
\begin{subfigure}[]{0.45\textwidth}
    \centering
    \includegraphics[width=1\linewidth]{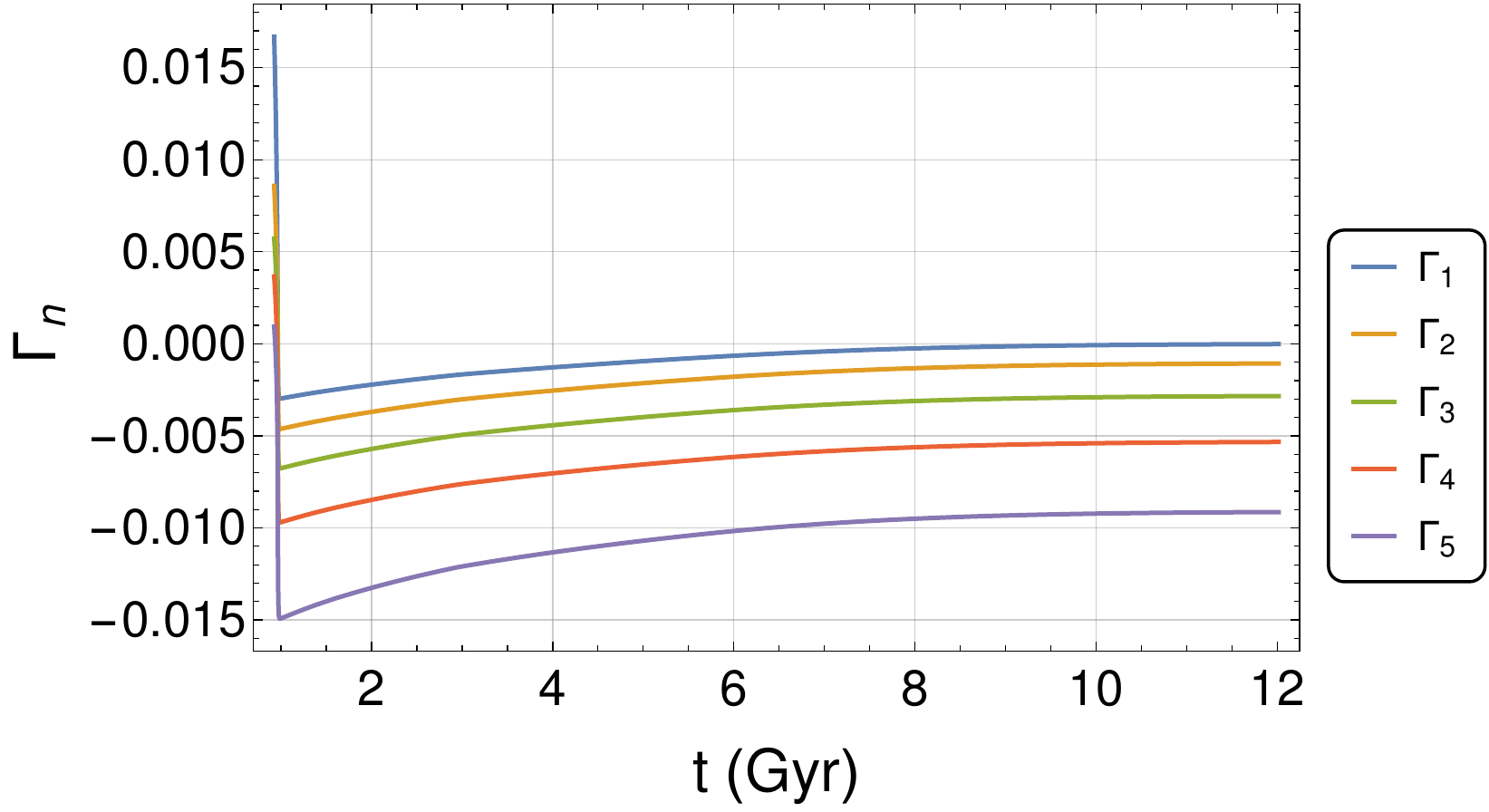}
    \caption{}
    \label{Gntplt}
  \end{subfigure}
\quad
    \begin{subfigure}[]{0.45\textwidth}
    \centering
    \includegraphics[width=1\linewidth]{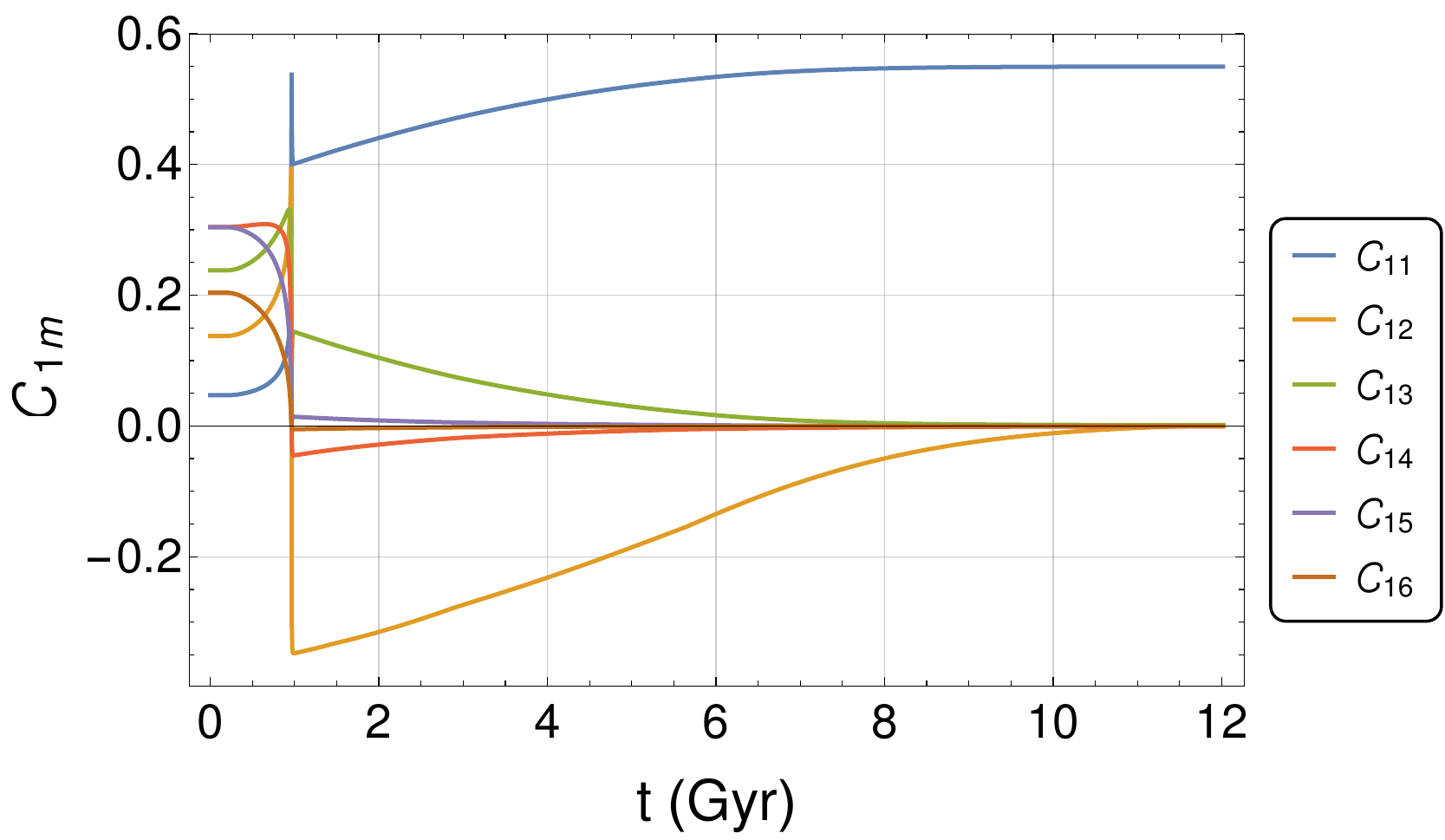}
    \caption{}
    \label{cnmplt}
 \end{subfigure}
\quad
\begin{subfigure}[]{0.45\textwidth}
    \centering
    \includegraphics[width=1\linewidth]{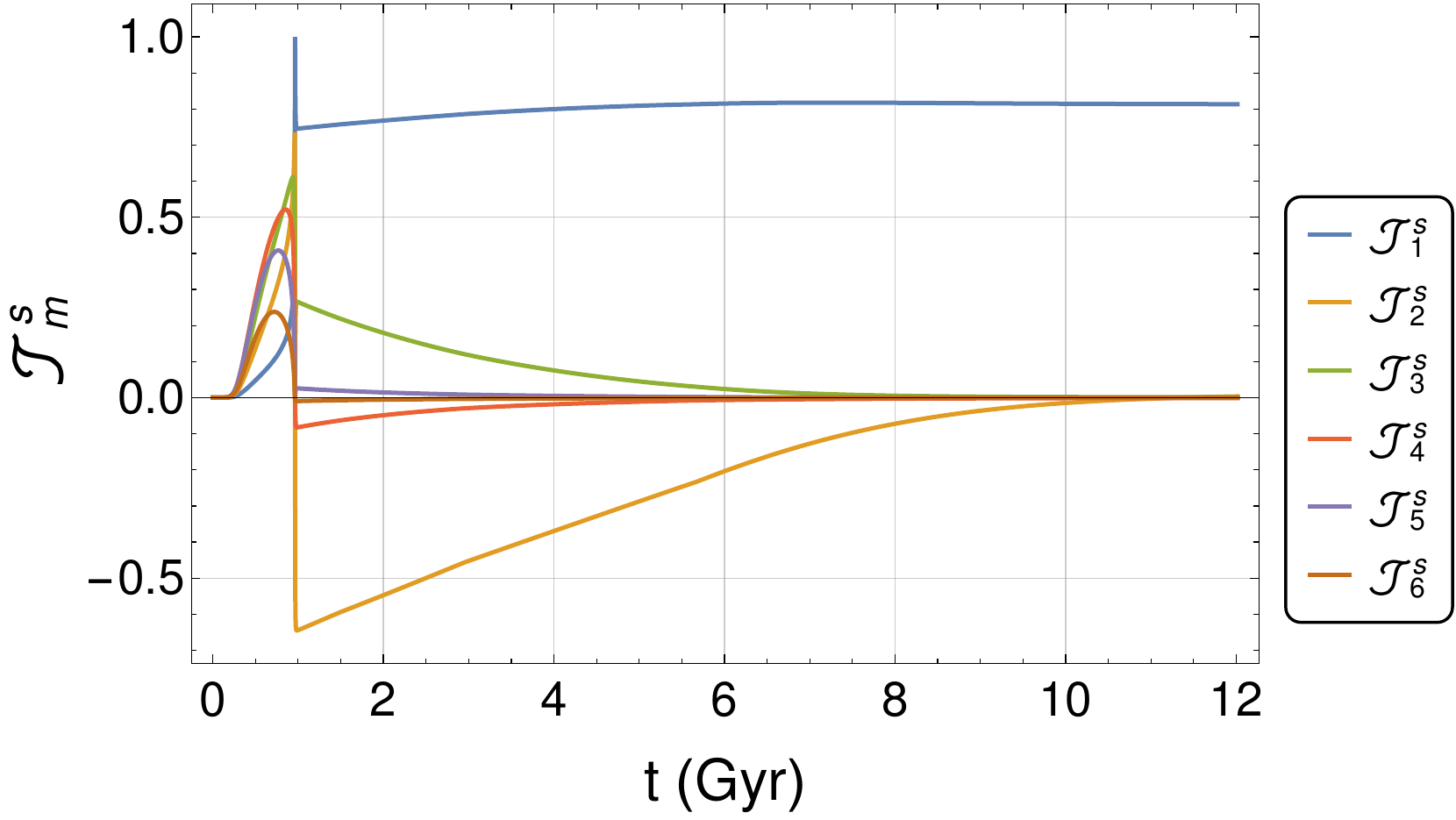}
    \caption{}
    \label{jplt}
  \end{subfigure}
  \caption{(a) The variation of the relative strength of different coefficients $w_n$ scaled with respect to the maximum value of $w_1$ as a function of time. (b) The variation of different eigenvalues $\Gamma_n$ with time $t$.  (c) The variation of different eigenvectors $\mathcal{C}_{1m}$ corresponding to eigenvalue $\Gamma_1$ with time. (d) The temporal evolution of different Bessel modes $\mathcal{J}_m^s$ normalized with respect to the maximum value of $\mathcal{J}_1^s$. All the plots correspond to $R_U=2$ and $R_\kappa=0$. The initial time is set at $t=0.96$ Gyr for panels (a) and (b) and $t=0$ Gyr for panels (c) and (d).}
  \label{f:gwc}
\end{figure}
\begin{figure}[p]
  \centering
  \includegraphics[width=0.6\linewidth]{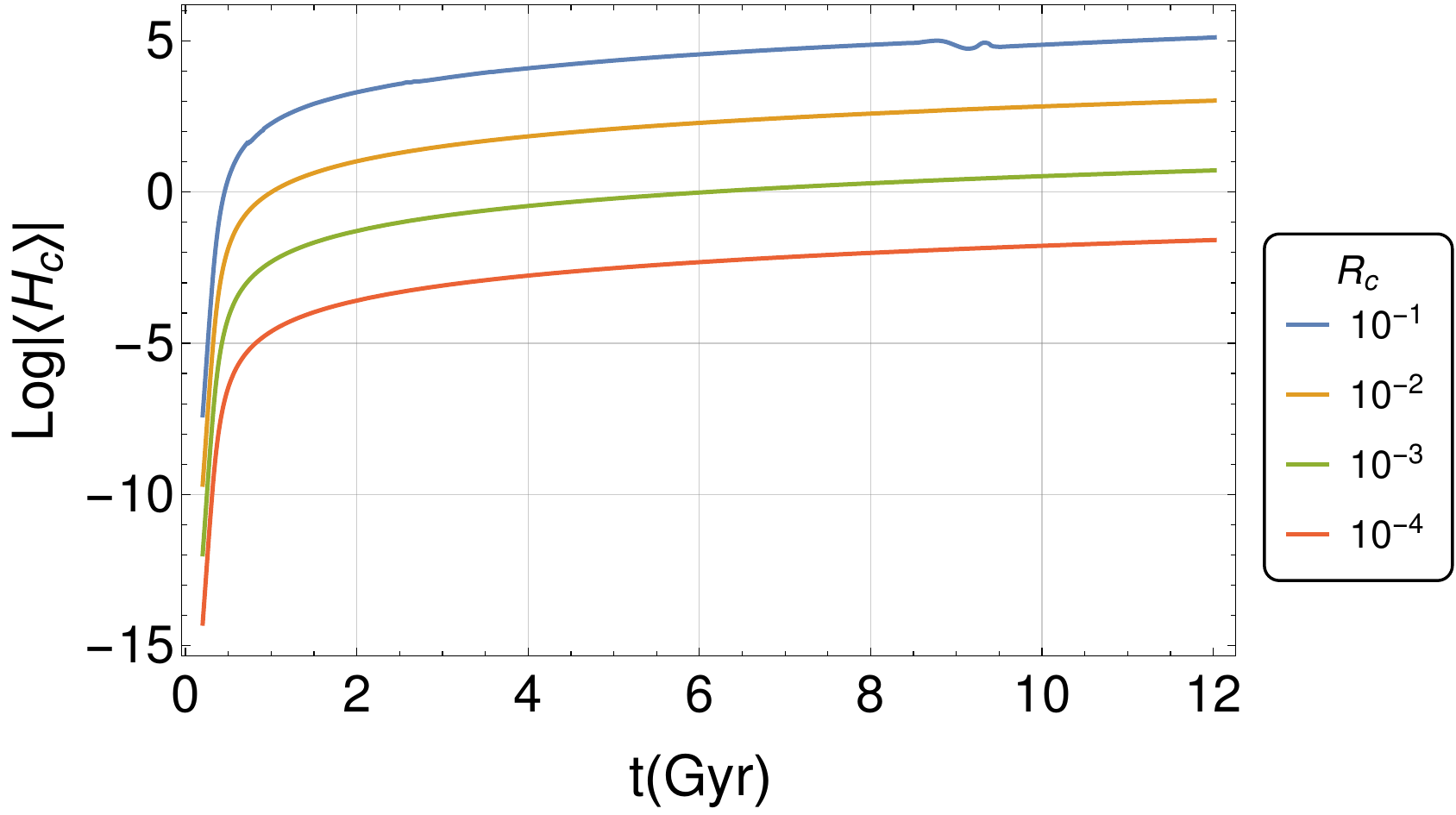}
  \captionof{figure}{The log of absolute mean coronal helicity plotted for different fractions $R_c$, varying between $10^{-4}$ and $10^{-1}$, for the case of $R_U=0$ and $R_\kappa=0.3$.}
  \label{hcffig}
  \centering
  \includegraphics[width=0.6\linewidth]{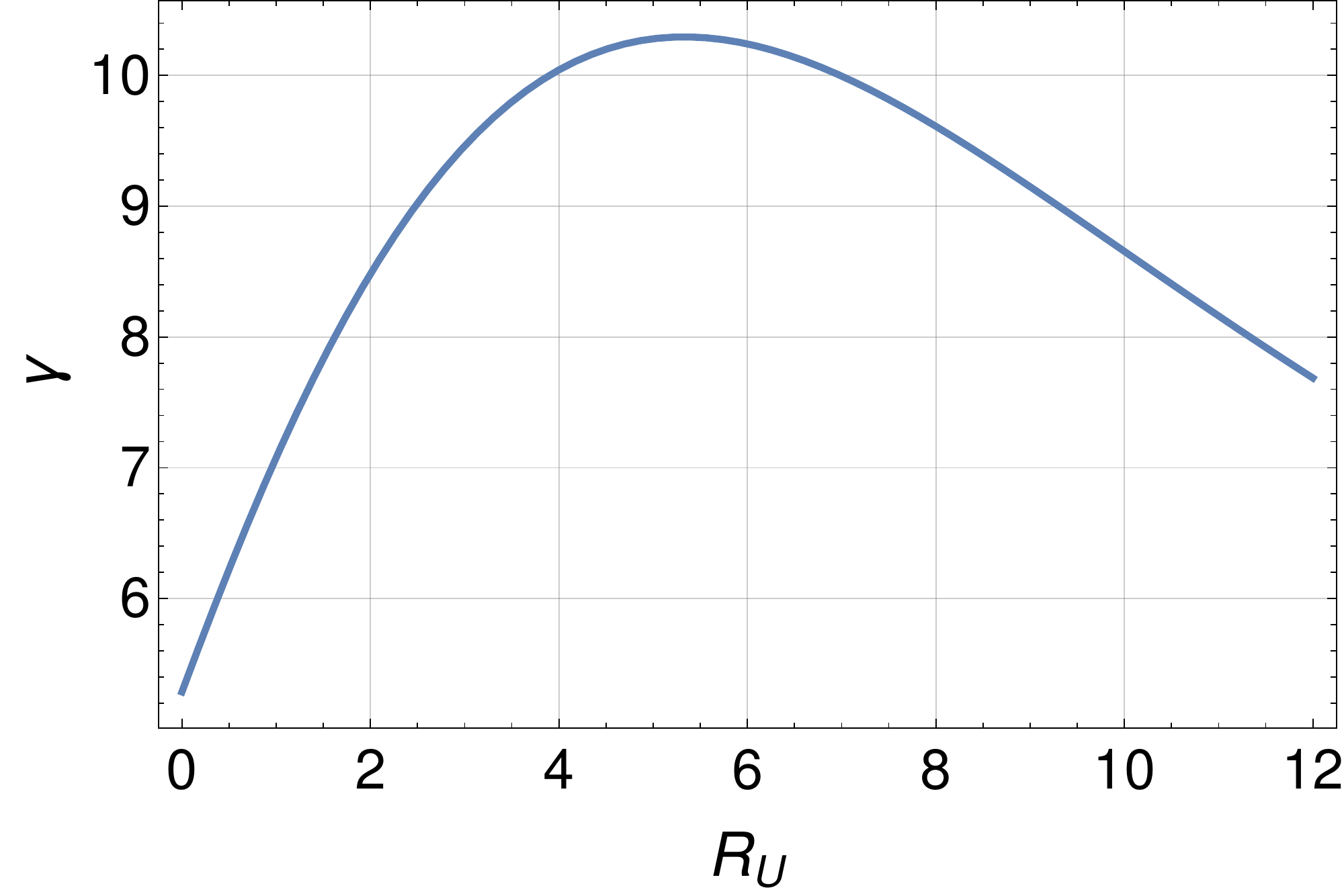}
  \captionof{figure}{The variation of the local growth rate $\gamma$ shown as a function of the advective flux $R_U$, for $\alpha_m=0$ and $\mu=\sqrt{\gamma_1^s}$. The range for $R_U$ in our simulations is 0-2, where $\gamma$ increases linearly with $R_U$.}
  \label{grufig}
\end{figure}

\item Figure \ref{wnplt} shows that the mode $w_1$ is about $10^5$ times stronger than the second highest mode $w_2$. Thus for all practical purposes, we need to follow only the behavior of eigenvalue $\Gamma_1$ and its corresponding eigenvector $\mathcal{C}_{1m}$. The variation of different eigenvalues $\Gamma_n$ with time is shown in Figure \ref{Gntplt} for the case of $R_U=2$ and $R_\kappa=0$. The initial value of $\Gamma_1$ for $\alpha_m=0$ is around 1.2 (see Figure \ref{G1amuplt}); it then decreases to zero at $\alpha_m=\alpha_m^s$. The behavior of $\Gamma_1$ mimics that of $\alpha_m$ as shown in Figure \ref{aplt}, whereby it initially overshoots the point $\Gamma_1=0$ and then asymptotically tends toward this point. The variation of different elements of $\mathcal{C}_{1m}$ is depicted in Figure \ref{cnmplt}. We notice that even though the initial magnetic field starts as a random mixture of different Bessel modes, after $t=0.96$ Gyr all other modes except $\mathcal{C}_{11}$ decay in strength. Finally, around $t=10$ Gyr, only $\mathcal{C}_{11}$ is dominant, while all other $\mathcal{C}_{1m}$ are nearly zero. Thus the asymptotic radial configuration is a pure Bessel mode $\mathcal{J}_1^s$. This can also be observed in Figure \ref{jplt}, where the temporal  evolution of Bessel modes $\mathcal{J}_m^s=\sum_{n=1}^N w_n \mathcal{C}_{nm}$ normalized with respect to the maximum value of $\mathcal{J}_1^s$ is shown.

\item The steady-state configuration is independent of the relative strengths of different Bessel modes $q_n$, taken  at $\tau=0$. We have explored different choices of seed fields, such as taking a pure $n=2$ seed field ($q_2=1$, all other $q_n=0$) or a seed field with a mixture of modes. These different choices of seed field only shift the time required to reach the steady state depending upon its closeness to the final configuration. For example, a seed field of pure $n=1$ mode reaches saturation faster (in about 0.1 Gyr) than a pure $n=2$ mode, because the final configuration in this case is the pure Bessel mode $\mathcal{J}_1^s$. In Figure \ref{jplt} we find that, since the initial seed field was given as a pure $\mathcal{J}_3^s$ mode, it initially dominates, but after $t=1$ Gyr, $\mathcal{J}_1^s$ remains the most dominant mode. We also see that $N=6$ is a good approximation as the contribution from the higher modes is quite small and for the most part only the first two modes $\mathcal{J}_1^s$ and $\mathcal{J}_2^s$ are dominant. We have also checked the solutions with higher values of $N$ ($N=12$ for a smaller range in $R_U$ and $R_\kappa$), and the results were found to be qualitatively similar because the higher orders do not contribute much to the final saturated field and decay at a much faster rate. Since we are computationally constrained, we have restricted our calculations to $N=6$.

\item The $r$, $z$, and $\phi$ components of the large-scale magnetic field in the corona $\mathbf{B_c}$, can be written by combining Equations (\ref{poleq}), (\ref{toreq}) and (\ref{e:psicor}) as
\begin{subequations}
\begin{align}
 (\mathbf{B_c})_r&=-\frac{1}{r}\fpar{\psi_c}{z}=\sum_{n,m=1}^N \sqrt{\gamma_m^s-\mu^2} w_n \mathcal{C}_{nm} a(1)(Q_m^s/r) \exp\left(\sqrt{\gamma_m^s-\mu^2}(1-z)\right)\label{bcr}\\
 (\mathbf{B_c})_z&=\frac{1}{r}\fpar{\psi_c}{r}=\sum_{n,m=1}^N w_n \mathcal{C}_{nm} a(1)(Q_m^{'s}/r) \exp\left(\sqrt{\gamma_m^s-\mu^2}(1-z)\right)\label{bcz}\\
 (\mathbf{B_c})_\phi&=\frac{1}{r}\mu \psi_c=\sum_{n,m=1}^N \mu w_n \mathcal{C}_{nm} a(1)(Q_m^s/r) \exp\left(\sqrt{\gamma_m^s-\mu^2}(1-z)\right)\label{bcp}.
\end{align}
\label{bcor}
\end{subequations}
Near saturation, only $w_1$ and $\mathcal{C}_{11}$ are the dominant terms (from Figures \ref{wnplt} and \ref{cnmplt}), and we find from Equation \eqref{bcr} that $(\mathbf{B_c})_r\sim0$ in the corona (since $\mu^s\sim \sqrt{\gamma_1^s}$; see Table \ref{t:sat}). Due to the small value of $|\mu^s|$, the strength of $(\mathbf{B_c})_\phi$ is also much weaker in the corona than in the disk (see Section \ref{s:dist} for more details). We infer from Equation \eqref{bcor} that the strength of the large-scale magnetic field in the corona depends only on the saturated values of $\mu$, $w_n$, and $\mathcal{C}_{nm}$, and these parameters are found from our simulations to be nearly independent of $R_c$. The parameter $R_c$ only changes the rate at which large-scale magnetic helicity accumulates in the corona during the course of dynamo operation. For higher values of $R_c$, the final value of $\mu$ will then approach its asymptotic value of $\sqrt{\gamma_1^s}$ faster (see Figure \ref{muplt}). This leads to an increase in the vertical length scale of the coronal field, which is effectively proportional to $1/\sqrt{\gamma_1^s-\mu^2}$ (from Equation \ref{e:psicor}). Thus from Equation (\ref{a:Hcf}), higher values of $R_c$ result in higher values of $H_c$ as shown in Figure \ref{hcffig} purely because of the increase in the extent of the large-scale field while its strength does not change significantly. This seems to be in agreement with our simulations done up to $R_c=0.1$, which is the higher limit allowed from reconnection arguments. For all the values of $R_c$ shown in Figure \ref{hcffig}, the values of $\langle B_{sat}\rangle$, $\alpha_m^s$ and $\mu^s$ are nearly the same. This is because the magnetic field in the corona is quite weak compared to that within the disk and the solutions within the disk are only weakly dependent on $\mu$.

\item We also find in Figure \ref{bbfiga} that the growth rate, $\gamma$ of the magnetic field is proportional to $R_U$ (for a given $R_\kappa$). This is true even in the kinematic regime. To illustrate the dependence of $\gamma$ on the advective flux $R_U$, we use the kinematic solutions of the dynamo equation (obtained by solving only Equations (\ref{qeq})-(\ref{bneq}) with $\alpha_m=0$ and $\mu=\sqrt{\gamma_1^s}$) for a larger range of $R_U$ (up to $R_U=20$). This is shown in Figure \ref{grufig}. We find that the growth rate increases with $R_U$ for smaller values ($R_U= 0-5$) until a maximum value of $R_U=5$ is reached, and then it decreases monotonically for higher values of $R_U$. Since, we have used only values of $R_U$ between 0 and 2, we find that in our cases that higher values of $R_U$ help the dynamo to operate faster. \citet{1992A&A...259..453B,1993A&A...271...36B} have also reported similar results in their numerical simulations where the dynamo action is enhanced by the aid of galactic winds.
\end{enumerate}

\subsection{Distribution of the saturated magnetic field across the disk}
\label{s:dist}
Here we discuss the structure of the steady-state magnetic field and its distribution across the disk and the corona. The complete radial and vertical dependences of the fields are shown through contour plots with respect to $r$ and $z$. The following are the key results.
\begin{enumerate}
 \item In Figure \ref{psifig}, we show the meridional contour plots of $\psi$ corresponding to $R_U=2$ and $R_\kappa=0$ at $t =t_{99} = 0.96$ Gyr (when the volume-averaged magnetic field strength $\langle B \rangle$ reaches 99\% of its final strength), $t=5$ Gyr (an intermediate period during the evolution of magnetic field) and $t=t_{sat}= 9.1$ Gyr, when the magnetic field almost reaches its steady-state configuration. These plots depict the shape of the poloidal component of the magnetic field. From Figure \ref{psi1}, we find that initially at $t=0.96$ Gyr, the magnetic field is primarily confined to the radius $r=0-8$ kpc with the field being strongest at around $r=4$ kpc. Subsequently, the magnetic field diffuses across the disk, and at $t=5$ Gyr (see Figure \ref{psi5}), it is predominantly confined to $r=4-16$ kpc. Finally, as shown in Figure \ref{psi10}, at $t=9.1$ Gyr, when the magnetic field has reached a near steady-state configuration, the field is spread out across the disk. The radial profile of $\psi$ is now proportional to $r \mathcal{J}_1^s$ as discussed previously in Section \ref{s:para}. \citet{1993MNRAS.264..285P}, in a similar analytical approach for a nonlinear thin disk galactic dynamo, found magnetic field reversals that occur in a quasi-stationary states for certain choices of seed fields.
\begin{figure}[hp]
 \centering
\begin{subfigure}[]{0.48\textwidth}
    \centering
    \includegraphics[width=1\linewidth]{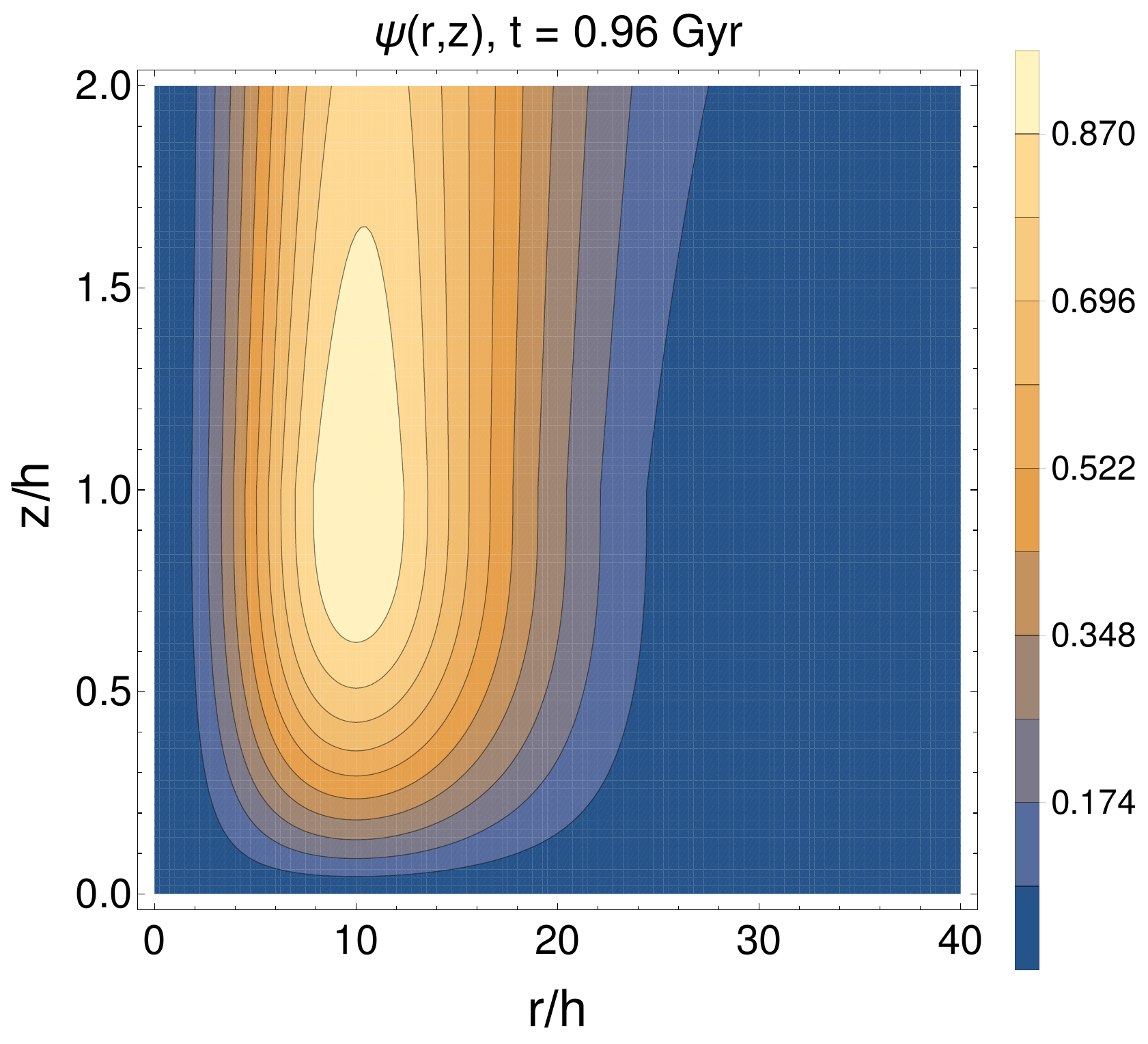}
    \caption{}
    \label{psi1}
  \end{subfigure}
\quad
\begin{subfigure}[]{0.48\textwidth}
    \centering
    \includegraphics[width=1\linewidth]{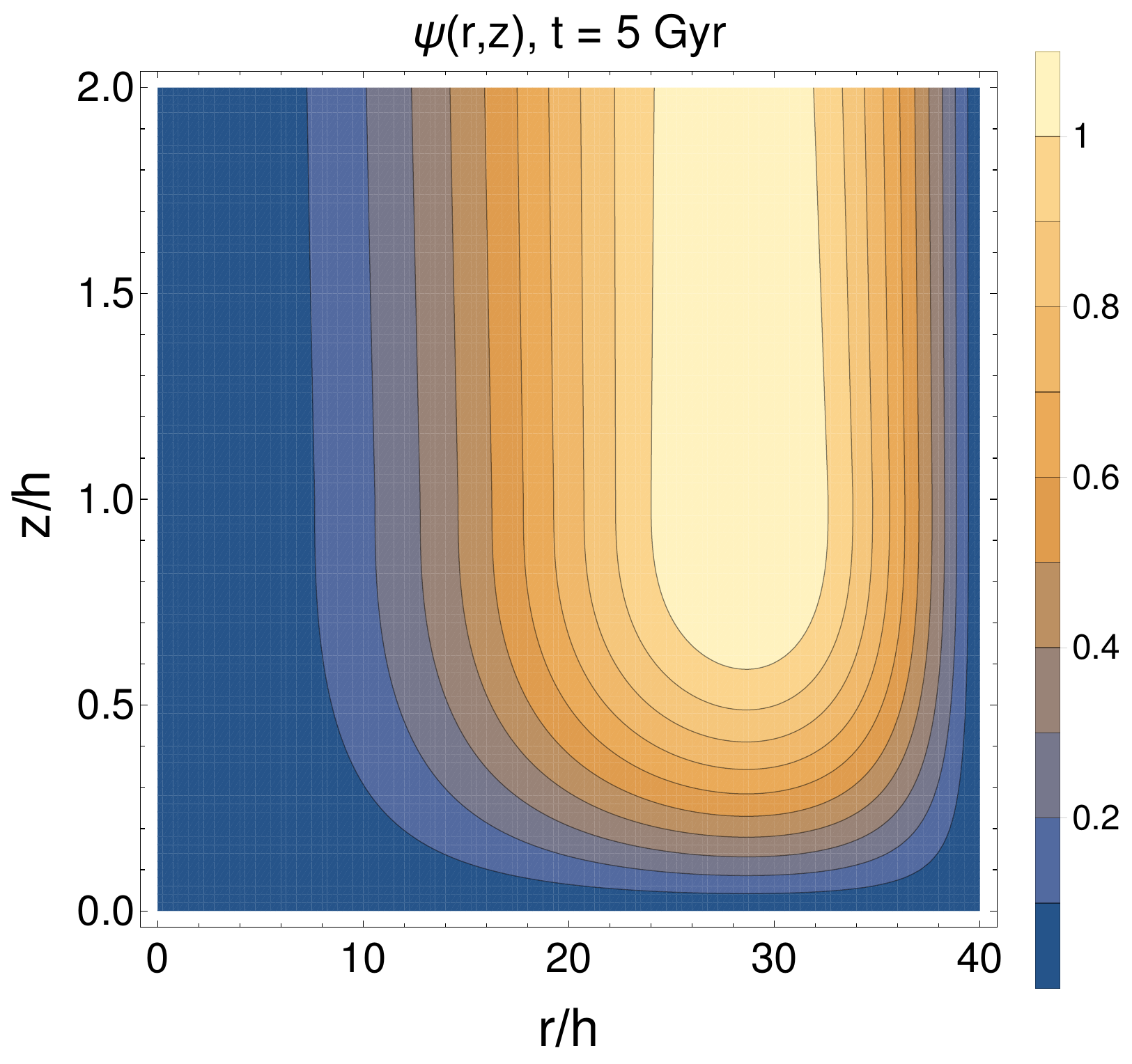}
    \caption{}
    \label{psi5}
  \end{subfigure}
\quad
  \begin{subfigure}[]{0.6\textwidth}
    \centering
    \includegraphics[width=1\linewidth]{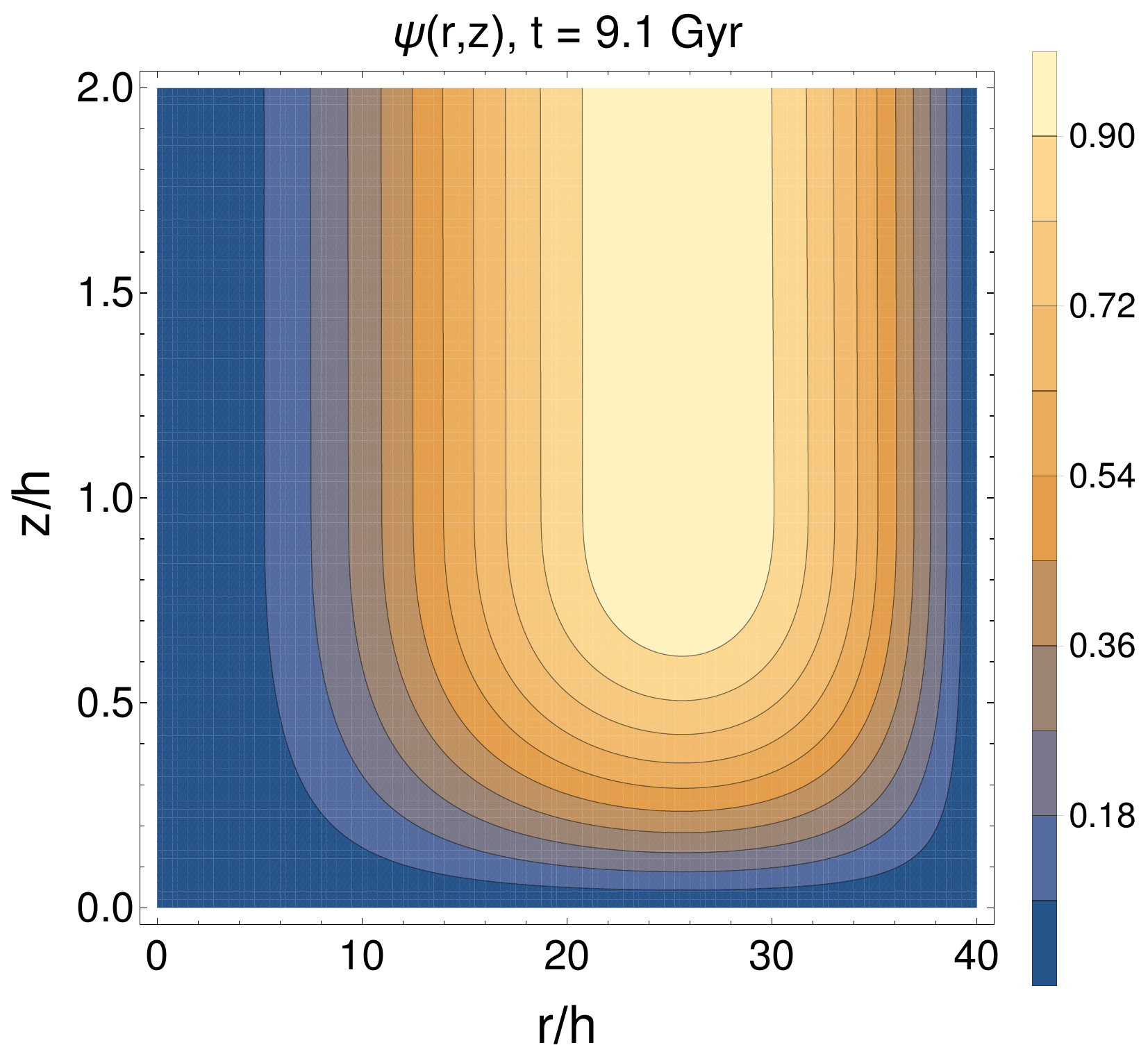}
    \caption{}
    \label{psi10}
  \end{subfigure}
  \caption{The meridional contour plots of $\psi$ corresponding to $R_U=2$ and $R_\kappa=0$ shown at different instants of time. (a) $t =t_{99} = 0.96$ Gyr when the volume-averaged magnetic field strength $\langle B \rangle$ reaches 99\% of its final strength. (b) $t=5$ Gyr, an intermediate period during the evolution of magnetic field. (c) $t=t_{sat}= 9.1$ Gyr, when the magnetic field almost reaches its steady-state configuration. Here $h=400$ pc corresponds to the half-width of the galactic disk.}
  \label{psifig}
\end{figure}
\item In Figure \ref{Tfig}, we show the meridional contour plots of $T$ corresponding to $R_U=2$ and $R_\kappa=0$ at $t =t_{99} = 0.96$ Gyr, $t=5$ Gyr and $t=t_{sat}= 9.1$ Gyr. The poloidal current becomes negligibly small outside the disk ($z>1$). The structural evolution of $T$ is similar to $\psi$, whereby we find that at $t=0.96$ Gyr (see Figure \ref{T1g}), the field is primarily confined to $r=0-4$ kpc, and then diffuses across the disk over time (see Figure \ref{T5g}). The near steady-state configuration is then shown in Figure \ref{T10g} at $t=9.1$ Gyr. The contour plots are shown only up to $z=1$ since the value of $T$ is very small outside the disk. Due to the small value of $\mu^s$, these results are not very different than what we have found under vacuum boundary conditions of $\mu=0$, which is generally considered for the disk dynamo \citep{1988ASSL..133.....R,2007MNRAS.377..874S,2014MNRAS.443.1867C}. The individual components of magnetic field inside and outside the disk are discussed below.
\begin{figure}[hp]
 \centering
\begin{subfigure}[]{0.48\textwidth}
    \centering
    \includegraphics[width=1\linewidth]{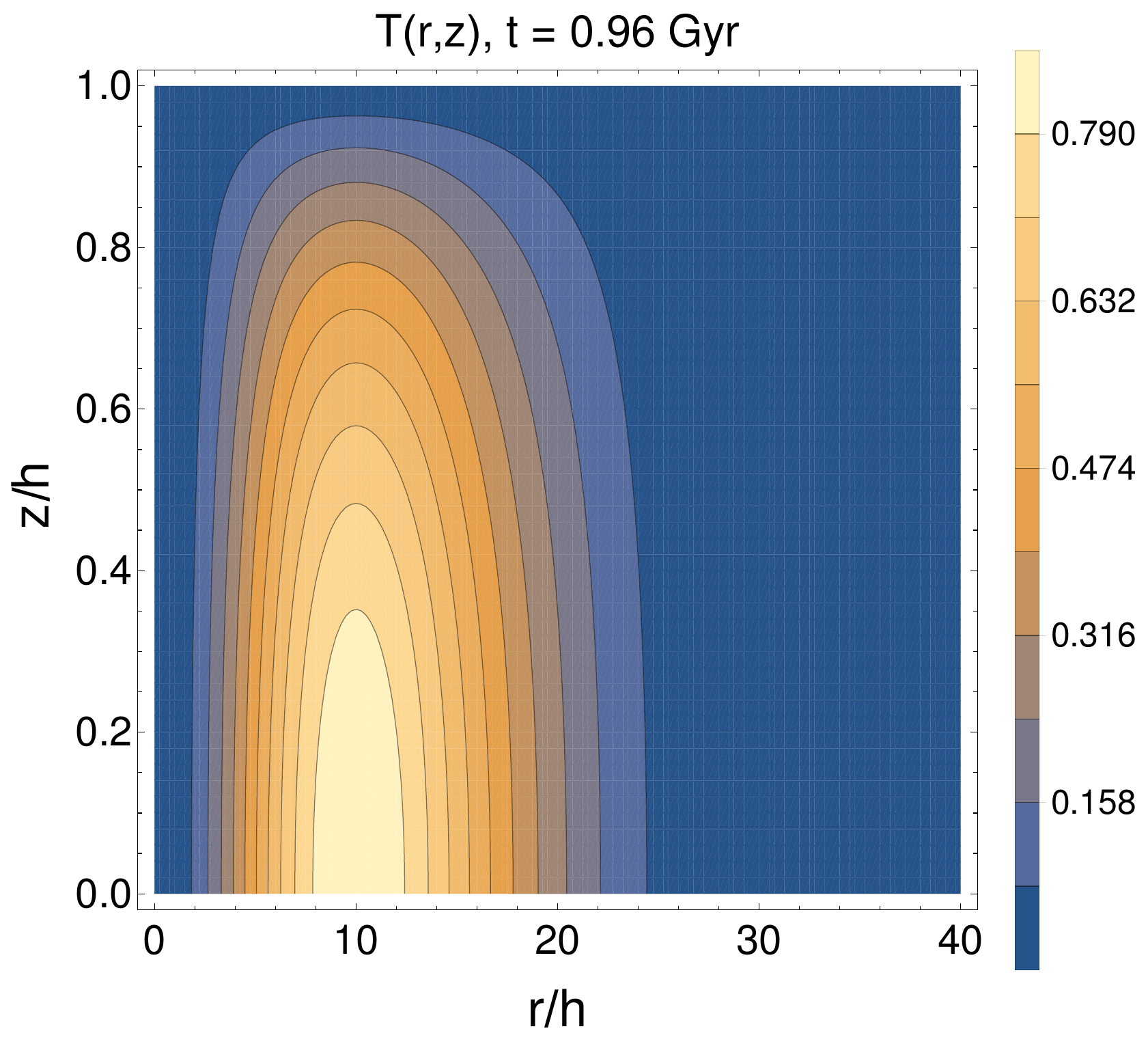}
    \caption{}
    \label{T1g}
  \end{subfigure}
\quad
\begin{subfigure}[]{0.48\textwidth}
    \centering
    \includegraphics[width=1\linewidth]{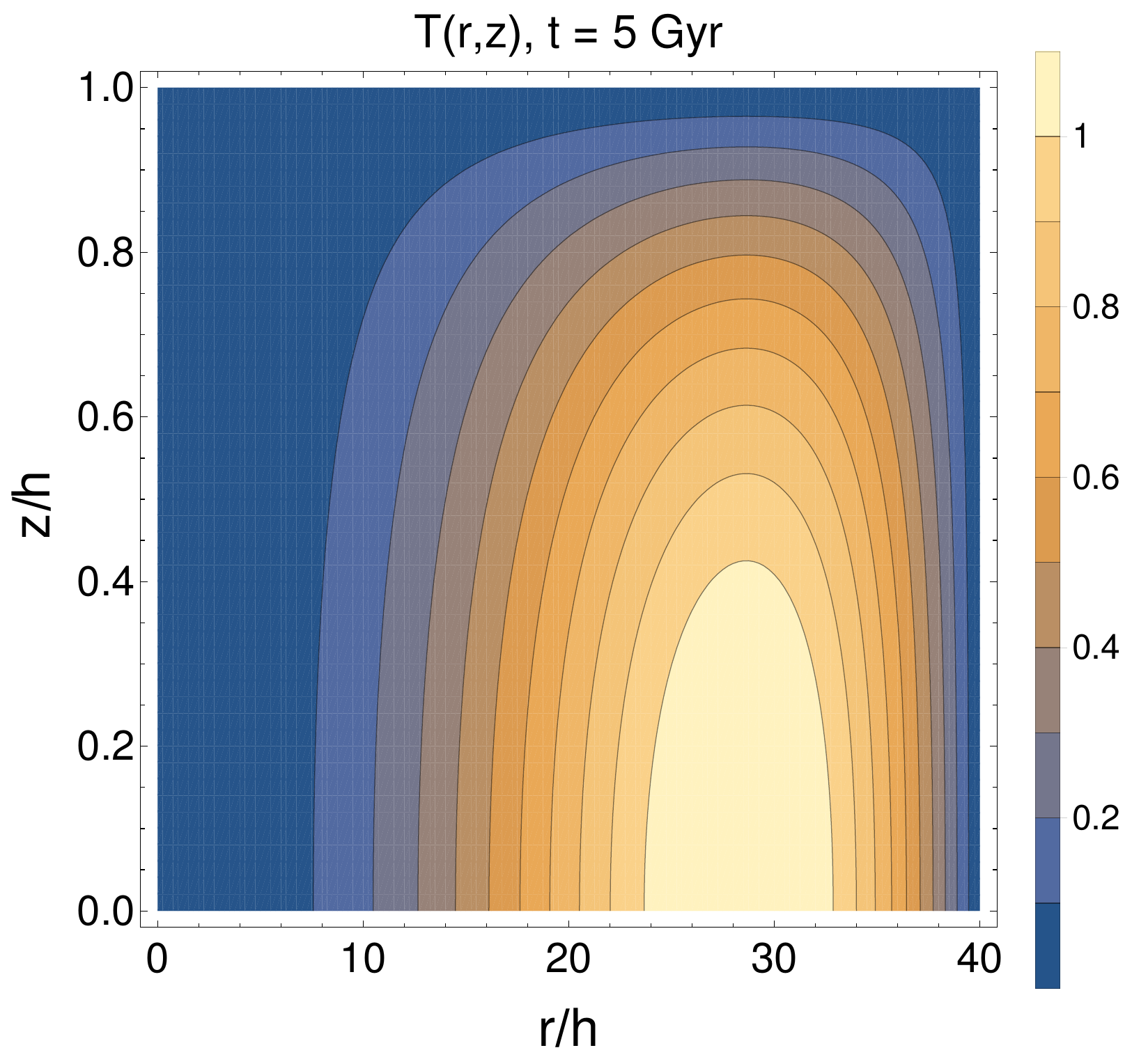}
    \caption{}
    \label{T5g}
  \end{subfigure}
\quad
  \begin{subfigure}[]{0.6\textwidth}
    \centering
    \includegraphics[width=1\linewidth]{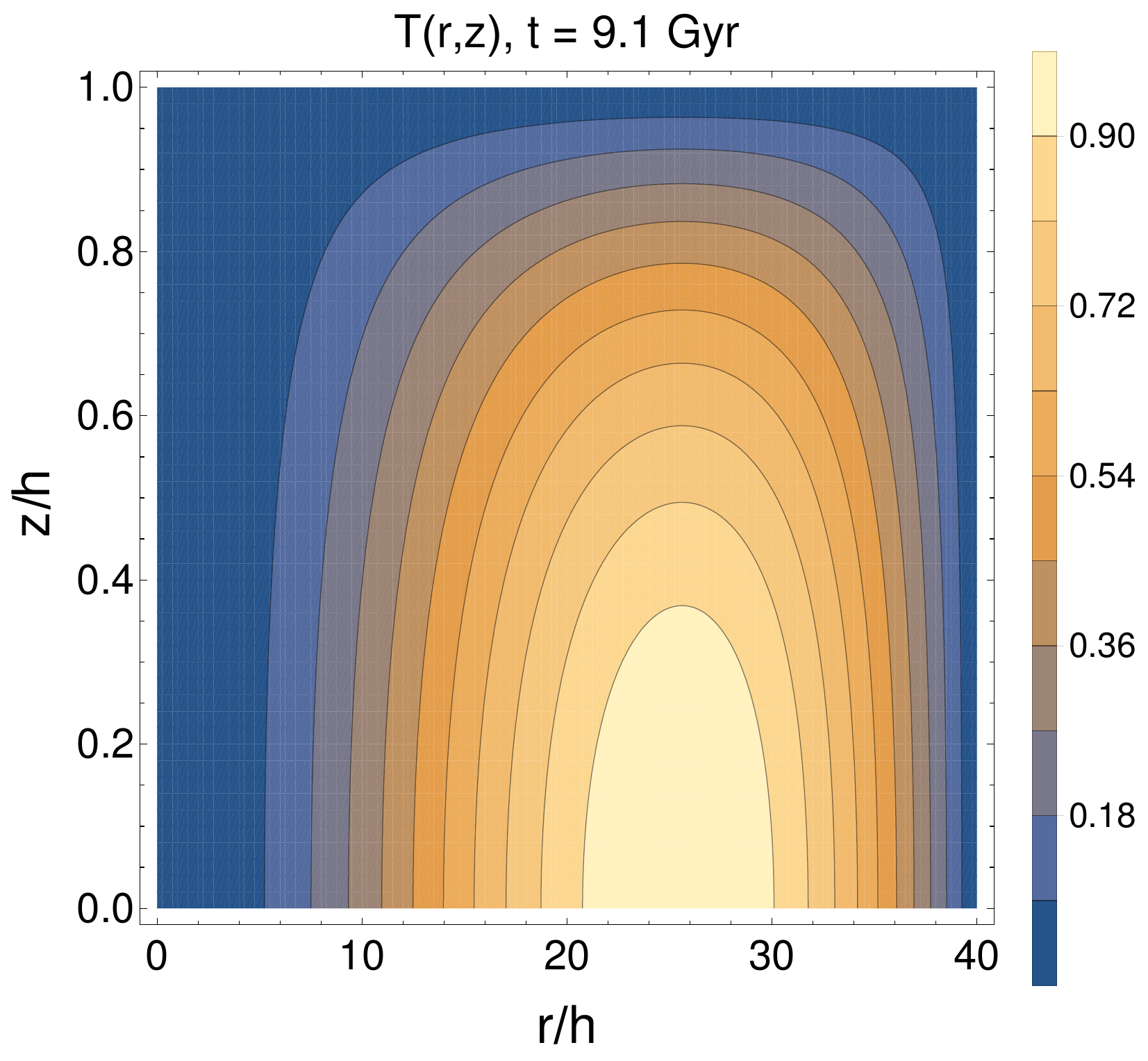}
    \caption{}
    \label{T10g}
  \end{subfigure}
  \caption{The meridional contour plots of $T$ corresponding to $R_U=2$ and $R_\kappa=0$ shown at different instants of time. (a) $t =t_{99} = 0.96$ Gyr when the volume-averaged magnetic field strength $\langle B \rangle$ reaches 99\% of its final strength. (b) $t=5$ Gyr, an intermediate period during the evolution of magnetic field. (c) $t=t_{sat}= 9.1$ Gyr, when the magnetic field almost reaches its steady-state configuration. Here $h=400$ pc corresponds to the half-width of the galactic disk.}
  \label{Tfig}
\end{figure}
\item In Figure \ref{f:brp}, we show the contour plots of magnetic field components $B_r$ and $B_\phi$ as functions of $r$ and $z$ for $R_U=2$ and $R_\kappa =0$ at $t=t_{sat}= 9.1$ Gyr where the magnetic field reaches a near steady-state configuration. Figures \ref{f:br1} and \ref{f:bp1} show the variations of $B_r$ and $B_\phi$ respectively within the disk for heights $z/h=0-1$ while Figures \ref{f:br2} and \ref{f:bp2} show the variations of these fields in the corona for heights $z/h=1-2$. The ratio of strength of $B_\phi$ and $B_r$ inside the disk is of the order $\sim [R_\omega/R_\alpha(1+\alpha_m)]^{1/2}$. As expected from the boundary condition given in Equation \eqref{e:bc12}, $B_r$ changes sign near the disk, which is a necessary condition for the dynamo to operate \citep{1988ASSL..133.....R}, so that the sign of the flux leaving through the surface is opposite to that of the flux in the mid-plane. The strength of the azimuthal field decreases with height and tends to zero near the disk surface. 
Both $B_\phi$ and $B_r$ are negligibly small in the corona (roughly two orders of magnitude less than their strengths inside the disk), which also gives the appearance of discontinuity in these functions at $z=1$. But their continuity is implied from the continuity of $\psi$ and $T$ which is guaranteed from the boundary conditions (Equations (\ref{e:bc12}) and (\ref{e:bc3})). The contour plot of $B_z$ for the same configuration is shown in Figure \ref{f:bz}. We find that although $B_z$ is much weaker than $B_r$ and $B_\phi$ inside the disk, it is the most dominant component in the corona. It is reported from observations that in general the strength of the large-scale magnetic field in the halo is comparable to that in the disk field \citep{2014arXiv1401.1317K}. We plan to consider in the future a halo model with a dynamo and/or a stronger galactic wind that can transport magnetic field from the disk in order to achieve higher magnetic field strengths \citep{1993A&A...271...36B,2010A&A...512A..61M}.
\begin{figure}[hp]
  \centering
  \begin{subfigure}[b]{0.45\textwidth}
    \centering
    \includegraphics[width=1\linewidth]{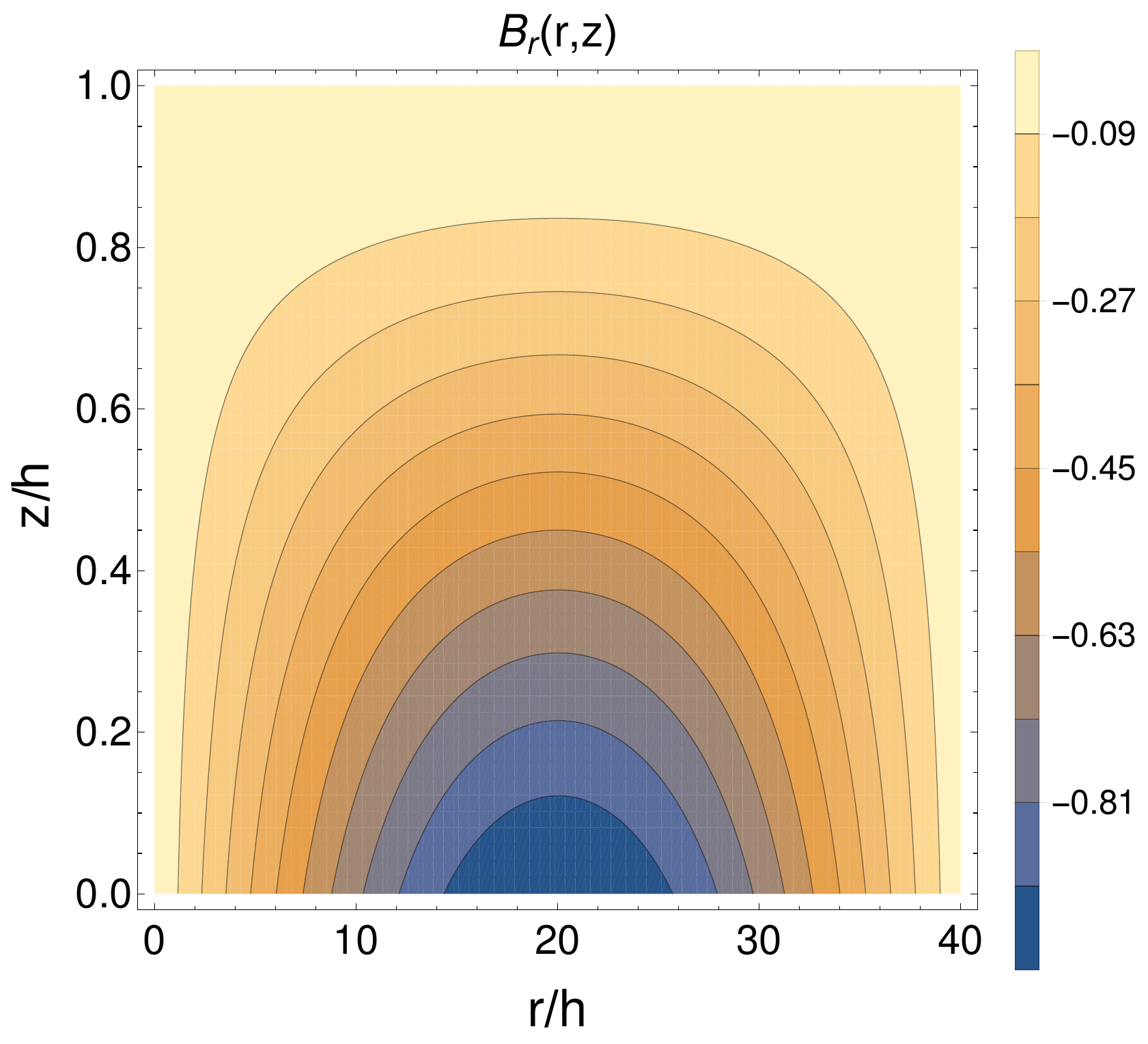}
    \caption{}
    \label{f:br1}
  \end{subfigure}
\quad
  \begin{subfigure}[b]{0.45\textwidth}
    \centering
    \includegraphics[width=1\linewidth]{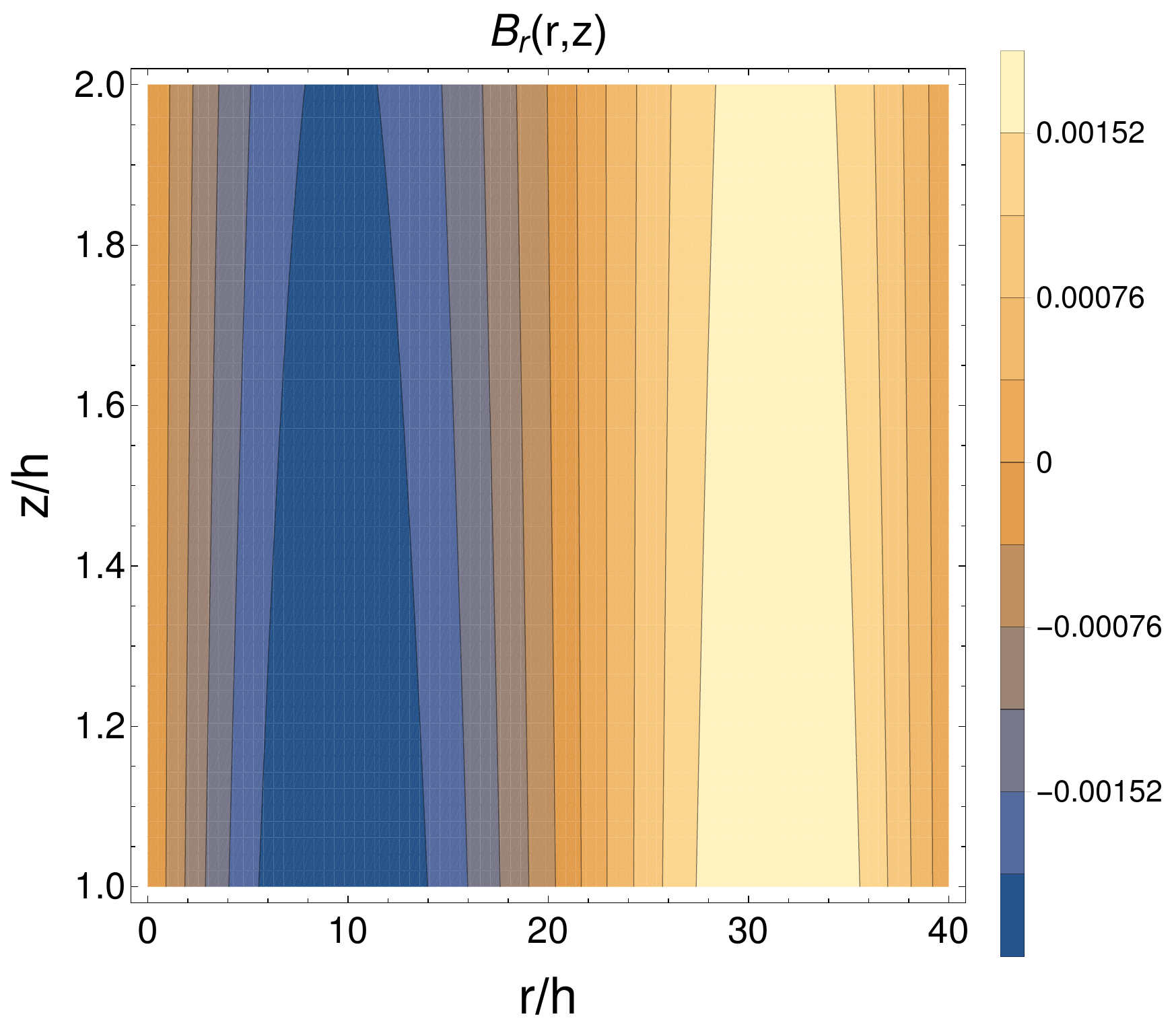}
    \caption{}
    \label{f:br2}
  \end{subfigure}
\quad
 \begin{subfigure}[b]{0.45\textwidth}
    \centering
    \includegraphics[width=1\linewidth]{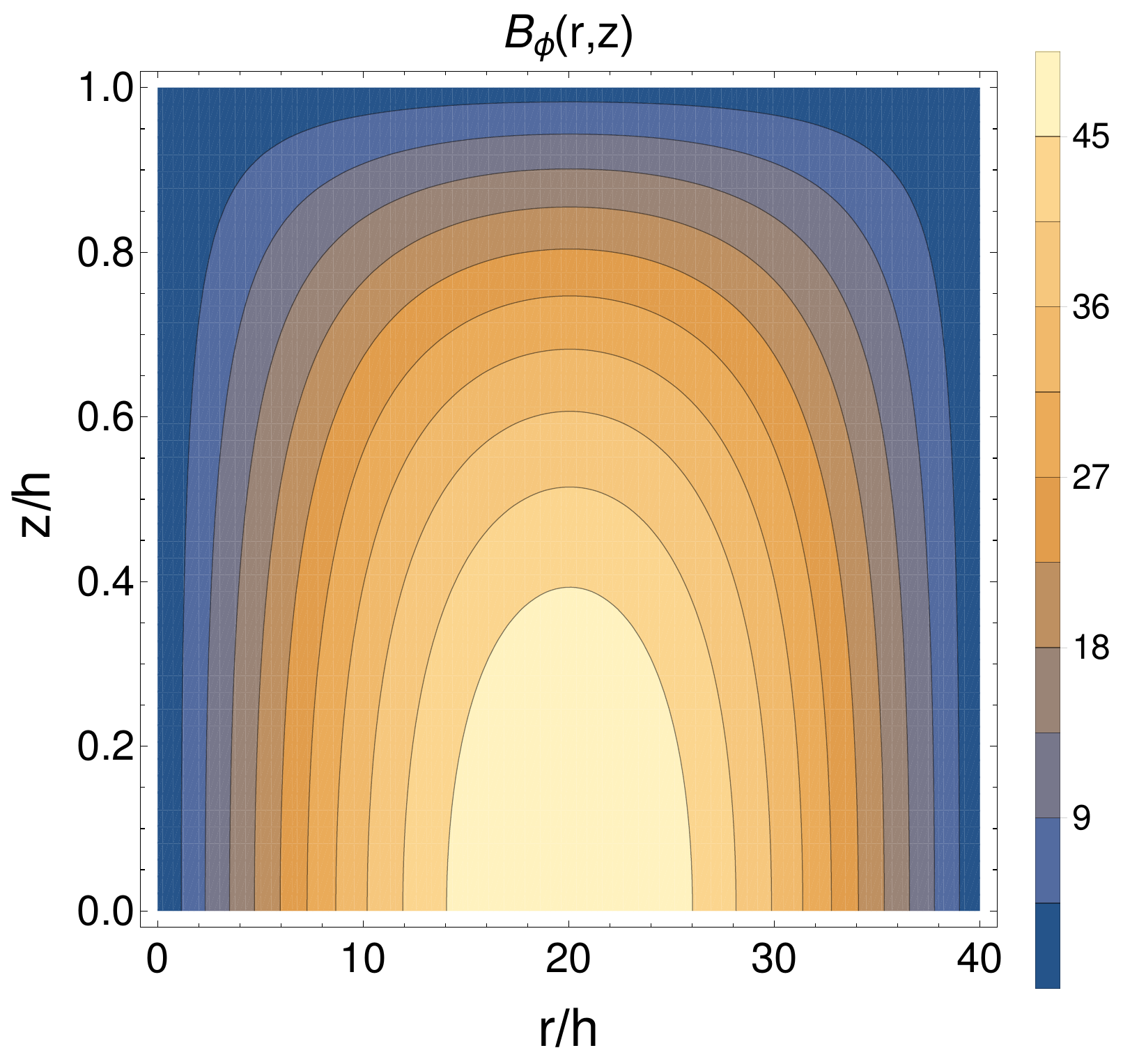}
    \caption{}
    \label{f:bp1}
  \end{subfigure}
\quad
\begin{subfigure}[b]{0.45\textwidth}
    \centering
    \includegraphics[width=1\linewidth]{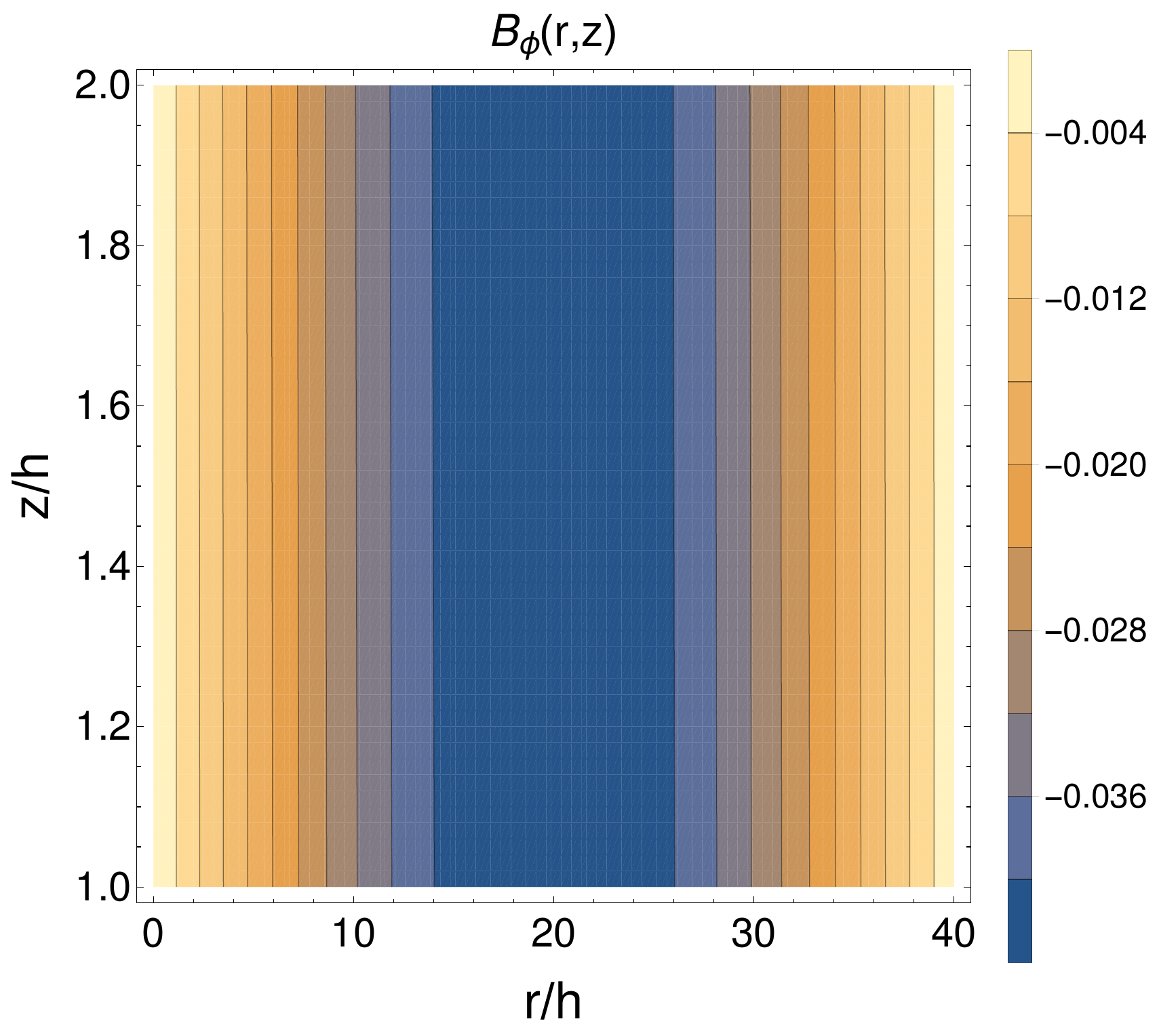}
    \caption{}
    \label{f:bp2}
  \end{subfigure}
  \caption{Contour plots of magnetic field components $B_r$ and $B_\phi$ as functions of $r$ and $z$ for $R_U=2$ and $R_\kappa =0$ at $t=t_{sat}= 9.1$ Gyr. Panels (a) and (c) show the variation of $B_r$ and $B_\phi$ respectively within the disk for heights $z/h=0-1$ while panels (b) and (d) show the variation of these fields in the corona for heights $z/h=1-2$. The contours in all the panels have been scaled with respect to the maximum value of $|B_r|$ within the disk.}
  \label{f:brp}
\end{figure}
\begin{figure}[hp]
\centering
    \includegraphics[width=0.6\linewidth]{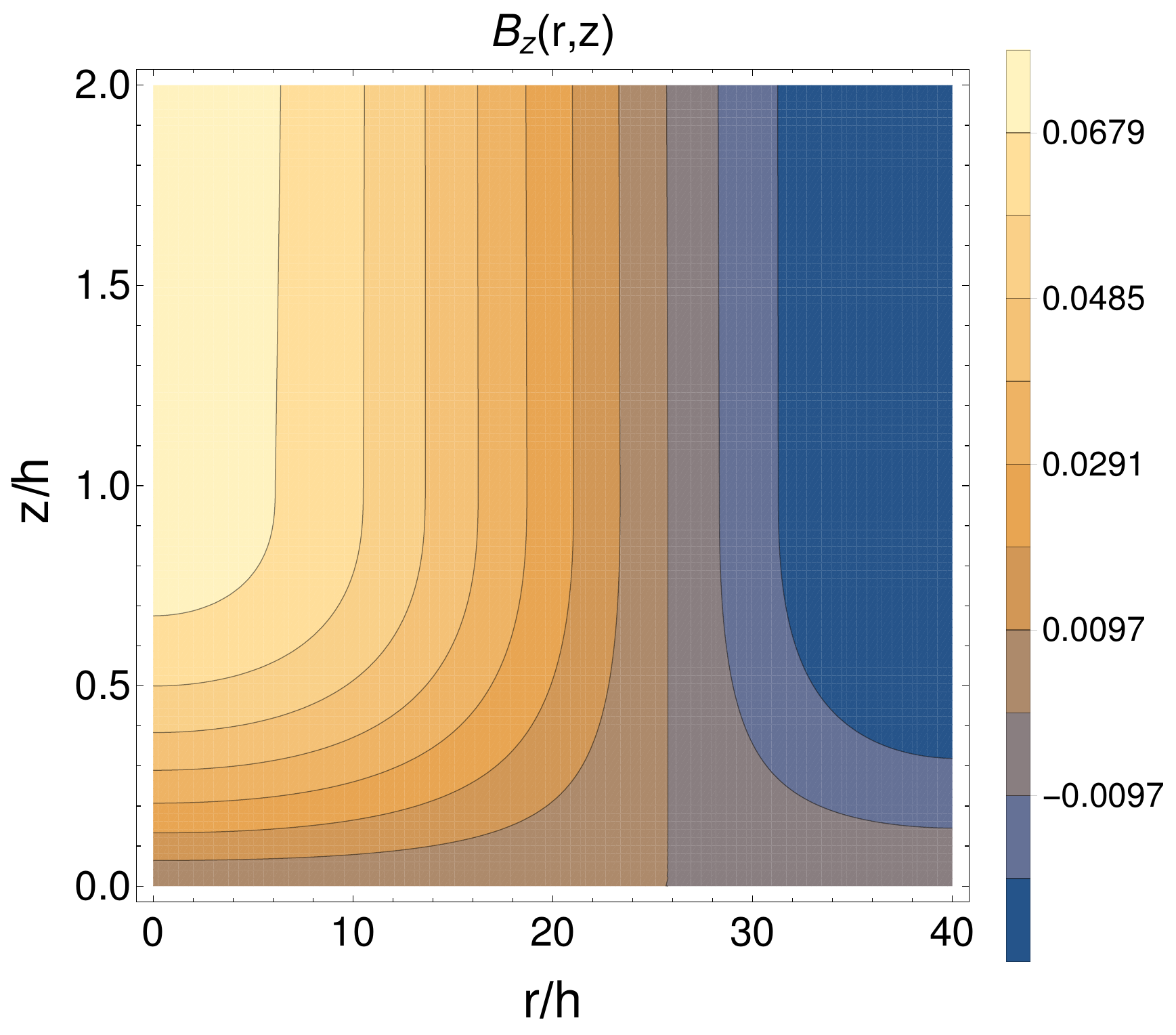}
     \caption{Contour plot of magnetic field component $B_z$ as a function of $r$ and $z$ for $R_U=2$ and $R_\kappa =0$ at $t=t_{sat}= 9.1$ Gyr. The contours in the figure have been scaled with respect to the maximum value of $B_r$ as shown in Figure \ref{f:br1}.}
    \label{f:bz}
\centering
\begin{subfigure}[]{0.55\textwidth}
    \centering
    \includegraphics[width=1\linewidth]{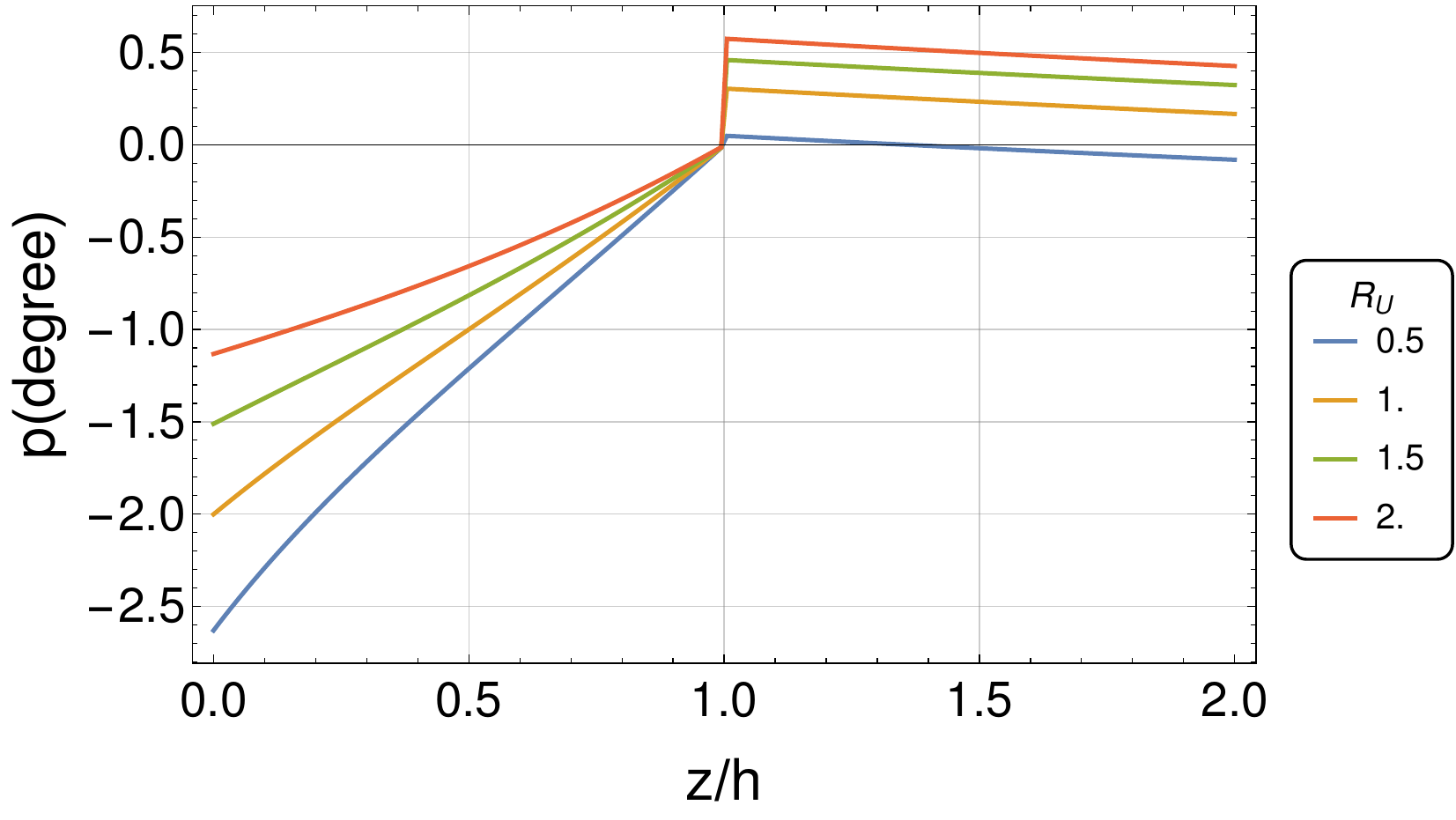}
    \caption{}
    \label{f:pita}
  \end{subfigure}
\quad
\begin{subfigure}[]{0.38\textwidth}
    \centering
    \includegraphics[width=1\linewidth]{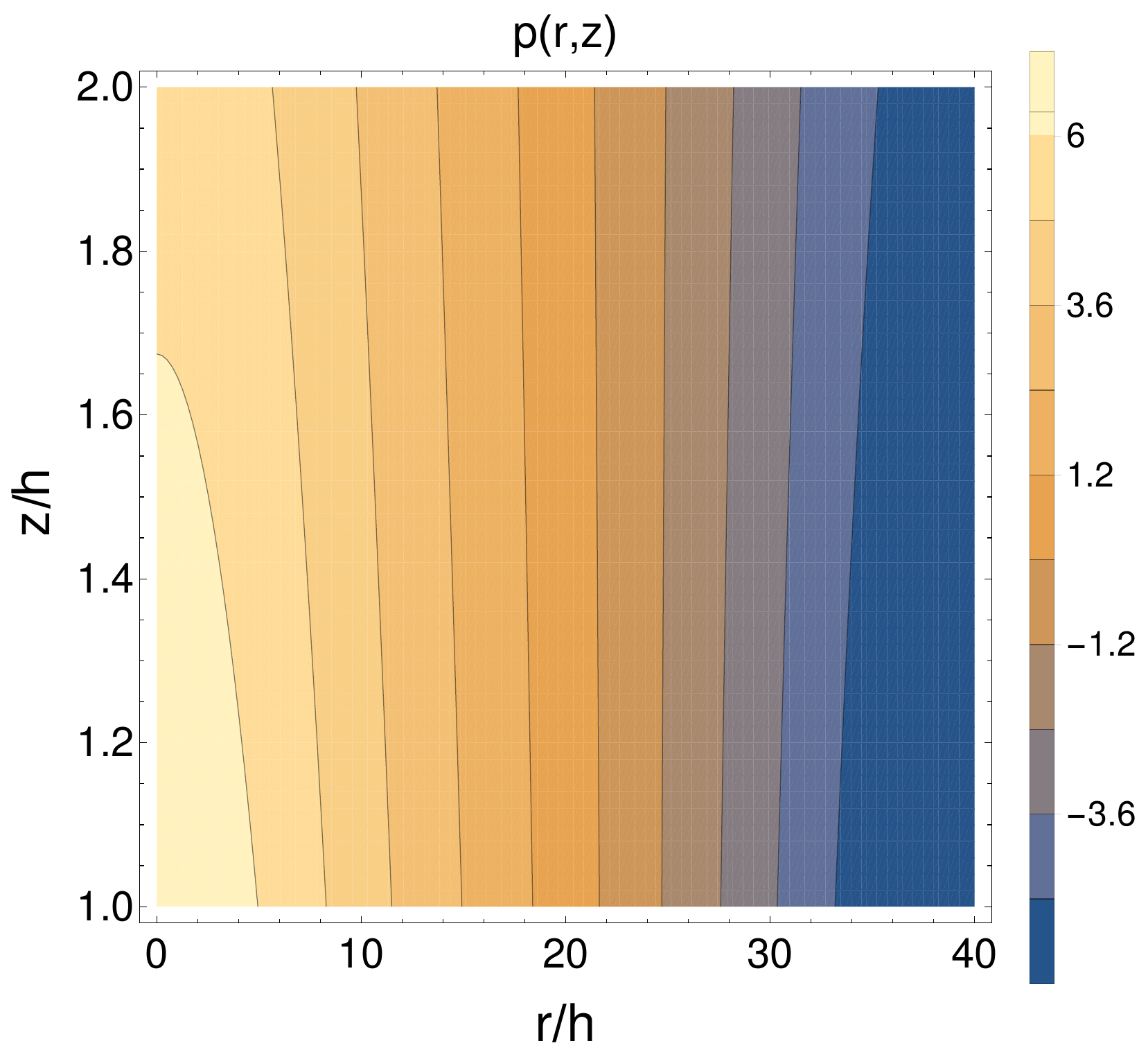}
    \caption{}
    \label{f:pitb}
  \end{subfigure}
    \caption{(a) The variation of the pitch angle within the disk (which is nearly in independent of $r$) with $z$ for different values of $R_U$ and $R_\kappa =0$. (b) The pitch angle in the corona as a function of $r$ and $z$ for $R_U=2$ and $R_\kappa =0$.}
    \end{figure}
\item The magnetic pitch angle is defined by $\disp{p=\tan^{-1}\left(B_r/B_\phi\right)}$. Using Equations (\ref{poleq}), (\ref{toreq}), and (\ref{tdsb}), we can write the pitch angle inside the disk as
\begin{equation}
p=\frac{-\sum_{n=1}^N w_n(t)Q^s_n(r)a'_n(z)}{\sum_{n=1}^N w_n(t)Q^s_n(r)b_n(z)}.
\end{equation}
Since a single mode $w_1$ is dominant over all other values of $w_n$ (see Figure \ref{wnplt}), the pitch angle within the disk $\disp{p\sim \frac{-a'_1(z)}{b_1(z)}}$, is nearly independent of $r$. The variation of magnetic pitch angle $p$ in degrees as a function of $z$ for different values of $R_U$ and $R_\kappa =0$ at $t=t_{sat}$ (corresponding to the value of $R_U$) is shown in Figure \ref{f:pita}. We find that $p$ varies from $-2.5^\circ$ to $0^\circ$ within the disk depending on the value of $R_U$. Since $B_r$ and $B_\phi$ have opposite signs inside the disk (see Figures \ref{f:br1} and \ref{f:bp1}), the pitch angle is negative inside the disk, decreases in magnitude with height, and becomes positive near the surface (when $B_r$ changes sign). This means that the magnetic spiral that is trailing within the disk starts leading near the surface. This is in agreement with what has been previously reported in \citet{2014MNRAS.443.1867C} and is expected in a model with outflows and corona \citep{1979Ap&SS..66..369R,2014GApFD.108..568J}. The pitch angle is found to varying between $-3.6^\circ$ to $6^\circ$ in the corona, as shown in Figure \ref{f:pitb}. The difference in the sign of pitch angle between the disk and the corona also implies that the large-scale magnetic helicities in the disk and corona are of opposite sign. This is expected in our model because the magnetic helicity in the corona grows through the small-scale helicity of the disk, which has opposite sign to that of the large-scale field. This inference can be verified through observations to further validate the role of small-scale magnetic helicity fluxes in dynamo action. The observed values of the pitch angle are close to $-20^{\circ}$ \citep{2010ASPC..438..197F}. It might be possible to obtain higher values for the pitch angle by incorporating mean radial flows \citep{2000A&A...358.1142M} or by invoking spiral shocks \citep{2015ApJ...799...35V} or using non-standard parameter values for the dynamo \citep{2015ApJ...808...28C}. We plan to investigate these effects in future work.
\end{enumerate}
 \section{Summary and conclusions}
We have developed a global semi-analytic 3D model for the dynamo operation in a galaxy with a corona. The model includes small-scale (advective and diffusive) magnetic helicity fluxes that transfer magnetic helicity from the disk to the corona and prevent the catastrophic quenching of the dynamo. The effect of these small-scale magnetic helicity fluxes on the nonlinear saturation of the dynamo is also demonstrated from the strength and structure of the global saturated magnetic field. Here we summarize and highlight the novel features of this work.
\begin{enumerate}
\item We have incorporated the radial dependence in the SNe- (and MRI-) driven turbulence parameters and have shown that all these parameters, $\alpha$, $\Omega$, $\eta_t$, and $U_z$ have similar radial variations ($\propto 1/r$). Thus  the dynamo parameters $R_\alpha$, $R_\omega$, and $R_U$ defined in Equation \eqref{dim} are nearly independent of $r$. This leads to a great simplification in our formulation, and the dynamo equations (\ref{poleq3} and \ref{toreq3}) take the same dimensionless form for both SNe- and MRI-driven turbulence.
  \item A comparison of different parameters for the cases of SNe- and MRI-driven turbulence is presented in Table \ref{t:comp}. We found that the SNe-driven dynamo operates at a much faster rate than the MRI-driven dynamo and hence the magnetic field generation in the disk is likely to be dominated by SNe-driven turbulence. As a combined treatment of both SNe- and MRI-driven  turbulence is beyond the scope of this paper, we have used only the SNe-driven turbulence parameters for our analysis.
  \item We have solved the dynamo equations inside the disk to obtain the global steady-state solutions, which are matched to a linear force-free field in the corona (see Section \ref{s:steady}). These global analytic solutions allowed us to calculate the global relative helicity for both the disk and the corona. We have presented an analysis of the relative helicity flux terms in Appendix \ref{AppA}. We have included the advective and diffusive fluxes for the work presented in this paper and plan to explore the contribution from other terms in the future.
 \item We have solved the full time-dependent problem in Section \ref{s:time} by writing the time-dependent magnetic field in a separable form (see Equation (\ref{tds})), where the radial solution is expressed in terms of the steady-state solutions already obtained in Section \ref{s:steady}. By studying the parametric dependence of the time-dependent solutions on $\alpha_m$ and $\mu$, we obtained the saturation value of $\alpha_m$ that enabled us estimate the corresponding global steady-state magnetic field geometry. 

\item To summarize our approach to solve for the saturation of the nonlinear dynamo, the overall time-dependent solution to Equation \eqref{tdeq} given by Equations (\ref{wneq}), (\ref{mat2}), and (\ref{tdsb}) is built using an expansion of the steady-state solutions whose time dependence is  parameterized through the local growth rate $\gamma(t)$. The radial part, given by Equation \eqref{e:wn}, provides connection between the global growth rate $\Gamma(t)$, radial diffusion, and the local growth rate $\gamma(t)$  that represents other effects of vertical advection, diffusion, shear, and the $\alpha-$effect.  In essence, the time-dependent solution describes a dynamo of large-scale magnetic field in the disk with vertical components $a(z)$ and $b(z)$ solved from the steady-state solution for a given $\gamma(t)$. The flux transport is introduced by the helicity Equation \eqref{hc7} which includes the efflux of {\em only} the small helicity and its conversion by reconnection to large scale magnetic helicity in the corona. The boundary conditions provide the connection of the external large-scale field to the corresponding disk field through the force-free parameter $\mu(t)$. Thus, in our model, the large-scale flux is not transported out of the disk or destroyed (see Section \ref{s:alp}). A more detailed treatment, including the large-and small-scale helicity fluxes (as given in Equations  (\ref{H}) and (\ref{h}) is beyond the scope of this paper and will be taken up in the future.

\item The analysis of the radial solution in terms of its eigenvectors (as shown in Section \ref{s:para}) also gives a clear understanding of the evolution of the global structure of magnetic field with time.
\item We have obtained quadrupolar solutions for the saturated magnetic field strength, $\sim\langle B_{eq}\rangle$, which is proportional to the advective and diffusive fluxes leaving the surface, see Table \ref{t:sat}. For the case of vertical outflow $U_z = 2$ kms$^{-1}$ at a radius of 4 kpc, we obtained a mean-field strength of $\langle B_{sat}\rangle=5-7\mu$G, which is close to what was reported in numerical simulations \citep{2013MNRAS.429..967G,2014MNRAS.443.1867C} and in observations \citep{2012SSRv..166..215B,2015ApJ...799...35V}. The dynamo was found to reach equipartition strength and 99\% of its saturation value in about 1 Gyr, which is faster than the timescales previously reported in \citet{2007MNRAS.377..874S} and \citet{2014MNRAS.443.1867C}. 
\item We found that during the dynamo operation, the small-scale magnetic helicity fluxes slowly build a corona with a magnetic helicity that carries the same sign as that of the small-scale magnetic helicity fluxes, see Figure \ref{mufiga}. We also found that the magnetic field inside the disk is not very sensitive to the fraction of large-scale helicity in the corona given by $R_c$ (shown in Figure \ref{hcffig}). In the absence of the flux terms ($R_U=R_\kappa=0$), we found that the mean magnetic field initially grew to a maximum value of $\sim 0.4~ \langle B_{eq}\rangle$ in the kinematic phase and then catastrophically quenched (see Figure \ref{bbfiga}). This confirms the crucial role of the helicity fluxes in the dynamo operation \citep{2007MNRAS.377..874S}.
\item We have shown the evolution of the global structure of the magnetic field in the disk as well as the corona, as a function of time (see Section \ref{s:dist}). The magnetic fields in the corona are found to be much weaker than those inside the disk and are dominated by $B_z$ (see Figures \ref{f:brp} and \ref{f:bz}). This indicates that the halo may require dynamo action or a stronger galactic wind to account for the much stronger magnetic fields  reported in observations \citep{2014arXiv1401.1317K}. We plan to take this up in the future.

\item We have improved upon previous work by introducing the following novelties: building a 3D model of the global field of the disk and corona using a simplified treatment of reconnecting the small-scale field to describe a large-scale force-free coronal field and balancing the global helicity by the use of gauge-free descriptions of absolute helicity. 

\end{enumerate}
In the future, we plan to work on a hybrid model for the dynamo with a simultaneous treatment of both SNe- and MRI-driven turbulence. We also plan to include a more realistic model for the coronal field that involves details of the helicity dissipation by reconnection in the corona. The contribution from the remaining small- and large-scale magnetic helicity flux terms (apart from advective and diffusive fluxes) in Equations (\ref{H}) and (\ref{h}) need to be explored in order to study its effect on the saturation of the dynamo. The magnetic pitch angle obtained in this model is much less than the observed values; we plan to investigate this further by expanding our parameter space and incorporating other effects in the model as discussed in the last paragraph of Section \ref{s:dist}.

\textit{Acknowledgements:}
We thank Kandaswamy Subramanian, Sharanya Sur, and Luke Chamandy for useful discussions and the anonymous referee for insightful comments and helpful suggestions. A.P. acknowledges CSIR for the SPM fellowship. We also thank the support staff of the IIA HPC facility and Sandra Rajiva for proof
reading the manuscript.

\appendix
\section{Magnetic helicity dynamics}
\label{AppA}
The induction equation is given by
\begin{equation}
 \partial_t \mathbf{B}=\fpar{\mathbf{B}}{t}=\nabla \times\left(\mathbf{U}\times \mathbf{B}-\eta \mathbf{J}\right). \label{e:ind}
\end{equation}
The mean-field component of the induction equation is given by
\begin{equation}
\partial_t\mbf{B} = \fpar{\mbf{B}}{t}=\nabla \times\left(\mbf{U}\times \mbf{B}-\eta \mbf{J}+\ve\right). \label{ady}
\end{equation}
Uncurling Equation \eqref{ady}, we get
\begin{equation}
 \partial_t \mbf{A}=\fpar{\mbf{A}}{t}=\mbf{U}\times \mbf{B}-\eta \mbf{J}+\ve+\nabla\varphi_1
\end{equation}
where $\varphi_1$ is a scalar function that depends only on spatial coordinates.
In order to calculate the temporal evolution of large-scale helicity $\disp{\ol{H}_d=\int_V\mbf{A}\cdot\mbf{B}~\dif V}$,
we take the partial time derivative of its density given by
\begin{equation}
 \partial_t (\mbf{A}\cdot\mbf{B})=2(\partial_t\mbf{A})\cdot\mbf{B}+\nabla\cdot[(\partial_t\mbf{A})\times\mbf{A}]
=-2\eta \mbf{J}\cdot\mbf{B}+2\ve\cdot\mbf{B}+2\nabla \cdot (\varphi_1\mbf{B})+ \nabla\cdot[(\partial_t \mbf{A})\times\mbf{A}],
\label{e:Hden}
\end{equation}
where $(\partial_t \mbf{A})\times\mbf{A}=\left(\mbf{U}\times\mbf{B}-\eta \mbf{J}+\ve+\nabla \varphi_1\right)\times \mbf{A}$.
The volume average of Equation \eqref{e:Hden} gives the equation for the temporal evolution of large-scale magnetic helicity, $\ol{H}_d$ \citep[][p. 69]{2008tdad.conf...69M} as
\begin{equation}
 \frac{\dif \ol{H}_d}{\dif  t}=\int_V \partial_t (\mbf{A}\cdot\mbf{B})~ \dif V=2\int_V \ve\cdot\mbf{B}~ \dif V -2 \int_V \eta \mbf{J}\cdot\mbf{B}~ \dif V-
\oint_S \mathbf{F}\cdot\hat{n}~\dif S,
\end{equation}
where $\hat{n}$ represents the normal to the surface $S$ enclosing volume $V$ and
\begin{equation}
\mathbf{F}=(\eta \mbf{J}-\mbf{U}\times\mbf{B}-\ve-\nabla \varphi_1)\times\mbf{A}-2\varphi_1\mbf{B}, 
\end{equation}for cross-referencing
is the large-scale magnetic helicity flux. 
Similarly, for the temporal evolution of the mean small-scale magnetic helicity $\disp{h_d=\int_V\ol{\mathbf{a}\cdot\mathbf{b}}~\dif V}$, 
we note that the time derivative of the small-scale magnetic field obtained by subtracting Equation \eqref{ady} from Equation \eqref{e:ind}
is given by
\begin{equation}
 \partial_t \bf{b}=\nabla\times(\bf{u}\times\mbf{B}+\mbf{U}\times\bf{b}+\bf{u}\times\bf{b}-\ve-\eta\bf{j}) \label{e:dbdt}
\end{equation}
Uncurling Equation \eqref{e:dbdt}, we get 
\begin{equation}
 \partial_t \bf{a}=\bf{u}\times\mbf{B}+\mbf{U}\times\bf{b}+\bf{u}\times\bf{b}-\ve-\eta\bf{j}+\nabla\varphi_2. \label{e:dadt}
\end{equation}
The time derivative of the mean small-scale magnetic helicity density, $\chi=\ol{\mathbf{a}\cdot\mathbf{b}}$, is then given by
\begin{equation}
 \partial_t(\ol{\mathbf{a}\cdot\mathbf{b}})=2\ol{(\partial_t \bf{a})\cdot\bf{b}}+\nabla \cdot \ol{(\partial_t \bf{a})\times\bf{a}}.
\end{equation}
Using Equation \eqref{e:dadt}, we get
\begin{equation}
 \partial_t(\ol{\mathbf{a}\cdot\mathbf{b}})=-2\ve\cdot\mbf{B}-2\eta\ol{\bf{j}\cdot\bf{b}}+2\nabla\cdot\ol{(\varphi_2 \bf{b})}+\nabla\cdot\ol{(\partial_t \bf{a})\times\bf{a}}. \label{e:dabdt}
 \end{equation}
The volume average of Equation \eqref{e:dabdt} now gives us the equation for the evolution 
of the mean small-scale helicity, $h_d$, as
\begin{equation}
 \frac{\dif h_d}{\dif t}=-2\int_V \ve \cdot \mbf{B}~ \dif V- 2 \int_V \eta \ol{\mathbf{j}\cdot\mathbf{b}}~\dif V -\oint_S \mathbf{f}\cdot \hat{n}~\dif S,
\end{equation}
where the $\mathbf{f}=-\ol{(\partial_t \bf{a})\times \bf{a}}-2\ol{\varphi_2\mathbf{b}}$ represents the surface flux terms, which can be expanded in detail as
\begin{eqnarray}
 \mathbf{f}&=&\ol{(\bf{a}\cdot\mbf{B})\bf{u}}-\ol{(\bf{a}\cdot\bf{u})\mbf{B}}-\ol{(\bf{a}\cdot\mbf{U})\bf{b}}
+\ol{(\bf{a}\cdot\bf{b})\mbf{U}}-\ol{(\bf{a}\cdot\bf{u})\bf{b}}+\ol{(\bf{a}\cdot\bf{b})\bf{u}} \nonumber \\ 
&&+\ol{\ve\times\bf{a}}+\eta \ol{\bf{j}\times\bf{a}}-\ol{\nabla\varphi_2\times\bf{a}}-2\ol{\varphi_2\mathbf{b}}.
\end{eqnarray}
\section {Derivation of Equation \eqref{e:dim}}
\label{a:dim}
We start with Equation \eqref{poleq2} given by
\begin{equation}
\left(\fpar{}{t}+\ol{U}_z\fpar{}{z}-\eta_t(r) \Lambda\right)\ol{\psi}=\alpha \ol{T}
\end{equation}
where we have explicitly mentioned the $r-$dependence of $\eta_t$.
Now substituting the variables using the transformations given in Equation \eqref{scale}, we get
\begin{equation}
 \left(\frac{\eta_t(h)}{h^2}\fpar{}{\tau}+\frac{\ol{U}_z}{h}\fpar{}{\tilde{z}}-\frac{\eta_t(r)}{h^2}\tilde{\Lambda}\right)\psi_0\tilde{\psi}=\tilde{\alpha}\alpha_0 \frac{\psi_0 \tilde{T}}{h}, \label{e:dim0}
\end{equation}
where we have used $t_d=h^2/\eta_t(h)$. Dividing the above equation throughout by $\disp{\frac{\psi_0 \eta_t(r)}{h^2}}$, we get
\begin{equation}
 \left[\left(\frac{\eta_t(h)}{ \eta_t(r)}\right)\fpar{}{\tau}+\left(\frac{\ol{U}_z h}{\eta_t(r)}\right)\fpar{}{\tilde{z}}-\tilde{\Lambda}\right]\tilde{\psi}=\left(\frac{\alpha_0 h}{\eta_t(r)}\right)\tilde{\alpha}\tilde{T}. \label{e:dim1}
\end{equation}
Since $\eta_t(r)\propto1/r$ (from Equations (\ref{eamri}) and (\ref{easn})), we can write
\begin{equation}
\left[\frac{\eta_t(h)}{ \eta_t(r)}\right]=\frac{r}{h}=\tilde{r}.\label{e:erat}
\end{equation}
Using definitions of $R_U$ and $R_\alpha$ from Equation \eqref{dim}, we write Equation \eqref{e:dim1} as
\begin{equation}
 \left(\tilde{r}\fpar{}{\tau}+R_U\fpar{}{\tilde{z}}-\tilde{\Lambda}\right)\tilde{\psi}=R_\alpha \tilde{\alpha}\tilde{T}.
\end{equation}
Similarly, we rewrite Equation \eqref{toreq2} as
\begin{equation}
\left(\fpar{}{t}+\ol{U}_z\fpar{}{z}-\eta_t \Lambda\right)\ol{T}=\Omega \fpar{\ol{\psi}}{z}
\end{equation}
where we have used $\disp{\Omega(r)=\frac{r_0 \Omega_0}{r}}$. Substituting the dimensionless variables from Equation \eqref{scale} into the above equation, we get 
\begin{equation}
\left(\frac{\eta_t(h)}{h^2}\fpar{}{\tau}+\frac{\ol{U}_z}{h}\fpar{}{\tilde{z}}-\frac{\eta_t(r)}{h^2}\tilde{\Lambda}\right)\frac{\psi_0\tilde{T}}{h}=\frac{\Omega \psi_0}{h}\fpar{\tilde{\psi}}{\tilde{z}}.
 \end{equation}
Following the same steps as taken after Equation \eqref{e:dim0}, we get the final form as
\begin{equation}
 \left(\tilde{r}\fpar{}{\tau}+R_U\fpar{}{\tilde{z}}-\tilde{\Lambda}\right)\tilde{T}=R_\omega \fpar{\tilde{\psi}}{\tilde{z}}.
\end{equation}

\section{Derivation of quadrupolar boundary conditions given in Equation \eqref{quadbc}}
\label{a:bound}
The functions $\psi$ and $T$ have the same radial dependence both outside and inside the disk, as given by Equation \eqref{bes}. Thus, in writing the boundary conditions for quadrupolar symmetry (Equations (\ref{e:bc12})-(\ref{e:bc45})), the radial part of the solution cancels out and we obtain a set of four equations relating the eigenvalues and eigenfunctions of $a_n^s$.
Substituting  Equation \eqref{e:ans1} into Equation \eqref{e:bc12} we get 
\begin{equation}
  \sum_{j=1}^4 c_{nj} \lambda_{nj} \exp (\lambda_{nj})=0 \label{e:BC1}
\end{equation}
We rewrite Equation \eqref{aneq} as
\begin{equation}
 b_n^s=\frac{1}{R_\alpha (1+\alpha_m^s)} \left(\gamma_n^s a_n^s +R_U \frac{\dif a_n^s}{\dif z} -\frac{\dif^2 a_n^s}{\dif z^2}\right) \label{e:bsn2}
\end{equation}
Combining Equations (\ref{e:ans1}), (\ref{e:bc3}) and (\ref{e:bsn2}), we get
\begin{equation}
  \sum_{j=1}^4 \left(\gamma_n^s+R_U \lambda_{nj}-\lambda_{nj}^2\right)c_{nj}\exp(\lambda_{nj})= \sum_{j=1}^4 \mu^s R_\alpha (1+\alpha_m^s) c_{nj}\exp (\lambda_{nj}).
  \end{equation}
Rearranging terms in the above equation, we get
\begin{equation}
  \sum_{j=1}^4 \left[\mu^s R_\alpha (1+\alpha_m^s) +\lambda_{nj}^2-R_U \lambda_{nj}-\gamma_n^s\right]c_{nj}\exp(\lambda_{nj})=0.  \label{e:BC2}
\end{equation}
For quadrupolar boundary conditions, we substitute Equation \eqref{e:ans1} into Equation \eqref{e:bc4} to get
\begin{equation}
 \sum_{j=1}^4 c_{nj}=0. \label{e:BC3}
\end{equation}
Differentiating Equation \eqref{e:bsn2} with respect to $z$, we get
\begin{equation}
 \frac{\dif b^s_n}{\dif z}=\frac{1}{R_\alpha (1+\alpha_m^s)}\left(\gamma_n^s\frac{\dif a_n^s}{\dif z}+R_U \frac{\dif^2 a_n^s}{\dif z^2}-\frac{\dif^3 a_n^s}{\dif z^3}\right). \label{e:bsn3}
\end{equation}
Substituting Equation \eqref{e:bsn3} into Equation \eqref{e:bc5}, we get
\begin{equation}
 \sum_{j=1}^4 \left(\gamma_n^s \lambda_{nj} +R_U\lambda_{nj}^2 - \lambda_{nj}^3\right)c_{nj}=0. \label{e:BC4}
\end{equation}

\section{Equation for evolution of $\alpha_m$}
\label{a:alp}
In order to derive an equation for the evolution of $\alpha_m$ with time, we first
calculate the divergence of the small-scale magnetic helicity fluxes given in Equations (\ref{ru}) and (\ref{rk}).
The divergence of the advective flux density obtained using Equations (\ref{ru}), (\ref{e:uphi}) and (\ref{e:uz}) is given by
\begin{equation}
 \nabla \cdot \calf_a = \nabla\cdot(\mbf{U} \alpha_m)=\frac{\partial}{r\partial \phi}(r \Omega \alpha_m )+\fpar{}{z}(U_0 \alpha_m).\label{e:calfa}
\end{equation}
The first term on the rhs of the Equation \eqref{e:calfa} goes to zero due to axisymmetry, and since $U_0$ is assumed to be independent of $z$, we obtain
\begin{equation}
 \nabla \cdot \calf_a = U_0 \fpar{\alpha_m}{z}. \label{e:ruf}
\end{equation}
From Equation \eqref{rk}, we can write
\begin{equation}
 \nabla \cdot \calf_\kappa = -\nabla \cdot (\kappa \nabla \alpha_m)=-0.3 \nabla \cdot (\eta_t \nabla \alpha_m). \label{e:fd0}
\end{equation}
Evaluating $\nabla \cdot (\eta_t \nabla \alpha_m)$ separately, we write
\begin{equation}
 \nabla \cdot (\eta_t \nabla \alpha_m)=\eta_t \nabla^2 \alpha_m + \nabla \alpha_m \cdot \nabla \eta_t. \label{e:fd1}
\end{equation}
Since $\eta_t$ depends only on $r$ (from Equations (\ref{eamri}) and (\ref{easn})), we can write $\disp{\nabla \eta_t \sim \left(\frac{\eta_t}{r_d}\right)}$. Also as the $z$ derivatives dominate over the $r$ derivatives, we can write the first term on the rhs of Equation \eqref{e:fd1} as $\disp{\eta_t \nabla^2 \alpha_m}$ $\disp{\approx \eta_t \frac{\partial^2 \alpha_m}{\partial z^2}}$. Also, the second term on the rhs of Equation \eqref{e:fd1}, $\disp{\nabla\alpha_m \cdot \nabla \eta_t}$ $\disp{\left(\sim\frac{\alpha_m}{h} \frac{\eta_t}{r_d}\right)}$, is small compared to the first term, $\disp{\eta_t \nabla^2 \alpha_m \left(\sim\frac{\alpha_m \eta_t}{h^2}\right)}$, and can be neglected. Thus Equation \eqref{e:fd0} can now be written as
\begin{equation}
 \nabla \cdot \calf_\kappa = -0.3 \eta_t \frac{\partial^2 \alpha_m}{\partial z^2}=-\kappa \frac{\partial^2 \alpha_m}{\partial z^2}.\label{e:rkf}
 \end{equation}
The small-scale magnetic helicity transport equation along with the flux terms can now be written by combining Equations (\ref{quench}), (\ref{e:ruf}) and (\ref{e:rkf}) as
\begin{equation}
 \fpar{\alpha_m}{t}=\frac{-2 \eta_t}{l^2_0}\left(\frac{\alpha \ol{B^2}-\eta_t\ol{\bf{J}\cdot\bf{B}}}{B_{eq}^2} +\frac{\alpha_m}{R_m}\right)
-U_0\fpar{\alpha_m}{z}+\kappa\frac{\partial^2 \alpha_m}{\partial z^2}.\label{alp}
\end{equation}
Rescaling Equation \eqref{alp} using the relations given in Equation \eqref{scale}, we write
\begin{equation}
 \frac{\eta_t(h)\alpha_0}{h^2} \frac{\dif \tilde{\alpha}_m}{\dif \tau}=-\frac{2\eta_t(r)}{l_0^2}\left(\alpha_0 \tilde{\alpha}\overline{\tilde{B}^2} -\frac{\eta_t(r)}{h}\overline{\mathbf{\tilde{J}}\cdot\mathbf{\tilde{B}}}+\frac{\alpha_0 \tilde{\alpha}_m}{R_m}\right )-\frac{U_0 \alpha_0 \tilde{\alpha}_m}{h}-\frac{\kappa \alpha_0\tilde{\alpha}_m}{h^2} \label{e:als1}
 \end{equation}
where $\disp{\tilde{\mathbf{B}}=\frac{\mathbf{B}}{B_{eq}}}$ and $\disp{\tilde{\mathbf{J}}=\frac{h\mathbf{J}}{B_{eq}}}$. Here (and in the following sections), we use the `no-$z$' approximation \citep{1993MNRAS.265..649S,1995MNRAS.275..191M,2014MNRAS.443.1867C}
for obtaining the $z$ derivatives of $\alpha$ by setting $\disp{\frac{\partial^2}{\partial z^2}\rightarrow\frac{-1}{h^2}}$ and $\disp{\fpar{}{z}\rightarrow\pm\frac{1}{h}}$, with the sign chosen appropriately. Multiplying Equation \eqref{e:als1} by the factor $\disp{\frac{h^2}{\alpha_0 \eta_t(r)}}$, we obtain
\begin{equation}
 \left[\frac{\eta_t(h)}{\eta_t(r)}\right]\frac{\dif \tilde{\alpha_m}}{\dif \tau}=-2 \left(\frac{h}{l_0}\right)^2\left[\tilde{\alpha}\ol{\tilde{B}^2}-\left(\frac{\eta_t(r)}{\alpha_0 h}\right)\overline{\mathbf{\tilde{J}}\cdot\mathbf{\tilde{B}}}+\frac{\tilde{\alpha}_m}{R_m}\right]-\left(\frac{U_0 h}{\eta_t(r)}\right)\tilde{\alpha}_m-\left(\frac{\kappa}{\eta_t(r)}\right)\tilde{\alpha}_m. \label{e:als2}
\end{equation}
Using $\tilde{\alpha}=1+\tilde{\alpha_m}$, the definitions given in Equation \eqref{dim}, and Equation \eqref{e:erat}, we write Equation \eqref{e:als2} as
\begin{equation}
 r \frac{\dif  \alpha_m}{\dif  \tau}= -C\left[(1+\alpha_m)\ol{B^2}-R_\alpha^{-1}\ol{\bf{J}\cdot\bf{B}}\right]-(R_U+R_\kappa)\alpha_m,\label{a:alp2}
\end{equation}
where we have dropped the tilde for clarity and
\begin{equation}
C=2\left(\frac{h}{l_0}\right)^2, \quad R_\kappa=\frac{\kappa}{\eta_t}.\label{a:cru}
\end{equation}

\section{Derivation of $\langle \ol{B^2}\rangle$ and $\langle\ol{\mathbf {J}\cdot \mathbf{B}}\rangle$}
\label{a:avg}
Using Equations (\ref{poleq}) and (\ref{toreq}), we can write 
\begin{equation}
 \mbf{B}=\mbf{B}_P+\mbf{B}_\phi; \quad \mbf{B}_P=\mathbf{\hat{P}}\psi;\quad \mbf{B}_\phi=\frac{T}{r}\hat{\phi}.  \label{e:b1}
\end{equation}
The expression for the energy of the mean magnetic field can be written as
\begin{equation}
 \ol{B^2}=\frac{1}{r^2}\left[(\partial_z \psi)^2+(\partial_r \psi)^2+T^2\right] \label{e:b2}
\end{equation}
where $\disp{\partial_z=\fpar{}{z}}$, $\disp{\partial_r=\fpar{}{r}}$. 
From Equation \eqref{tds}, we can write 
\begin{equation}
\partial_z \psi =\sum_{n=1}^N q_n Q_n a' w_n; \quad \partial_r \psi =\sum_{n=1}^N q_n Q'_n a w_n \label{pt1}
\end{equation}
where $\disp{a'=\frac{\dif a}{\dif z}}$ and $\disp{Q'_n=\frac{\dif Q_n}{\dif r}}$.
Thus substituting Equation \eqref{pt1} into \eqref{e:b2}, we obtain
\begin{align}
 \ol{B^2}&= \sum_{n,m=1}^N  \frac{1}{r^2}\left[ Q_n Q_m \left( {a'}^{ 2} + {b'}^2 \right)+ Q'_n Q'_m {a'}^2 \right] q_n q_m w_n w_m\nonumber\\
 &=\sum_{n,m,l,k=1}^N  \frac{1}{r^2}\left[ Q_l^s Q_k^s \left( {a'}^2 + {b'}^2 \right)+ Q_l^{'s} Q_k^{'s} {a'}^2 \right] \mathcal{C}_{nl} \mathcal{C}_{mk} w_n w_m  \label{mb2}
\end{align}
For the mean current density, we can write
\begin{equation}
 \mbf{J}=\nabla\times\mbf{B}=\nabla\times \mathbf{\hat{P}} \psi + \nabla\times\left(\frac{T}{r}\hat{\phi}\right)=-\left(\frac{1}{r}\Lambda\psi\right)\hat{\phi} +\mathbf{\hat{P}}T, \label{e:j}
 \end{equation}
where we have used $\disp{\nabla\times \mathbf{\hat{P}}=-\hat{\phi}\frac{\Lambda}{r}}$ and $\disp{\nabla\times\left(\frac{T}{r}\hat{\phi}\right)=-\frac{1}{r}\fpar{T}{z}\hat{r}+\frac{1}{r}\fpar{T}{r}\hat{z}}=\mathbf{\hat{P}} T$. Combining Equations (\ref{e:b1}) and (\ref{e:j}), we get
\begin{equation}
 \ol{\mathbf {J}\cdot \mathbf{B}}= \mathbf{\hat{P}} T \cdot \mathbf{\hat{P}} \psi -\frac{1}{r^2} (\Lambda \psi)T. \label{j2}
\end{equation}
The first term on the rhs of Equation \eqref{j2} is given by
\begin{align}
\mathbf{\hat{P}} T \cdot \mathbf{\hat{P}} \psi&=\frac{1}{r^2}\left(\partial_z T \partial_z \psi + \partial_r T \partial_r \psi\right)= \sum_{n,m=1}^N  \frac{1}{r^2} \left(Q_n Q_m a' b'+ Q'_n Q'_m a b\right)q_n q_m w_n w_m \nonumber\\
&=\sum_{n,m,l,k=1}^N  \frac{1}{r^2} \left(Q_l^s Q_k^s a' b'+ Q_l^{'s} Q_k^{'s} a b\right)\mathcal{C}_{nl} \mathcal{C}_{mk} w_n w_m \label{p0}
\end{align}
The second term on the rhs of Equation \eqref{j2} is given by
\begin{equation}
-\frac{1}{r^2}(\Lambda \psi)T=\frac{1}{r^2}\left[r \partial_r \left(\frac{1}{r} \partial_r \psi\right)+\partial_z^2 \psi \right]T.\label{j3}
\end{equation}
The first term inside the brackets on the rhs of Equation \eqref{j3} can be written as
\begin{equation}
r\partial_r \left(\frac{1}{r}\partial_r \psi \right)=\sum_{n=1}^N r \frac{\dif}{\dif r} \left(\frac{1}{r}\frac{\dif Q_n}{\dif r}\right)a q_n w_n = \sum_{n=1}^N (\Lambda_r Q_n) a w_n=\sum_{n,l=1}^N -\gamma_l^s \mathcal{C}_{nl} Q_l^s a  w_n. \label{j4}
\end{equation}
 Noting that $\disp{\partial_z^2 \psi = \sum_{n=1}^N q_n Q_n a''w_n}$ and
substituting Equation \eqref{j4} into Equation \eqref{j3}, we get
 \begin{equation}
-\frac{1}{r^2}(\Lambda \psi)T=\sum_{n,m,l,k=1}^N \frac{1}{r^2}\left[\mathcal{C}_{nl} \mathcal{C}_{mk} Q_l^s Q_k^s  w_n w_m b (a''-\gamma_l^s a)\right].\label{j5}
 \end{equation}
Substituting Equations (\ref{p0}) and (\ref{j5}) into Equation \eqref{j2}, we obtain
\begin{equation}
 \ol{\mathbf {J}\cdot \mathbf{B}}= \sum_{n,m,l,k=1}^N \frac{1}{r^2}\left[Q_l^s Q_k^s \left(a' b'_m + a'' b -\gamma_l^s a b\right)
 + Q_l^{'s} Q_k^{'s} a b \right]\mathcal{C}_{nl} \mathcal{C}_{mk} w_n w_m. \label{j6}
\end{equation}

In order to obtain the volume-averaged quantities $\langle \ol{B^2}\rangle$ and $\langle\ol{\mathbf {J}\cdot \mathbf{B}}\rangle$, we note that, since the quantities in Equations (\ref{mb2}) and (\ref{j6}) are separable in variables $r$ and $z$, we can split the volume average into radial averages on functions related to $Q(r)$ multiplied by vertical averages on functions of $a(z)$ and $b(z)$. Using the above relations, we can write the volume-averaged quantities as
\begin{align}
\langle \ol{B^2}\rangle &=\sum_{n,m,l,k=1}^N w_n w_m \mathcal{C}_{nl} \mathcal{C}_{mk} \left[\left \langle \frac{Q_l^s Q_k^s}{r^2}  \right \rangle \langle {a'}^2 + {b}^2 \rangle
+\left\langle \frac{Q_l^{'s} Q_k^{'s}}{r^2} \right\rangle \langle {a}^2 \rangle \right]  \nonumber\\
&=\sum_{n,m,l=1}^N w_n w_m \mathcal{C}_{nl} \mathcal{C}_{ml}\left[J_2^2(\sqrt{\gamma_l^s} r_d) \langle {a'}^2 + {b}^2 \rangle+ \gamma_l^s J_0^2 (\sqrt{\gamma_l^s} r_d) \langle {a}^2 \rangle \right].
\end{align}
\begin{align}
\langle\ol{\mathbf {J}\cdot \mathbf{B}}\rangle&= \sum_{n,m,l,k=1}^N w_n w_m \mathcal{C}_{nl} \mathcal{C}_{mk}\left[\left\langle \frac{Q_l^s Q_k^s}{r^2} \right\rangle \langle a' b' + a'' b -\gamma_l^s a b\rangle +\left\langle \frac{Q_l^{'s} Q_k^{'s}}{r^2} \right \rangle \langle a b \rangle\right] \nonumber\\
 &=\sum_{n,m,l=1}^N w_n w_m \mathcal{C}_{nl} \mathcal{C}_{ml} \left[J_2^2(\sqrt{\gamma_l^s} r_d)\langle a' b' + a'' b -\gamma_l^s a b\rangle +\gamma_l^s J_0^2 (\sqrt{\gamma_l^s} r_d) \langle a b \rangle\right]
\end{align}
where we have used Equation (\ref{bes}) and the orthogonality properties of Bessel functions to write
\begin{align}
 \left \langle \frac{Q_l^s Q_k^s}{r^2}  \right \rangle &= \sum_{l,k=1}^N \frac{2}{r_d^2} \int_0^{r_d} \frac{Q_l^s Q_k^s}{r^2} r\dif r= \sum_{l,k=1}^N \delta_{lk} J_2^2 (\sqrt{\gamma_k^s}r_d)= \sum_{l=1}^N J_2^2 (\sqrt{\gamma_l^s}r_d)\\
 \left\langle \frac{Q_l^{'s} Q_k^{'s}}{r^2} \right \rangle &= \sum_{l,k=1}^N \frac{2}{r_d^2} \int_0^{r_d} \frac{Q_l^{'s} Q_k^{'s}}{r^2} r\dif r= \sum_{l,k=1}^N \delta_{lk}\gamma_k^s J_0^2 (\sqrt{\gamma_k^s}r_d)=\sum_{l=1}^N\gamma_l^sJ_0^2 (\sqrt{\gamma_l^s}r_d)
  \end{align}
and the vertical averaging is defined in the following manner: $\disp{\langle a \rangle =\int_0^1 a(z)~ \dif z}$.
\section{Gauge invariant description of helicity in cylindrical geometry}
\label{AppB}

The Chandrasekhar--Kendall representation of magnetic fields in cylindrical geometry in terms of generating functions $\phi$ and $\psi$ is given by \citep{2006ApJ...646.1288L,2011PhPl...18e2901L}
\begin{eqnarray}
 \mathbf{B}&=&\mathbf{B_\phi}+\mathbf{B_\psi}\\
\mathbf{B_\phi}&=&\nabla\times\phi \hat{z};
\quad\mathbf{B_\psi}= \nabla\times(\nabla\times\psi\hat{z}).\label{bck}
\end{eqnarray}
Then the absolute magnetic helicity density, defined as
\begin{equation}
 h_{abs}(\psi,\phi)=(\nabla\times\psi \hat{z})\cdot[\nabla\times(\nabla\times\psi\hat{z})+2(\nabla\times\phi\hat{z})] \label{habs}
\end{equation}
is a gauge-invariant measure of magnetic helicity density. The magnetic vector potential given as
$ \mathbf{A}=\nabla\times\psi \hat{z}+\phi\hat{z}$
is also well defined. For the case of axisymmetry, we get $\disp{\nabla\times\psi \hat{z}=-\fpar{\psi}{r}\hat{\phi}}$.
Thus, we can write the $\phi$ component of $\mathbf{A}$ as
\begin{equation}
 \mathbf{A}_\phi=\nabla\times\psi\hat{z}=-\fpar{\psi}{r}\hat{\phi}, \label{aphi}
\end{equation}
and rewrite Equation \eqref{habs} as
\begin{equation}
 h_{abs}=\mathbf{A}_\phi \cdot\left(\mathbf{B}_\psi+2\mathbf{B}_\phi\right). \label{habs2}
\end{equation}
Also under axisymmetry, Equation \eqref{bck} can be rewritten as
\begin{equation}
 \mathbf{B}_\phi=-\fpar{\phi}{r}\hat{\phi},\quad \mathbf{B}_\psi=\frac{\partial^2 \psi}{\partial r \partial z} \hat{r}-\frac{1}{r}\fpar{}{r}\left(r\fpar{\psi}{r}\right)\hat{z}\label{psieq2}.
\end{equation}
Thus combining Equations (\ref{aphi}), (\ref{habs2}) and (\ref{psieq2}), we get
\begin{equation}
 \mathbf{A}_\phi\cdot\mathbf{B}_\psi=0; \quad \mathbf{A}_\phi\cdot\mathbf{B}_\phi=A_\phi B_\phi.
\end{equation}
So, we get the final expression for absolute magnetic helicity density for an axisymmetric field in cylindrical geometry as
\begin{equation}
 h_{abs}=2A_\phi B_\phi. \label{habs3}
\end{equation}
Comparing the definition for the field in Equation \eqref{bck} with our definition in Equations (\ref{poleq}) and (\ref{toreq}),
we get $A_\phi=\disp{\frac{\psi}{r}}$ and $B_\phi=\disp{\frac{T}{r}}$. Thus Equation \eqref{habs3}
in our notation takes the following form:
\begin{equation}
 h_{abs}=\frac{2 \psi T}{r^2}.
\end{equation}
The mean magnetic helicity can then be defined as
\begin{equation}
 \ol{H}=\int_V \frac{2 \psi T}{r^2}~ \dif V. \label{hellow}
\end{equation}

\section{Balance of magnetic helicity fluxes and evolution of coronal helicity}
\label{a:corhel}
The mean magnetic helicity within the galactic disk can be written using Equation \eqref{hellow} as
\begin{equation}
\ol{H}_d=\int_V\frac{2\psi T}{r^2} ~\dif V. \label{a:hd1}
\end{equation}
Substituting for $\psi$ and $T$ in Equation \eqref{a:hd1} using Equation \eqref{tds}, we obtain
\begin{align}
\ol{H}_d&=2\pi \int_V\frac{2}{r^2}\left(\sum_{n,m,l,k=1}^N w_n w_m \mathcal{C}_{nl} \mathcal{C}_{mk} Q_l^s Q_k^s a b\right)r~\dif r~\dif z\nonumber\\
&=4\pi\sum_{n,m,l,k=1}^N w_n w_m \mathcal{C}_{nl} \mathcal{C}_{mk} \int_{r=0}^{r_d}r~\mathcal{J}_l^s J_k^s  ~\dif r \int_{z=0}^1 a b ~\dif z\nonumber\\
&=2\pi r_d^2 \sum_{n,m,l=1}^N w_n w_m \mathcal{C}_{nl} \mathcal{C}_{ml} J_2^2(\sqrt{\gamma_l^s} r_d)\langle a b \rangle.\label{a:Hdf}
\end{align}
For calculating the mean magnetic helicity in the corona, we substitute Equation \eqref{e:psicor} into Equation \eqref{corhel}, 
\begin{align}
 \ol{H}_c=\sum_{n,m,l,k=1}^N &4 \pi \mu w_n w_m \mathcal{C}_{nl} \mathcal{C}_{mk} a^2(1)\int_0^{r_d} J_1(\sqrt{\gamma_l^s}r) J_1(\sqrt{\gamma_k^s}r) r \dif r \nonumber\\
 &\int_1^\infty \exp\left[\left(\sqrt{\gamma_l^s-\mu^2}+\sqrt{\gamma_k^s-\mu^2}\right)(1-z)\right] \dif z  \label{e:hceq1}
\end{align}
Using the orthogonality property of Bessel functions given in Equation \eqref{besf}, we can simplify Equation \eqref{e:hceq1} as
\begin{align}
 \ol{H}_c=\sum_{n,m,l=1}^N 2 \pi \mu r_d^2 J_2^2(\sqrt{\gamma_l^s}r_d) w_n w_m \mathcal{C}_{nl} \mathcal{C}_{ml} a^2(1)\exp[2\sqrt{\gamma_l^s-\mu^2}] \int_1^\infty \exp[-2\sqrt{\gamma_l^s-\mu^2}z] \dif z.  \label{e:hceq2}
\end{align}
Upon evaluating the $z$ integral in Equation \eqref{e:hceq2}, we obtain
\begin{equation}
 \ol{H}_c=\sum_{n,m,l=1}^N \pi \mu r_d^2 J_2^2(\sqrt{\gamma_l^s}r_d)w_n w_m \mathcal{C}_{nl}\mathcal{C}_{ml}\frac{a^2(1)}{\sqrt{\gamma_l^s-\mu^2}}. \label{e:hc2}
\end{equation}
The expression for the small-scale magnetic helicity density in the disk is \citep{2007MNRAS.377..874S}
$\disp{ \chi= \frac{l_0^2 B_{eq}^2 \alpha_m}{\eta_t}}.$ Rescaling the expression for $\chi$, we can write
\begin{align}
 B_{eq}^2 h \tilde{\chi}&= \frac{l_0^2 B_{eq}^2 \alpha_0}{\eta_t}\tilde{\alpha_m}\nonumber\\
\Rightarrow \tilde{\chi}&=\frac{ l_0^2}{h^2}\left(\frac{\alpha_0 h}{\eta_t}\right)\tilde{\alpha_m}
=\frac{2}{C}  R_\alpha\tilde{\alpha_m}.
\end{align}
Dropping the tilde for simplicity, we get $\disp{\chi= \frac{2}{C} R_\alpha\alpha_m.}$
The small-scale magnetic helicity within the disk can be estimated by integrating $\chi$ over the volume:
\begin{equation}
 h_d=\int_V \chi ~\dif V=\frac{2}{C} R_\alpha\alpha_m V=\frac{2\pi r_d^2 }{C}R_\alpha\alpha_m. \label{a:hcf}
\end{equation}

The conservation of total magnetic helicity for disk and corona combined together can be written using Equations (\ref{e:htot1}), (\ref{a:Hdf}), (\ref{a:Hcf}), and (\ref{a:hcf}) as
\begin{align}
&H_0=\ol{H}_d+h_d+\frac{\ol{H}_c}{ R_c}\label{a:htot1}\\
&=2\pi r_d^2\left[\sum_{n,k,l=1}^N w_n w_m \mathcal{C}_{nl} \mathcal{C}_{kl} J_2^2(\sqrt{\gamma_l^s} r_d)\left(\langle a b \rangle+\frac{\mu a^2(1)}{R_c \sqrt{\gamma_l^s-\mu^2}} \right)+\frac{1}{C} R_\alpha\alpha_m\right]
\end{align}
where $H_0$ is the initial magnetic helicity of the system contributed entirely by the
mean field in the disk. Differentiating Equation \eqref{a:htot1} with respect to time, we get the following equation for the rate of change of large-scale magnetic helicity in the corona:
\begin{equation}
 \frac{\dif H_c}{\dif t}=-R_c\left(\frac{\dif \ol{H}_d}{\dif t}+\frac{\dif h_d}{\dif t}\right) \label{a:dhdt1}
\end{equation}
The terms inside the bracket on the rhs of Equation \eqref{a:dhdt1} represent the rate of change of total magnetic helicity of the disk and can be written using Equations (\ref{H}) and (\ref{h}) as 
\begin{equation}
\frac{\dif H_d}{\dif t}= \frac{\dif \ol{H}_d}{\dif t}+\frac{\dif h_d}{\dif t}=-2 \int_V \eta \mbf{J}\cdot\mbf{B}~ \dif V- 2 \int_V \eta \ol{\mathbf{j}\cdot\mathbf{b}}~ \dif V -\oint_S \mathbf{f}\cdot \hat{n}~\dif S, \label{a:dhdt2}
\end{equation}
where we have neglected the large-scale magnetic helicity flux. 
Since the microscopic resistivity $\eta$ is small, the first two terms on the rhs of Equation \eqref{a:dhdt2} are negligible compared to the third term and can be dropped from the equation. The rate of change of the large-scale magnetic helicity in the corona can then be written by combining Equations (\ref{a:dhdt1}) and (\ref{a:dhdt2}) and using Equations (\ref{calf})$-$(\ref{rk}) to write the flux terms. We then get 
\begin{equation}
 \frac{\dif \ol{H}_c}{\dif t}= R_c l_0^2 B_{eq}^2 \int_V \left(\frac{1}{\eta_t}\nabla\cdot\calf\right)\dif V. \label{hce1}
\end{equation}
Now we can write 
\begin{equation}
 \frac{1}{\eta_t}\nabla\cdot \calf=\nabla\cdot\left(\frac{\calf}{\eta_t}\right)-\calf\cdot\nabla\left(\frac{1}{\eta_t}\right)=\nabla\cdot\left(\frac{\calf}{\eta_t}\right) \label{hce11}
\end{equation}
where we have dropped the term containing the radial derivative of $1/\eta_t$, as it is negligible compared to the other term containing $z$ derivatives of $\calf$. Now we can write Equation \eqref{hce1} using Equation \eqref{hce11} as
\begin{equation}
  \frac{\dif \ol{H}_c}{\dif t}= R_c l_0^2 B_{eq}^2 \int_V \nabla \cdot \left(\frac{\calf}{\eta_t}\right)~\dif V
=R_c l_0^2 B_{eq}^2\int_S \left(\frac{\calf}{\eta_t}\right)\cdot \hat{z}\dif S \label{hce2}
\end{equation}
where $S$ represents the top surface of the disk. Using Equations (\ref{calf}), (\ref{ru}) and (\ref{rk}), we can write
\begin{equation}
\calf \cdot \hat{z}=U_z \alpha_m-\kappa\partial_z\alpha_m. 
\end{equation}
 Also under the no-$z$ approximation, we write $\partial_z\alpha_m=-\alpha_m/h $; Equation \eqref{hce2} now becomes
\begin{equation}
  \frac{\dif \ol{H}_c}{\dif t}= R_c l_0^2 B_{eq}^2 \int_S \left(\frac{U_z \alpha_m}{\eta_t}+\frac{\kappa \alpha_m}{\eta_t h}\right) \dif S. \label{hce3}
\end{equation}
Rescaling Equation \eqref{hce3} using the transformations given in Equation \eqref{scale}, we write
\begin{equation}
 \frac{\eta_t(h)}{h^2} B_{eq}^2h^4 \frac{\dif \ol{\tilde{H}_c}}{\dif \tau}=R_c l_0^2 B_{eq}^2 \int_S \left[\frac{U_z h}{\eta_t(r)} +\frac{\kappa}{\eta_t(r)}\right]\left(\frac{\alpha_0 h}{\eta_t(r)}\right)\tilde{\alpha}_m\eta_t(r)\dif \tilde{S} \label{hce5}
\end{equation}
where $\tilde{S}=S/h^2$ and $\ol{\tilde{H}_c}=\ol{H}_c/(B_{eq}^2 h^4)$. Simplifying Equation \eqref{hce5} and using the definitions given in Equations (\ref{dim}) and (\ref{cru}), we get
\begin{equation}
 \frac{\dif \ol{\tilde{H}_c}}{\dif \tau}= \left(\frac{2}{C}\right) R_c  \int_S \left(\frac{\eta_t(r)}{\eta_t (h)}\right) R_\alpha (R_U+R_\kappa)\tilde{\alpha}_m \dif \tilde{S} \label{hce6}
\end{equation}
Dropping the tilde in Equation \eqref{hce6} for clarity and using Equation \eqref{e:erat}, we get the final equation for the rate of change of large-scale helicity in the corona as
\begin{align}
\frac{\dif \ol{H}_c}{\dif \tau}=-R_c\frac{\dif H_d}{\dif \tau}=R_c\frac{4\pi  r_d}{C} R_\alpha (R_U + R_\kappa) \alpha_m.
\end{align}

\bibliography{ms5a}
\end{document}